\input amstex
\input amsppt.sty
\magnification 1200

\voffset=-1cm
\hoffset=0.5cm
%\vsize 19truecm

\TagsOnRight
\NoBlackBoxes
\NoRunningHeads

\define\R{\Bbb R}
\define\Z{\Bbb Z}
\define\C{\Bbb C}
\define\h{\Bbb H}

\define\al{\alpha}
\define\be{\beta}  
\define\ga{\gamma}
\define\la{\lambda}
\define\Om{\Omega}
\define\om{\omega}
\define\x{\goth X}

\define\inr{\operatorname{in}}
\define\out{\operatorname{out}}
\define\xin{\x_{\inr}}
\define\xout{\x_{\out}}
\define\Dim{\operatorname{Dim}}
\define\psiin{\psi_{\inr}}
\define\psiout{\psi_{\out}}
\define\Rout{R_{\out}}
\define\Sout{S_{\out}}
\define\Rin{R_{\inr}}
\define\Sin{S_{\inr}}
\define\wt{\widetilde}

\define\const{\operatorname{const}}

\define\Ma{\operatorname{Mat}}

\define\Ga{\Gamma}
\define\s{\frak S}
\define\ka{\varkappa}

\define\sgn{\operatorname{sgn}}

\define\tht{\thetag}
\define\A{\Cal A}

\define\ze{\zeta}
\define\F{\Cal F}
\define\G{\Cal G}
\define\tr{\operatorname{Tr}}
\define\bb{\Cal B}%{B_{-1/2}}
\define\ba{\Cal A}%{B_{1/2}}
\define\bc{\Cal C}%{B_s}

\define\sig{\sigma_3}

\define\si{\sigma}
\define\tha{\theta_{\ba}}
\define\thb{\theta_{\bb}}
\define\xa{x_{\ba}}
\define\ya{y_{\ba}}
\define\za{z_{\ba}}
\define\xc{x_{\bc}}
\define\yc{y_{\bc}}
\define\zc{z_{\bc}}
\define\hsi{\hat\si}
\define\up{\Upsilon}
\define\diag{\operatorname{diag}}
\define\r{\operatorname{Res}\limits_{\zeta=x}}

\topmatter

\author Alexei Borodin and Percy Deift
\endauthor

\title Fredholm determinants, Jimbo-Miwa-Ueno tau-functions, and
representation theory
\endtitle

\abstract
The authors show that a wide class of Fredholm determinants arising in the representation theory of ``big'' groups such as the infinite--dimensional unitary group, solve Painlev\'e equations. Their methods are based on the theory of integrable operators and the theory of Riemann--Hilbert problems. 
\endabstract

%\date Preliminary version. December 5, 2000 \enddate

\toc
\widestnumber\head{??}
\head {} Introduction \endhead
\head 1. Harmonic analysis on the infinite-dimensional unitary group\endhead
\head 2. Continuous ${}_2F_1$ kernel. Setting of the problem \endhead
\head 3. The resolvent kernel and the corresponding Riemann-Hilbert
         problem\endhead 
\head 4. System of linear differential equations with rational coefficients
         \endhead 
\head 5. General setting \endhead
\head 6. Isomonodromy deformations. Jimbo-Miwa-Ueno $\tau$-function. \endhead
\head 7. Painlev\'e VI.\endhead
\head 8. Other kernels \endhead
\subhead 8.1. The Jacobi kernel \endsubhead
\subhead 8.2. The Whittaker kernel \endsubhead
\subhead 8.3. The confluent hypergeometric kernel \endsubhead
\head 9. Differential equations: a general approach\endhead
\head {} Appendix. Integrable operator and Riemann-Hilbert problems \endhead 
\head {} References \endhead
\endtoc

\endtopmatter
\NoRunningHeads
\document
\head Introduction 
\endhead

Consider the kernel
$$
K(x,y)=\frac{A(x)B(y)-B(x)A(y)}{x-y}\, \sqrt{\psi(x)\psi(y)},\quad
x,y\in\left(\tfrac 12,+\infty\right), 
\tag 0.1
$$
where 
$$
\gathered
\psi(x)=\frac{\sin \pi z\sin\pi z'}{\pi^2}\cdot 
\left(x-\tfrac 12\right)^{-z-z'}\left(x+\tfrac 12\right)^{-w-w'},\\
A(x)=\left(\frac{x+\frac 12}{x-\frac 12}\right)^{w'}\,
{}_2F_1\left[\matrix 
z+w',\,z'+w'\\ z+z'+w+w'\endmatrix\,\Biggl|\,
\frac 1{\frac 12-x}\right]\,,\\
B(x)=\frac{\Ga( z+w+1)\, 
\Ga(z+w'+1)\,\Ga(z'+w+1)\,\Ga(z'+w'+1)}{\Ga(z+z'+w+w'+1)\,\Ga(z+z'+w+w'+2)}\\
\times  \frac 1{x-\frac 12}\,\left(\frac{x+\frac 12}{x-\frac 12}\right)^{w'}\,
{}_2F_1\left[\matrix z+w'+1,\,z'+w'+1\\ 
z+z'+w+w'+2\endmatrix\,\Biggl|\,\frac 1{\frac 12 -x}\right]\,.
\endgathered
$$
Here ${}_2F_1\left[\matrix a,\,b\\ c\endmatrix\,\Bigl|\,\ze\right]$ stands for the Gauss hypergeometric function, and $z,z',w,w'$ are some complex numbers. We call $K(x,y)$ the {\it continuous ${}_2F_1$ kernel} or simply the {\it ${}_2F_1$ kernel}.

The basic problem considered in this paper is the derivation of an ordinary differential equation for the
Fredholm determinant $D(s)=\det(1-K|_{(s,+\infty)})$.

This kernel originates in the representation theory of the 
infinite--dimensional unitary group $U(\infty)$. Briefly, decomposition of a certain natural representation of $U(\infty)$ into irreducibles is described by a
probability measure on the infinite-dimensional space of all irreducible
representations; a projection of this measure onto a 1--dimensional subspace has
the distribution function equal to $D(s)=\det(1-K|_{(s,+\infty)})$, where $K$ is as
above. The study of this representation theoretic problem is the main subject
of the two recent papers \cite{Ol2}, \cite{BO5}. For a more detailed
description of the problem and the results in these papers the reader is referred to \S1 below.

The problem of deriving differential equations for determinants of the form $D(s)$ as above, has a long history. In their pioneering work \cite{JMMS} in 1980, M.~Jimbo, T.~Miwa, Y.~Mori,
and M.~Sato considered the so-called sine kernel which has the form \tht{0.1}
with $\psi(x)=1/\pi$, $A(x)=\sin x$, $B(x)=\cos x$. They showed that the
determinant of the identity operator minus this kernel restricted to an
interval of varying length $s$  can be
expressed through a solution of the {\it Painlev\'e V}
equation. Their proof was based
on the theory of isomonodromy deformations of linear systems of differential
equations with rational coefficients. This theory in turn goes back to the work of
Riemann, Schlesinger, Fuchs, Garnier, and others. \cite{JMMS} used the results of
\cite{JMU} and \cite{JM}, where the theory of isomonodromy deformations was
developed in a setting more general than in the classic papers mentioned above. Along with the one interval case, \cite{JMMS} also considered the
restriction of the sine kernel to a union of a finite number of intervals.
They showed that the corresponding Fredholm determinant, as a function of the
endpoints of the intervals, is a {\it $\tau$-function} (in the sense of
\cite{JMU}) of the corresponding isomonodromy problem. In other words, it can
be expressed through a solution of a ``completely integrable'' system of partial
differential equations called the {\it Schlesinger equations}. 

Kernels of the form \tht{0.1} are of great interest in random matrix
theory. Indeed, the Fredholm determinant related to the kernel
\tht{0.1} restricted to a domain $J$, with $A$ and $B$ being $n$th and
$(n-1)$st orthogonal polynomials with the weight function $\psi$, measures the
probability of having no particles in $J$ for certain $n$-particle
systems called {\it orthogonal polynomial ensembles}. Such systems describe
the spectra of random unitary and Hermitian matrices. We refer the reader to
\cite{Me1} for details.

The results of \cite{JMMS} attracted considerable attention in the random
matrix community. In 1992 M.~L.~Mehta \cite{Me2} rederived the Painlev\'e V
equation for the sine kernel. Approximately at the same time, C.~Tracy and H.~Widom
\cite{TW1} gave their own derivation of this result. Moreover, they
produced a general algorithm (see \cite{TW4}) to obtain a system of partial differential equations for a Fredholm
determinant associated with a kernel of type \tht{0.1} restricted to a union of intervals, in the case that
the functions $\psi,A,B$ satisfy a differential equation of the form
$$
\frac{d}{dx}\bmatrix \sqrt{\psi(x)} A(x)\\\sqrt{\psi(x)} B(x)\endbmatrix=
R(x)\bmatrix \sqrt{\psi(x)} A(x)\\\sqrt{\psi(x)} B(x)\endbmatrix,
\tag 0.2
$$
where $R(x)$ is a traceless rational $2\times 2$ matrix.
Using their method, they derived different Painlev\'e equations for a number
of kernels relevant to random matrix theory \cite{TW1}--\cite{TW4}.

Shortly after, J.~Palmer \cite{Pal} showed that the partial 
differential equations arising in the Tracy-Widom method are precisely
the Schlesinger equations for an associated isomonodromy problem.

Among more recent papers, we mention (in no particular order) the 
works \cite{AsvM}, \cite{AvM}, where a different approach to the kernels arising from matrix models can be found, the paper \cite{HS}, where the Painlev\'e VI equation for the Jacobi kernel was derived, the paper \cite{DIZ2}, where the theory of Riemann--Hilbert problems was applied to derive the Schlesinger
equations for certain kernels and to analyze the asymptotics of solutions, the paper \cite{HI}, where a multidimensional analog of the sine kernel was treated using the isomonodromy deformation method, and the papers \cite{FW}, \cite{WF}, \cite{WFC}, where, in particular, a two-interval situation was reduced to an ordinary differential equation in one variable.

Returning to our specific ${}_2F_1$ kernel, we find that
our functions $\psi, A, B$ satisfy an equation of the form \tht{0.2} (see
Remark 4.8 below). 

However, the method in \cite{TW4} leads in our case to considerable 
algebraic complexity, and we have not been able to see our way through the calculation. A similar situation arose in the case of the (simpler)
Jacobi kernel, for which the method in \cite{TW4} leads to a third order 
differential equation. This equation was shown to be equivalent to the
(second order) Painlev\'e VI equation only in the later work of Haine and
Semengue \cite{HS}. In the face of these difficulties, we decided to look
for a different approach.

The representation theoretic origin of the ${}_2F_1$ kernel suggests 
a new approach. It turns out that the construction of the kernel $K$, see \S1, strongly indicates that $K$ should have a ``simple'' resolvent kernel $L=K(1-K)^{-1}$. ``Simple'' in the sense that the formula for $L(x,y)$ should not involve any special functions! At the formal level
``$\det(1-K)=(\det(1+L))^{-1}$''. However, we are interested in the
restricted operator $K|_{(s,+\infty)}$, and it is not at all clear that the ``simple'' kernel $L$ can be used in any way to compute 
$D(s)=\det(1-K|_{(s,+\infty)})$. It is the {\it basic observation} of this paper that the kernel $L$ can indeed be used to compute $D(s)$, and 
this leads, as we will see, to the desired differential equations.

In the analysis that follows, a crucial fact is that both kernels 
$K$ and $L$ are {\it integrable} in the sense of \cite{IIKS}. We refer the reader to the Appendix for the definition and basic properties
of integrable operators and also for the definition of a Riemann--Hilbert Problem (RHP).
Our method is as follows (see \S5). 

\smallskip

\noindent{\bf Step 1.} The kernel $K(x,y)$ is expressed through an
explicit solution of a RHP $(\R,v)$, where $v$ comes from $L$ and is
``simple''.

\smallskip

\noindent{\bf Step 2.} $D(s)=\det(1-K|_{(s,+\infty)})$ is expressed through the solution $m_s$ of a normalized RHP $((s,+\infty),\hat v)$,
where $\hat v$ involves special functions as in \tht{0.1}.

\smallskip

\noindent{\bf Step 3.} The product $m_sm$ satisfies the RHP
$(\R\setminus (s,+\infty)=(-\infty,s],v)$, where $v$ is again the ``simple'' jump matrix occurring in Step 1. 

\smallskip

Step 3, which is the key fact, is a consequence of the theory of
integrable operators and the following elementary observation: 
let $\Sigma=\Sigma_1\cup\Sigma_2\subset \C$ be a union of two contours.
Let $m,m_1$ be solutions of the RHP's $(\Sigma, v)$, $(\Sigma_1,v)$,
respectively. Then $m_2\equiv m_1m^{-1}$ solves the RHP 
$(\Sigma_2, v_2=m_+v^{-1}m_+^{-1}=m_-v^{-1}m_-^{-1})$. Conversely, if $m,m_2$ are solutions of the RHP's $(\Sigma, v)$, $(\Sigma_2,v_2)$, then
$m_1=m_2m$ solves the RHP $(\Sigma_1,v)$.

As we will see, if $v$ is the jump matrix associated via the theory of
integrable operators with the kernel $L$ , then 
$m_+v^{-1}m_+^{-1}=m_-v^{-1}m_-^{-1}$ is the jump matrix $\hat v$ associated with the kernel $K$. 

In the RH framework, differential equations are deduced from the fact that the jump matrix for the problem at hand can be conjugated to a form
which does not depend on the parameters relevant for the problem. 
A prototypical calculation, which can be traced essentially to the
beginning of the inverse scattering theory, is as follows (see, e.g., 
\cite{DIZ1} and references therein). The defocusing Nonlinear Schroedinger 
(NLS) equation is associated with the RHP 
$(\Sigma=\R, v_{x,t}=e^{i\theta\sigma_3}\,v\, e^{-i\theta \sigma_3})$ where
$$
\theta=x\ze-t\ze^2,\quad \sigma_3=\bmatrix 1&0\\0&-1\endbmatrix,
\quad v=\bmatrix 1-|r(\ze)|^2&r(\ze)\\-\bar{r}(\ze)&1\endbmatrix
$$
for some {\it reflection coefficient} $r$. If $m$ is a solution
of $(\R,v_{x,t})$, then $\Psi=me^{i\theta\sigma_3}$ solves the RHP
$(\R,v)$ which is independent of $x$ and $t$. It follows that
$\frac{\partial \Psi}{\partial x}$ and $\frac{\partial \Psi}{\partial t}$
solve the same RHP and hence 
$\frac{\partial \Psi}{\partial x}\,\Psi^{-1}$ and 
$\frac{\partial \Psi}{\partial t}\,\Psi^{-1}$ have no jump across
$\Sigma=\R$. A short calculation then leads to the {\it Lax pair}
$\frac{\partial \Psi}{\partial x}=P\Psi$, 
$\frac{\partial \Psi}{\partial x}=L\Psi$ for some polynomial matrices
$P=P(\ze)$, $L=L(\ze)$. Cross--differentiation 
$\frac{\partial}{\partial t}\frac{\partial \Psi}{\partial x}=
\frac{\partial}{\partial x}\frac{\partial \Psi}{\partial t}$
then leads to the NLS equation.

As we will see in \S4, the jump matrix $v$ in Steps 1 and 3 is easily 
conjugated to a jump matrix $V$ which is piecewise constant. In the
spirit of the above calculation for NLS, this means that a solution
$M$ of the RHP $M_+=M_-V$ can be differentiated with respect to the 
variable $\ze$ on the contour, and also with respect to $s$, leading as 
above to the relations of the form 
$\frac{\partial M}{\partial \ze}=PM$, $\frac{\partial M}{\partial s}=LM$
where $P=P(\ze)$ and $L=L(\ze)$ are now rational. Cross--differentiation 
then leads to a set of differential relations. In order to extract
specific equations, such as PVI for $D(s)=\det(1-K|_{(s,+\infty)})$, we
recall the result in \cite{Pal}. As $V$ is piecewise constant, the
above equations 
$\frac{\partial M}{\partial \ze}=PM$, $\frac{\partial M}{\partial s}=LM$
describe an isomonodromy deformation, and hence one can construct an associated tau--function $\tau=\tau(s)$ as in \cite{JMU}. A separate
calculation (\S6) shows that in fact $D(s)=\tau(s)$, and PVI follows
using calculations similar to those as in \cite{JM, Appendix C}.

The above calculations generalize immediately to the case where the interval $(s,+\infty)$ is replaced by a union of intervals $J$.
 
The idea of reducing the Riemann-Hilbert problem for $m_s$ to a problem with a piecewise constant jump matrix has been used recently in \cite{Pal}, \cite{HI}, \cite{DIZ2}, \cite{KH}, see
also \cite{Its}. However, the method outlined above
 of performing the reduction seems to be new. 

As noted above, the property of the kernel $K$ which is important for us, is the existence of a simple resolvent kernel $L=K(1-K)^{-1}$. This property seems to be new and was first observed in the context of the 
representation theory of the infinite symmetric group $S(\infty)$ in
\cite{BO1}. In random matrix theory the operators $K$ which arise are
projection operators (of Christoffel--Darboux type), or their scaling limits. All these kernels have norm 1 and hence the operator $L=K(1-K)^{-1}$ is not defined. However, our problem has a
different origin
which makes it possible not only to define $L$, but also to express it
in an explicit way \cite{BO1}, \cite{BO5}. 

The method that we introduce can be used to recover the results in \cite{TW4} for integrable operators with entries satisfying equations
of type \tht{0.2}. We will illustrate the situation in the specific case of the Airy kernel in \S9.  

In the remainder of the paper we consider a variety of kernels similar to \tht{0.1}. 

Firstly, we apply our methods to the Jacobi kernel and we prove that the determinant of the identity minus the Jacobi kernel restricted to a finite union of intervals is the $\tau$-function of the corresponding isomonodromy problem. For the one interval case we again get the Painlev\'e VI equation, reproving the result of \cite{HS}.

Secondly, we  apply our formalism to the so-called Whittaker kernel and its
special case -- the Laguerre kernel. The Whittaker kernel appeared in the
works \cite{P.I-P.V}, \cite{BO1}, \cite{Bor1} on the representation theory of the infinite symmetric group. The calculations for the
${}_2F_1$ kernel are applicable to (the simpler case of) the Whittaker
kernel. We prove that the Fredholm determinant of
the Whittaker kernel on a union of intervals is a $\tau$-function of an
isomonodromy problem, and we derive Painlev\'e V in the one interval case. This last result was proved in \cite{Tr}, and in
\cite{TW4} for the special case of the Laguerre kernel. 

Finally, we observe that the ${}_2F_1$ kernel degenerates  in a certain limit to a
kernel which we call the {\it confluent hypergeometric kernel}. This kernel appears in
a problem of decomposing a remarkable family of probability measures on the
space of infinite Hermitian matrices on ergodic components, see
\cite{BO4}. It can also be obtained as a scaling limit of
Christoffel-Darboux kernels for the so-called pseudo-Jacobi orthogonal
polynomials, see \cite{WF}, \cite{BO4}. We show that the Fredholm
determinant in the one interval case for this kernel can be expressed in terms of 
a solution of the Painlev\'e V equation. The confluent hypergeometric kernel
depends on 1 complex parameter $r$, and for real values of $r$ the last
result was proved in \cite{WF}. For $r=0$ the kernel turns into the sine
kernel, which recovers the original result of \cite{JMMS}. 

The paper is organized as follows. In \S1 we describe the  representation
theoretic origin of the problem. In \S2 we introduce the ${}_2F_1$ kernel and
study its properties. In \S3 the resolvent kernel $L$ is defined, and the
matrix $m$ in Step 1 above is considered. In \S4 we derive the Lax
pair for $M$ as above. In \S5 we describe the general setting
in which our method is applicable. The reader interested primarily in the derivation of the differential equations might want to start
reading the paper with this section. In \S6 we prove that the Fredholm
determinants of kernels, satisfying the general conditions of \S5, are
$\tau$-functions of associated isomonodromy problems. In \S7 we solve our
initial problem: the Painlev\'e VI equation for $\det(1-K|_{(s,+\infty)})$ is
derived. \S8 deals with the applications of our method to the Jacobi,
Whittaker, and confluent hypergeometric kernels. \S9 presents a general approach to kernels of the form \tht{0.1} subject to \tht{0.2}, worked out in the case of the Airy kernel.
Finally, the Appendix
contains a brief description of the formalism of integrable operators and
Riemann-Hilbert problems. 

A discrete version of many of the results in this paper is given in \cite{Bor3}.

\subhead Acknowledgments \endsubhead
The authors would like to thank A.~Kitaev for important discussions about this work and O.~Costin and R.~D.~Costin for informing us of their calculations on Panilev\'e VI. The authors would also like to thank A.~Its and G.~Olshanski
for many useful discussions. 
This research was partially conducted during the period the first author
served as a Clay Mathematics Institute Long-Term Prize Fellow.
The work of the first author was also supported in part by the NSF grant
\#DMS-9729992, and the work of the second author was supported in part by
the NSF grant \#DMS-0003268.

\head 1. Harmonic analysis on the infinite-dimensional unitary group
\endhead

By a character of a (topological) group $K$ (in the sense of von Neumann) we
mean any central (continuous) positive definite function $\chi$ on $K$
normalized by the condition $\chi(e)=1$. Recall that centrality
means $\chi(gh)=\chi(hg)$ for any $g,h\in K$, and positive definiteness means
$\sum_{i,j} z_i\bar{z}_j\,\chi(g_ig_j^{-1})\ge 0$ for any $z_i\in \C$, $g_i\in K$, $i=1,\dots,n$. The characters form a convex set. The
extreme points of this set are called indecomposable characters, and the other
points are called decomposable characters.

The characters of $K$ give rise to representations in
two ways.

Through the Gelfand--Naimark--Segal construction each character $\chi$ determines a unitary representation of
$K$ which will be denoted as $\Pi(\chi)$. When $\chi$ is
indecomposable, $\Pi(\chi)$ is a {\it factor\/} representation of finite type in the sense of von Neumann, see \cite{Th}. Recall that
$\Pi(\chi)$ is a factor representation means that if $S$ commutes with $\{\Pi(\chi)(g),\ g\in K\}$ and $S$ lies in the weak closure $W$ of   
$\{\Pi(\chi)(g),\ g\in K\}$, then $S$ is a multiple of the identity. Finite type means that $W$ carries a finite trace function.

Alternatively (see \cite{Ol1}), set $G=K\times K$ and let $\diag K$
denote the diagonal subgroup in $G$, which is isomorphic to $K$. We
interpret $\chi$ as a function on the first copy of $K$ in $G$, and then extend
it to the whole group $G$ by the formula $$
\psi(g_1,g_2)=\chi(g_1g_2^{-1}), \qquad (g_1,g_2)\in K.
$$
Note that $\psi$ is the only extension of $\chi$ that is a
$\diag K$-biinvariant function on $G$. The function $\psi$ is also positive
definite, so the GNS construction assigns to it a unitary
representation which we will denote by $T(\chi)$. By its very construction, it
possesses a distinguished $\diag K$-invariant vector.

If $\chi$ is indecomposable
 then $T(\chi)=T(\chi^{(\omega)})$ is
irreducible. The representations of the form $T(\chi)$ with indecomposable
$\chi$'s are exactly the irreducible unitary representations of the group $G$
possessing a $K$-invariant vector. See \cite{Ol1} for details.

If $K$ is a finite or compact group then the indecomposable characters of $K$
are all of the form 
$$
\chi^{\pi}(g)=\frac {\tr(\pi(g))}{\dim \pi},
\tag 1.1
$$
where $\pi$ is an irreducible (finite-dimensional) representation of $K$, and
$\dim \pi$ is its dimension. Moreover, any character can be written in a unique
way as a convex linear combination of indecomposable ones:
$$
\chi(g)=\sum_{\pi\in\operatorname{Irr}(K)} P(\pi)\,\chi^{\pi}(g),\qquad
P(\pi)\ge 0,\quad  \sum_{\pi\in\operatorname{Irr}(K)} P(\pi)=1.
\tag 1.2
$$

If $\chi=\chi^\pi$ is of the form \tht{1.1} with an irreducible $\pi$ then
$\Pi(\chi)=\pi$, and $T(\chi)=\pi\otimes \overline{\pi}$, where
$\overline{\pi}$ denotes the representation conjugate to $\pi$.  If $\pi$ acts in $V$ then $\pi\otimes\overline{\pi}$ acts in $V\otimes V^*\sim\operatorname{End}(V)$, and
$\operatorname{Id}\in\operatorname{End}(V)$ is the $\diag K$-invariant
vector for $T(\chi)$. 

In particular, if $K=U(N)$, the group of $N\times N$ unitary matrices, then the irreducible representations of $K$ are
pa\-ra\-me\-trized by the highest weights (see, e.g., \cite{Zh})
$$
\la=(\la_1\ge\dots\ge\la_N), \quad \la_i\in\Z, \ i=1,\dots,N,
$$
and every character can be written in the form \tht{1.2}
$$
\chi=\sum_{\la_1\ge\dots\ge \la_N} P_N(\la)\chi^\la,
\tag 1.3
$$
where $\chi^\la$ is the normalized (as in \tht{1.1}) character of $U(N)$ corresponding to
$\la$. Note that the coordinates of $\la$ may be negative.

Now let $K=U(\infty)$ be the infinite-dimensional unitary group defined as the
inductive limit of the finite-dimensional unitary groups $U(N)$ with respect
to the natural embeddings $U(N)\hookrightarrow U(N+1)$. Equivalently,
$U(\infty)$ is the group of matrices $U=[u_{ij}]_{i,j=1}^\infty$ such that all
but finitely many off-diagonal entries are zero, all but finitely many
diagonal entries are equal to 1, and $U^*=U^{-1}$.

A fundamental result of the representation theory
of the group $U(\infty)$ is a complete description of indecomposable
characters. They are naturally parameterized by the points 
$$
\om=(\al^+,\be^+,\al^-,\be^-,\ga^+,\ga^-)\in \R^{4\infty+2}
$$
such that
$$
\gathered
\al_1^+\ge\al_2^+\ge\dots\ge 0, \quad \be_1^+\ge\be_2^+\ge\dots\ge 0,\\
\al_1^-\ge\al_2^-\ge\dots\ge 0, \quad \be_1^-\ge\be_2^-\ge\dots\ge 0,\\
\ga^+\ge 0, \quad \ga^-\ge 0,\\
\sum_{i=1}^\infty (\al_i^++\be_i^++\al_i^-+\be_i^-)<\infty, \quad
\be_1^++\be_1^-\le 1.
\endgathered
\tag 1.4
$$
The values of extreme characters are provided by Voiculescu's formulas 
\cite{Vo}. This classification result can be established in two ways:
by reduction to a deep theorem due to Edrei \cite{Ed} about two--sided totally positive sequences, see \cite{Boy} and \cite{VK}, and by applying Kerov--Vershik's asymptotic approach, see \cite{VK} and \cite{OkOl}.

We denote the set of all points $\om$ satisfying \tht{1.4} by $\Om$. The coordinates
$\al^+_i$, $\be^+_i$, $\al^-_i$, $\be^-_i$, $\ga^+$, $\ga^-$ are called the {\it Voiculescu parameters}.

Instead of giving a more detailed description of the indecomposable characters (which is
rather simple and can be found in \cite{Vo}), we will explain why such
parameterization is natural. 
It can be shown that every indecomposable
character $\chi^\om$ of $U(\infty)$ is a limit of indecomposable characters
$\chi^{\la(N)}$ of growing finite-dimensional unitary groups $U(N)$ as
$N\to\infty$. Here $\la(N)=\la_1(N)\ge \dots\ge \la_N(N)$ is a highest weight
of $U(N)$. The label $\om\in \Om$ of the character $\chi^\om$ can be viewed as
a limit of $\la(N)$'s as $N\to\infty$ in the following way. 

We write the set of nonzero coordinates of $\la(N)$ as a union of two sequences
of positive and negative coordinates:
$$
\gathered
\{\la_i(N)\ne 0\}=\la^+(N)\sqcup (-\la^-(N)),\\
\la^+(N)=\{\la_1^+(N)\ge \dots \ge\la_k^+(N)\},\quad
\la^-(N)=\{\la_1^-(N)\ge \dots \ge\la_l^-(N)\},
\endgathered
$$
where $\la_i^+>0$, $\la_i^->0$ for all $i$, and $k$ and $l$ are the numbers of
positive and negative coordinates in $\la(N)$, respectively. Note that $k+l\le
N$. We now regard $\la^+(N)$ and $\la^-(N)$ as Young diagrams (of length $k$
and $l$, respectively), and write them in the Frobenius notation
(see \cite{Mac, \S1} for the definition):
$$
\gathered
\la^+(N)=(p_1^+(N)>p_2^+(N)>\dots\,|\,q_1^+(N)>q_2^+(N)>\dots),\\ 
\la^-(N)=(p_1^-(N)>p_2^-(N)>\dots\,|\,q_1^-(N)>q_2^-(N)>\dots),
\endgathered
$$

Then, if $\chi^\om$ is a limit of $\chi^{\la(N)}$ as $N\to\infty$, we must have
$$
\gathered
\al_i^+=\lim_{N\to\infty}\frac {p_i^+(N)}{N},\quad 
\be_i^+=\lim_{N\to\infty}\frac {q_i^+(N)}{N}, \\
\al_i^-=\lim_{N\to\infty}\frac {p_i^-(N)}{N},\quad 
\be_i^-=\lim_{N\to\infty}\frac {q_i^-(N)}{N}.
\endgathered
\tag 1.5
$$
for all $i=1,2,\dots$, see \cite{VK}, \cite{OkOl}. The parameters $\ga^+$, $\ga^-$ can
also be described in a similar manner. Since we will not be 
concerned with
them, we refer an interested reader to \cite{VK}, \cite{OkOl} for the asymptotic meaning
of $\ga^+$ and $\ga^-$. 

Observe that the condition $\beta_1^++\beta_1^-\le 1$ in \tht{1.4} is now
easily explained --- it follows from the relation $q_1^++q_1^-=k+l-2\le N$. 

The next question that we address is how the characters of $U(\infty)$
decompose in terms of the indecomposable ones.

\proclaim{Theorem 1.1 \cite{Ol2}} Let $\chi$ be a character of $U(\infty)$.
Then there exists a unique probability measure $P$ on $\Om$ such that
$$
\chi=\int_\Om \chi^\om P_\chi(d\om),
\tag 1.6
$$
where $\chi^\om$ is the indecomposable character of $U(\infty)$ corresponding
to $\om\in \Om$. 
\endproclaim

The measure $P_\chi$ is called the {\it spectral measure } of the character
$\chi$. The problem of finding the spectral measure for a given character
$\chi$ is referred to as the {\it problem of harmonic analysis} for $\chi$.

The decomposition \tht{1.6} is the infinite-dimensional analog of \tht{1.3}.

Since the indecomposable characters $\chi^\om$ are limits of the normalized
characters $\chi^\la(N)$ of $U(N)$, it is natural to expect that the measure
$P_\chi$ from Theorem 1.1 can be approximated by discrete measures $P_N$ from
\tht{1.3} as $N\to\infty$. To formulate the exact result we need more notation.

Define $\Om^o$ as the set of points
$\om^o=(\al^+,\be^+,\al^-,\be^-)\in\R^{4\infty}$ satisfying the conditions
\tht{1.4}. There is a natural projection $\Om\to \Om^o$ which consists of
omitting the 2 gammas. Denote by $P^o$ the push-forward of the measure $P$ under this projection. As we will only be concerned with statistical
quantities depending on $\om^o$, and not on $\ga^+,\ga^-$, it is enough to consider $P^o$ instead of $P$.

For every $N=1,2,\dots$ define a map $i_N$ which embeds the set of all highest
weights $\la(N)$ of $U(N)$ into $\Om^o$ as follows.  For
$\la(N)=(\la_1(N)\ge\dots\ge \la_N(N))$, using the above notation, we set
$$
i_N(\la)=\left\{\al_i^+=\frac {p_i^+(N)}N, \ \be_i^+=\frac {q_i^+(N)}N, \ 
\al_i^-=\frac {p_i^-(N)}N, \ 
\be_i^-=\frac {q_i^-(N)}N \right\}\in \Om^o.
$$

\proclaim{Theorem 1.2 \cite{Ol2}} Let $\chi$ be a character of $U(\infty)$,
$\chi_N$ be its restriction to $U(N)$, and
$$
\chi|_{U(N)}=\sum_{\la_1\ge\dots\ge\la_N} P_N(\la)\,\chi^{\la},\qquad
P_N(\la)\ge 0,\quad  \sum_{\la_1\ge\dots\ge\la_N} P_N(\la)=1,
\tag 1.7
$$
be the decomposition of $\chi_N$ on indecomposable characters. Then the
projection $P^o_\chi$ of the spectral measure $P_\chi$ of $\chi$ is the weak
limit of push-forwards of the measures $P_N$ under the embeddings $i_N$. 
In other words, if $F$ is a bounded continuous function on $\Om^o$, then
$$
\lim_{N\to\infty} \sum_{\la_1\ge\dots\ge\la_N} F(i_N(\la))P_N(\la)=
\int_{\om\in\Om^o} F(\om) P^o(dw).
$$
\endproclaim

Now, following \cite{BO5}, we apply the above general theory to a
specific family of decomposable characters of $U(\infty)$ constructed in
\cite{Ol2}. The group $U(\infty)$ does not carry Haar measure, and hence the naive definition of the regular representation fails. The representations in \cite{Ol2} should be viewed as analogs of
the nonexisting regular representation of $U(\infty)$. A beautiful geometric
construction of these representations can be also found in \cite{Ol2}.

For every $N=1,2,\dots$ and a highest weight $\la=(\la_1\ge\dots\ge \la_N)$
set
$$
\gathered
P_N(\la)=c_N\cdot\Dim^2_N(\la)\cdot \prod_{i=1}^N f(\la_i-i),\\
f(x)=\frac
1{\Gamma(z-x)\Gamma(z'-x)\Gamma(w+N+1+x)\Gamma(w'+N+1+x)}\,,\\
c_N=\prod_{i=1}^N\frac{\Gamma(z+w+i)\Gamma(z+w'+i)
\Gamma(z'+w+i)\Gamma(z'+w'+i)\Gamma(i)}{\Gamma(z+z'+w+w'+i)
}\,,
\endgathered
\tag 1.8
$$
where $\Dim_N(\la)$ is the dimension of the irreducible representation of
$U(N)$ corresponding to $\la$,
$$
\Dim_N\la
=\prod_{i\le i<j\le N}\frac{\la_i-i-\la_j+j}{j-i}\,,
$$
see, e.g., \cite{Zh}.
Here $z,z',w,w'$ are complex parameters such that $P_N(\la)>0$ for all $N$ and
$\la$. This implies that

(1) $z'=\overline{z}\in \C\setminus \Z$ or $k<z,z'<k+1$ for some $k\in\Z$;

(2) $w'=\overline{w}\in \C\setminus \Z$ or $l<z,z'<l+1$ for some $l\in\Z$.

We also want the series $\sum_\la P_N(\la)$ to converge, and this condition is
equivalent to the additional inequality
 
(3) $z+z'+w+w'>-1$.

Under these conditions the choice of $c_N$ makes $P_N$ into a
probability distribution.

\proclaim{Theorem 1.3 \cite{Ol2}} Let $z,z',w,w'$ satisfy the conditions
(1)--(3) above. Then there exists a character $\chi=\chi^{(z,z',w,w')}$ of
$U(\infty)$ such that
$$
\chi|_{U(N)}=\sum_{\la_1\ge\dots\ge\la_N}P_N(\la)\chi^\la
$$
with $P_N(\la)$ given by \tht{1.8}.
\endproclaim

In order to describe the spectral measures for $\chi^{(z,z',w,w')}$
we need to switch to a different representation for the $\la$'s. First, we describe the measures
$P_N$ in a different way.

Consider the lattice  
$$
\x^{(N)}=\cases \Z, &N\text{ is odd},\\
                   \Z+\frac 12, &N \text{ is even},
		  \endcases
$$
and divide it into two parts
$$
\gathered
\x^{(N)}=\xin^{(N)}\sqcup\xout^{(N)},\\
\xin^{(N)}=\left\{-\tfrac{N-1}2,-\tfrac{N-3}2,\dots,\tfrac{N-3}2,
\tfrac{N-1}2\right\},\quad |\xin^{(N)}|=N,\\
\xout^{(N)}=\left\{\dots,-\tfrac{N+3}2,-
\tfrac{N+1}2\right\}\sqcup\left\{\tfrac{N+1}2,\tfrac{N+3}2,\dots\right\},\quad 
|\xout^{(N)}|=\infty.
\endgathered
$$

Let us associate to every highest weight $\la=\la_1\ge \dots\ge \la_N$ a finite
point configuration $X(\la)\subset \x^{(N)}$ as follows:
$$
X(\la)=\left\{p_i^++\tfrac{N+1}2\right\}\sqcup
\left\{\tfrac{N-1}2-q_i^+\right\}\sqcup
\left\{-p_j^--\tfrac{N+1}2\right\}\sqcup
\left\{-\tfrac{N-1}2+q_j^-\right\},
$$
where $p$'s and $q$'s are the Frobenius coordinates of the positive and
negative parts of $\la$ as explained above. Note that $\la$ can be
reconstructed if we know $X(\la)$.

The probability measure $P_N(\la)$ makes these point configurations random,
and, according to the usual terminology \cite{DVJ}, we obtain a random point
process. We will denote this process by $\Cal P_N$. 

Introduce a matrix $L^{(N)}$ on $\x^{(N)}\times\x^{(N)}$ which in block form 
corresponding to the splitting $\x^{(N)}=\xout^{(N)}\sqcup\xin^{(N)}$ is given by
$$
L^{(N)}=\left[\matrix 0&\A^{(N)}\\
-{(\A^{(N)})}^*&0\endmatrix\right]\,, $$
where $\A^{(N)}$ is a matrix on $\xout\times\xin$,
$$
\gathered
\A^{(N)}(a,b)=\frac{\sqrt{\psiout^{(N)}(a)\psiin^{(N)}(b)}}{a-b}\,, \qquad
a\in\xout^{(N)}, \quad b\in\xin^{(N)},\\
\psiin^{(N)}(x)=\frac{f(x)}{\left(\Gamma\left(-x+\frac{N+1}2\right)
\Ga\left(x+\frac{N+1}2\right)\right)^2}\,,\\
\psiout^{(N)}(x)=\cases \left(\dfrac{\Ga\left(x+\frac {N+1}2\right)}
{\Ga\left(x-\frac {N-1}2\right)}\right)^2 f(x),&x\ge\frac {N+1}2\,,\\
\left(\dfrac{\Ga\left(-x+\frac {N+1}2\right)}{\Ga\left(-x-\frac
{N-1}2\right)}\right)^2 f(x),&x\le-\frac {N+1}2\,,\endcases 
\endgathered 
$$ 
and $f(x)$ was introduced in \tht{1.8}.

\proclaim{Proposition 1.4 \cite{BO5}} For any highest weight
$\la=(\la_1\ge\dots\ge \la_N)$ 
$$
P_N(\la)=\frac{\det L^{(N)}_{X(\la)}}{\det(1+L^{(N)})}\,,
$$
where $L_{X(\la)}^{(N)}$ denotes the finite submatrix of $L^{(N)}$ on 
$X(\la)\times X(\la)$. Moreover, if a finite point configuration $X\subset
\x^{(N)}$ is not of the form $X=X(\la)$ for some highest weight $\la$, then
$\det L^{(N)}_{X}=0$. 
\endproclaim

Proposition 1.4 implies that $\Cal P_N$ is a {\it determinantal point
process} (see \cite{So}, \cite{BOO, Appendix}, \cite{BO5} for a general discussion of such processes). In
particular, this implies the following claim.

\proclaim{Corollary 1.5 \cite{BO5}} The matrix $L^{(N)}$ defines a
finite rank (and hence trace class) operator in $\ell^2(\x^{(N)})$. The correlation functions
$$
\rho_k^{(N)}(x_1,\dots,x_k)=P_N\{\la\,|\,\{x_1,\dots,x_k\}\subset X(\la)\}
$$
of the process $\Cal P_N$ have the determinantal form
$$
\rho_k^{(N)}(x_1,\dots,x_k)=\det[K^{(N)}(x_i,x_j)]_{i,j=1}^k, \quad
k=1,2,\dots, $$
where $K^{(N)}(x,y)$ is the matrix of the operator
$K^{(N)}=L^{(N)}/(1+L^{(N)})$ in  $\ell^2(\x^{(N)})$.
\endproclaim

Explicit formulas for $K^{(N)}$ can be found in \cite{BO5}.  

Now we will describe the limit situation as $N\to\infty$. 
Define the {\it continuous phase space} 
$$
\x=\x^{(\infty)}=\R\setminus\left\{\pm\frac 12\right\}
$$
and divide it into two parts
$$
\gathered
\x=\xin\sqcup\xout,\\ \xin=\left(-\tfrac 12,\tfrac 12\right),\quad\xout=\left(-
\infty,-\tfrac 12\right)\sqcup\left(\tfrac 12,+\infty\right).
\endgathered
$$

To each point $\om\in\Om^o$ we associate a 
point configuration in $\x$ as follows:
$$
\multline
\om=(\al^+,\be^+;\al^-,\be^-) \\ \mapsto
X(\om)=\left\{\al_i^++\tfrac{1}2\right\}\sqcup\left\{\tfrac{1}2-
\be_i^+\right\}\sqcup\left\{-\al_j^--\tfrac{1}2\right\}\sqcup\left\{-
\tfrac{1}2+\be_j^-\right\},
\endmultline
$$
where we omit possible zeros in $\al^+,\be^+,\al^-,\be^-$, and possible ones in $\be^+,\be^-$. 

Let us denote by $P=P^{(z,z',w,w')}$ the spectral measure for the character 
$\chi^{(z,z',w,w')}$ given by Theorem 1.3, and let $P^o$ be its push-forward to $\Om^o$. Then using the above
correspondence between points in $\Om^o$ and point configurations, $P^o$ can be interpreted as a measure on the space of locally finite point
configurations in $\x$, that is, as a point process. We will denote this
process by $\Cal P$.

Since the measures $P_N$ converge to the spectral measure $P^o$ as $N\to\infty$
(Theorem 1.2), we should expect the correlations functions $\rho_k^{(N)}$
to converge to the correlation functions of $\Cal P$ as $N\to\infty$. 

For any $x\in \x$ we will denote by $x_N$ the point of the
lattice $\x^{(N)}$ which is closest to $xN$. 

\proclaim{Theorem 1.6 \cite{BO5}} The correlation functions 
$$
\multline
\rho_k(x_1,\dots,x_k)\\
=\lim_{\Delta x_1,\dots,\,\Delta x_k\to +0}
\frac{P^o\{\om\, |\, X(\om) \text{ \rm intersects each interval  }
(x_i,x_i+\Delta x_i),\,i=1,\dots,k\}}
{\Delta x_1\cdots \Delta x_k}  
\endmultline
$$
of the process $\Cal P$ have determinantal form
$$
\rho_k(x_1,\dots,x_k)=\det[K(x_i,x_j)]_{i,j=1}^k, \quad
k=1,2,\dots, 
$$
where $K(x,y)$ is a kernel on $\x$ which is the scaling limit of the kernels
$K^{(N)}(x,y)$ introduced above:
$$
K(x,y)=\lim_{N\to\infty}N\cdot K^{(N)}(x_N,y_N), \quad x,y\in\x.
\tag 1.9
$$
\endproclaim

The kernel $K(x,y)$ is called the {\it continuous ${}_2F_1$ kernel} and is precisely the kernel in \tht{0.1} for $x,y>\frac 12$. Explicit
formulas for $K(x,y)$ can be found in the next section. This kernel is a
real-analytic function of the parameters $(z,z',w,w')$. We will use the same
notation for its natural analytic continuation.  

It is worth noting that the correlation functions $\rho_k(x_1,\dots,x_k)$
determine the process $\Cal P$ uniquely.

% Thus, Theorem 1.6 along with the explicit
% formulas for the kernel $K(x,y)$ can be viewed as a solution of the 
% problem of
% harmonic analysis for the characters $\chi^{(z,z',w,w')}$, up to the 
% distributions of $\ga^+$ and $\ga^-$. 

It is a well-known elementary observation that the probability that a
determinantal point process with a correlation kernel $\Cal K$ does not have
particles in a given part $J$ of the phase space is equal to
the Fredholm determinant $\det(1-\Cal K|_J)$, see, e.g., \cite{So}, \cite{TW1}.
\footnote{If the correlation kernel is self-adjoint and this probability is
nonzero then the integral operator defined by the kernel $\Cal K$ is of trace
class and the determinant is well-defined, see \cite{So, Theorem 4}. For kernels which are not self-adjoint, the existence of the determinant, generally speaking,
needs to be justified, see, e.g., the end of Section 2 below.}  

In what follows we study determinants of the form $\det(1-K|_J)$ where $K$
is the continuous ${}_2F_1$ kernel and $J$ is a union of finitely many
(possibly infinite) intervals.

\head 2. Continuous ${}_2F_1$ kernel. Setting of the problem
\endhead

Following \cite{BO5} we consider the {\it continuous ${}_2F_1$ kernel} with parameters satisfying the conditions (1)\,-(3) of \S1.

To avoid unnecessary complications (poles in certain formulas below), we
exclude the set where $z+z'+w+w'=0$ from our consideration. Most of the
results, however, can be extended to this set by analytic continuation in one
of the parameters. 

Recall that in \S1 we introduced the space 
$$
\x=\R\setminus\left\{\pm\tfrac 12\right\}
$$
and divided it into two parts
$$
\gathered
\x=\xout\sqcup\xin,\\ \xout=\left(-
\infty,-\tfrac 12\right)\sqcup\left(\tfrac 12,+\infty\right), 
\quad \xin=\left(-\tfrac 12,\tfrac 12\right).
\endgathered
$$

Introduce the functions
$$
\gathered
\psiout:\xout\to \R_+,\quad \psiin:\xin\to\R_+,\\
\psiout(x)=\cases C(z,z')\cdot\left(x-\frac 12\right)^{-z-
z'}\left(x+\frac 12\right)^{-w-w'},& x>\frac 12\,,\\
C(w,w')\cdot\left(-x-\tfrac 12\right)^{-w-w'}\left(-x+\tfrac
12\right)^{-z-z'}, & 
x<-\frac 12\,,
\endcases
\\
\psiin(x)=\left(\tfrac 12-x\right)^{z+z'}\left(\tfrac 
12+x\right)^{w+w'},\quad -\tfrac 12<x<\tfrac 12,
\\
C(z,z')=\frac{\sin(\pi z)\sin(\pi z')}{\pi^2},\quad C(w,w')=\frac{\sin(\pi 
w)\sin(\pi w')}{\pi^2}.
\endgathered
$$ 
Note that $C(z,z')>0$ and $C(w,w')>0$, so that $\psiout(x)$ and $\psiin(x)$ are positive.

We now define the ${}_2F_1$ kernel on $\x$. It is convenient to write it in 
block form corresponding to the splitting $\x=\xout\sqcup\xin$: 
$$
K=\bmatrix K_{\out,\out}&K_{\out,\inr}\\
K_{\inr,\out}& K_{\inr,\inr}\endbmatrix.
$$
We set
$$
\gathered
K_{\out,\out}(x,y)=\sqrt{\psiout(x)\psiout(y)}
\,\frac {\Rout(x)\Sout(y)-\Sout(x)\Rout(y)}{x-y}\,,\\
K_{\out,\inr}(x,y)=\sqrt{\psiout(x)\psiin(y)}
\,\frac {\Rout(x)\Rin(y)-\Sout(x)\Sin(y)}{x-y}\,,\\
K_{\inr,\out}(x,y)=\sqrt{\psiin(x)\psiout(y)}
\,\frac {\Rin(x)\Rout(y)-\Sin(x)\Sout(y)}{x-y}\,,\\
K_{\inr,\inr}(x,y)=\sqrt{\psiin(x)\psiin(y)}
\,\frac {\Rin(x)\Sin(y)-\Sin(x)\Rin(y)}{x-y}\,,
\endgathered
$$
where
$$
\aligned
\Rout(x)&=\left(\frac{x+\frac 12}{x-\frac 12}\right)^{w'}\,{}_2F_1\left[\matrix 
z+w',\,z'+w'\\ z+z'+w+w'\endmatrix\,\Biggl|\,\frac 1{\frac 12 -x}\right]\,,\\
\Sout(x)&=\Gamma\left[\matrix z+w+1,\, 
z+w'+1,\,z'+w+1,\,z'+w'+1\\z+z'+w+w'+1,\,z+z'+w+w'+2\endmatrix\right]\\ &\times 
\frac 1{x-\frac 12}\,\left(\frac{x+\frac 12}{x-\frac 
12}\right)^{w'}\,{}_2F_1\left[\matrix z+w'+1,\,z'+w'+1\\ 
z+z'+w+w'+2\endmatrix\,\Biggl|\,\frac 1{\frac 12 -x}\right]\,,
\endaligned
$$
$$
\aligned
\Rin(x)=&-\frac {\sin \pi z}{\pi}\,\Gamma\left[\matrix z'-
z,\,z+w+1,\,z+w'+1\\z+w+z'+w'+1\endmatrix\right]\\ &\times
\left(\frac 12 +x\right)^{-w}\left(\frac 12 -x\right)^{-z'}{}_2F_1\left[\matrix 
z+w'+1,\, -z'-w\\z-z'+1\endmatrix\,\Biggl|\,\frac 12 -x\right]\\ &-
\frac {\sin \pi z'}{\pi}\,\Gamma\left[\matrix z-
z',\,z'+w+1,\,z'+w'+1\\z+w+z'+w'+1\endmatrix\right]\\ &\times
\left(\frac 12 +x\right)^{-w}\left(\frac 12 -x\right)^{-z}{}_2F_1\left[\matrix 
z'+w'+1,\, -z-w\\z'-z+1\endmatrix\,\Biggl|\,\frac 12 -x\right]\,,
\endaligned
$$
$$
\aligned
\Sin(x)=&-\frac{\sin\pi z}{\pi}\, \Gamma\left[\matrix z'-z,\,z+z'+w+w'\\
z'+w,\,z'+w'\endmatrix \right]\\ &\times
\left(\frac 12 +x\right)^{-w}\left(\frac 12 -x\right)^{-z'}{}_2F_1\left[\matrix 
z+w',\, -z'-w+1\\z-z'+1\endmatrix\,\Biggl|\,\frac 12 -x\right]\\ &-
\frac{\sin\pi z'}{\pi}\, \Gamma\left[\matrix z-z',\,z+z'+w+w'\\
z+w,\,z+w'\endmatrix \right]\\ &\times
\left(\frac 12 +x\right)^{-w}\left(\frac 12 -x\right)^{-z}{}_2F_1\left[\matrix 
z'+w',\, -z-w+1\\z'-z+1\endmatrix\,\Biggl|\,\frac 12 -x\right].
\endaligned
$$
Here ${}_2F_1\left[\matrix a,\, b\\ c\endmatrix\,\Bigl|\,x\right]$ is the {\it
Gauss hypergeometric function}, see, e.g.,  \cite{Er, Ch. 2}, and the 
notation $\Gamma\bmatrix a,\,b,\,\dots\\ c,\,d,\,\dots\endbmatrix$
means $\dfrac {\Ga(a)\Ga(b)\cdots}{\Ga(c)\Ga(d)\cdots}\,$.

Note that for $z=z'$, the functions $\Rin$ and $\Sin$ are, formally speaking,
not defined because of the presence of factors $\Ga(z-z')$ and $\Ga(z'-z)$. 
However, the formulas have a well-defined limit as
$z\to z'$, because the second summands in the formulas for $\Rin$ and $\Sin$
are equal to the first summands with $z$ and $z'$ interchanged. 

In the sequel we will need to know certain analytic properties of the ${}_2F_1$
kernel. We discuss these properties below.

\subhead 2.1. Smoothness \endsubhead
$K(x,y)$ is a real-analytic function in 2 variables defined on $\x\times\x$.
Its values on the diagonal are determined by the L'H\^opital rule:
$$
K(x,x)=\cases
\psiout(x)(\Rout'(x)\Sout(x)-\Sout'(x)\Rout(x)),&x\in\xout,\\
\psiin(x)(\Rin'(x)\Sin(x)-\Sin'(x)\Rin(x)),&x\in\xin.
\endcases
$$

\subhead 2.2 Symmetries of $\Rout,\,\Sout,\,\Rin,\,\Sin$ 
\endsubhead
All four functions 
$\Rout,\,\Sout,\,\Rin,\,\Sin$ are invariant with respect to the transpositions
$z\leftrightarrow z'$ and $w\leftrightarrow w'$. This follows easily from
the above formulas and the identities
$$
{}_2F_1\left[\matrix a,\, b\\c\endmatrix\Bigl |\, \ze\right]=
(1-\ze)^{c-a-b}{}_2F_1\left[\matrix c-a,\,c-b\\c\endmatrix\Bigl |\,
\ze\right],\quad
{}_2F_1\left[\matrix a,\, b\\c\endmatrix\Bigl |\, \ze\right]=
{}_2F_1\left[\matrix b,\, a\\c\endmatrix\Bigl |\, \ze\right].
$$
Since
$$
\overline{{}_2F_1\left[\matrix a,\, b\\c\endmatrix\Bigl |\,
\ze\right]}=
{}_2F_1\left[\matrix \bar a,\, \bar b\\ \bar c\endmatrix\Bigl |\,\bar
\ze\right], $$
where the bar means complex conjugation, and the parameters $(z,z')$, as well
as $(w,w')$, are either real or complex conjugate, the functions $\Rout$,
$\Sout$ and $\Rin$, $\Sin$ take real  values on $\xout$ and $\xin$,
respectively.

Further, let us denote by $\Cal C$ the following change of the parameters and independent variable: $(z,z',w,w',x)\longleftrightarrow (w,w',z,z',-x)$. Then
$$
\gather
\Cal C(\psiout)=\psiout,\quad \Cal C(\psiin)=\psiin,\\
\Cal C(\Rout)=\Rout, \quad \Cal C(\Sout)=-\Sout,\quad \Cal C(\Rin)=\Rin, \quad
\Cal C(\Sin)=-\Sin.
\endgather
$$
For $\psiout$ and $\psiin$ the claim is obvious from the definition. For
$\Rout$ and $\Sout$ the symmetry relation follows from the identity
$$
{}_2F_1\left[\matrix a,\, b\\c\endmatrix\Bigl |\,
\ze\right]=(1-\ze)^{-a}{}_2F_1\left[\matrix a,\, c-b\\c\endmatrix\Bigl |\,
\frac\ze{\ze-1}\right]=(1-\ze)^{-b}{}_2F_1\left[\matrix c-a,\,
b\\c\endmatrix\Bigl |\, \frac\ze{\ze-1}\right].
$$
For $\Rin$ and $\Sin$ the symmetry is a corollary of the symmetries of
$\psiin$, $\Rout$, $\Sout$, and the branching relation \tht{2.1} below.

\subhead 2.3. Symmetries of the kernel
\endsubhead
Since the functions $\Rout,\,\Sout,\,\Rin,\,\Sin$ take real values, the kernel
$K(x,y)$ is real. Moreover, from the explicit formulas for the kernel it
follows that
$$
\gather
K_{\out,\out}(x,y)=K_{\out,\out}(y,x), \quad
K_{\inr,\inr}(x,y)=K_{\inr,\inr}(y,x),\\
K_{\inr,\out}(x,y)=-K_{\out,\inr}(y,x).
\endgather
$$
This means that the kernel $K(x,y)$ is (formally) symmetric with respect to
the indefinite metric $\operatorname{id}\oplus (-\operatorname{id})$ on
$L^2(\x,dx)=L^2(\xout,dx)\oplus L^2(\xin,dx)$. 

\subhead 2.4. Branching of analytic continuations
\endsubhead 
The formulas for $\Rout,\,\Sout,$ $\Rin,$ $\Sin$ above provide analytic
continuations of these functions. We can view $\Rout$ and $\Sout$ as functions
which are analytic and single-valued on $\C\setminus\xin$, and $\Rin$ and
$\Sin$ as functions which are analytic and single-valued on $\C\setminus\xout$.
(Recall that the Gauss hypergeometric function can be viewed as an analytic
and single valued function on $\C\setminus [1,+\infty)$.)

For a function $F(\ze)$ defined on $\C\setminus \R$ we will denote by $F^+$
and $F^-$ its boundary values:
$$
F^+(x)=F(x+i0), \qquad F^-(x)=F(x-i0).
$$
We will show below that
$$
\gather
\text {on  }\xin\qquad
\frac 1{\psiin}\,\frac{\Sout^--\Sout^+}{2\pi i}=\Rin\,,
\quad \frac1{\psiin}\,\frac {\Rout^--\Rout^+}{2\pi i}=\Sin\,,
\tag 2.1\\
\text {on  }\xout\qquad
\frac
1{\psiout}\,\frac{\Sin^--\Sin^+}{2\pi i}=\Rout\,,\quad
\frac1{\psiout}\,\frac {\Rin^--\Rin^+}{2\pi i}=\Sout\,. \tag 2.2 
\endgather
$$
We will use the following formula for the analytic continuation of the 
Gauss hypergeometric function, see \cite{Er, 2.1.4(17)},
$$
\gathered
{}_2F_1\left[\matrix a,\, b\\c\endmatrix\Bigl |\,
\ze\right]=\frac{\Ga(b-a)\Ga(c)}{\Ga(b)\Ga(c-a)}\,(-\ze)^{-a}
{}_2F_1\left[\matrix a,\, 1-c+a\\1-b+a\endmatrix\Bigl |\,
\ze^{-1}\right]\\+
\frac{\Ga(a-b)\Ga(c)}{\Ga(a)\Ga(c-b)}\,(-\ze)^{-b}
{}_2F_1\left[\matrix b,\, 1-c+b\\1-a+b\endmatrix\Bigl |\,
\ze^{-1}\right].
\endgathered
\tag 2.3
$$
This formula is valid if $b-a\notin \Z$, $c\notin\{0,-1,-2,\dots\}$, and
$\ze\notin \R_+$.

Both of the formulas in \tht{2.1} are direct consequences of \tht{2.3} and the trivial
relation
$$
\text{on    } \R_-\qquad \frac {(\ze^u)^{-}-(\ze^u)^+}{2\pi i}=-\frac{\sin(\pi
u)}{\pi}\,(-\ze)^u,\quad u\in\C.
$$

To verify the first formula of \tht{2.2}, we use the relation \tht{2.3} for
both hypergeometric functions in the definition of $\Sin$. Thus, we get 4
summands in total. After computing the jump $(\Sin^--\Sin^+)/2\pi i$, the
second and the fourth summands cancel out. As for the first and the third
summands, they produce exactly $\psiout\Rout$, which can be seen from the identities
$$
\gather
\Ga(s)\Ga(1-s)=\frac \pi{\sin(\pi s)}, \quad s\in\C,\\
\frac{\sin(\pi(z+w))\sin(\pi(z+w'))}{\sin(\pi(z+z'+w+w'))\sin(\pi(z-z'))}+
\frac{\sin(\pi(z'+w))\sin(\pi(z'+w'))}{\sin(\pi(z+z'+w+w'))\sin(\pi(z'-z))}=1.
\endgather
$$
The second part of \tht{2.2} is proved similarly. 

The restriction $b-a\notin \Z$ for \tht{2.3} in our situation means that our
proof works when $z'\ne z$. For $z'=z$ the result is obtained by the limit
transition $z'\to z$ in \tht{2.1} and \tht{2.2}.

\subhead 2.5. Differential equations (due to G.~Olshanski)\endsubhead 
We use Riemann's notation 
$$
P\left(\matrix t_1 & t_2 & t_3\\ a & b & c \\ a' & b' & c'\endmatrix \ \ze
\right) $$
to denote the two--dimensional space of solutions to the second order
Fuchs' equation with singular points $t_1,t_2,t_3$ and exponents
$a,a'$; $b,b'$; $c,c'$, see, e.g., \cite{Er, 2.6}. If $a-a'\notin \Z$ then this
means that about $t_1$, there are two solutions of the form
$$
(\ze-t_1)^a\times\text{\{a holomorphic function\}}, \quad
(\ze-t_1)^{a'}\times\text{\{a holomorphic function\}}.
$$
If $a=a'$ then the basis of the space of
solutions near $t_1$ has the form
$$
(\ze-t_1)^a\times\text{\{a holomorphic function\}}, \quad
\ln(\ze-t_1)(\ze-t_1)^{a}\times\text{\{a holomorphic function\}}.
$$
The holomorphic functions above must take nonzero values at $t_1$. 
For $t_2$ and $t_3$ the picture is similar.

We always have $a+a'+b+b'+c+c'=1$.

The Gauss hypergeometric function ${}_2F_1\left[\matrix a,\,b\\ c
\endmatrix\,\biggl |\,\ze\right]$ belongs to the space
$$
P\left(\matrix 0 & \infty & 1\\ 0 & a  & 0 \\ 1-c & b & c-a-b\endmatrix\
\ze \right)
$$
and, since it is holomorphic around the origin, it corresponds to the exponent
$0$  at the origin. 

Riemann showed (see \cite{Er, 2.6.1}) that
$$
\left(\frac{\ze-t_1}{\ze-t_2}\right)^\ka 
\left(\frac{\ze-t_3}{\ze-t_2}\right)^\mu 
P\left(\matrix t_1 & t_2 & t_3\\ a & b & c \\ a' & b' & c'\endmatrix\
\ze\right) =
P\left(\matrix t_1 & t_2 & t_3\\ a+\ka & b-\ka-\mu & c+\mu \\ a'+\ka &
b'-\ka-\mu & c'+\mu\endmatrix\ \ze\right),
\tag 2.4
$$
where if $t_n=\infty$ then the factor $\ze-t_n$ should be replaced by 1, and
$$
P\left(\matrix t_1 & t_2 & t_3\\ a & b & c \\ a' & b' & c'\endmatrix\
\ze\right)=P\left(\matrix s_1 & s_2 & s_3\\ a & b & c \\ a' & b' &
c'\endmatrix\ \eta\right),
$$
where
$$
\gathered
\eta=\frac{A\ze+B}{C\ze +D}, \qquad s_n=\frac{At_n+B}{Ct_n +D},\quad n=1,2,3,
\\
A,B,C,D\in\C, \quad AD-CB\ne 0.
\endgathered
$$ 

Using these facts, we immediately see that (denote 
$\s=z+z'+w+w'\ne 0$)
$$
\Rout(x)\in P\left(\matrix -\frac12 & \infty & \frac 12\\ w & 0 & z \\ w' &
1-\s & z'\endmatrix\ x \right).
\tag 2.5
$$
Moreover, $\Rout$ is the only element of this space which corresponds
to the exponent $0$ at the infinity and has asymptotics 1 there.

Similarly,
$$
\Sout(x)\in
P\left(\matrix -\frac12 & \infty & \frac 12\\ w & 1 & z \\ w' & -\s &
z'\endmatrix\ x\right),
\tag 2.6
$$
and this is the only element of this space, up to a multiplicative constant,
with the asymptotics $\const\cdot x^{-1}$ at infinity. 

Hence, by \tht{2.1} and \tht{2.4} we get 
$$
\Rin(x)\in P\left(\matrix -\frac12 & \infty & \frac 12\\ -w' & 0 & -z' \\ -w &
1+\s & -z \endmatrix\ x\right),\quad \Sin(x)\in P\left(\matrix -\frac12 &
\infty & \frac 12\\ -w' & 1 & -z' \\ -w & \s & -z\endmatrix\ x\right).
\tag 2.7
$$

\subhead 2.6. Asymptotics at singular points \endsubhead
The results of the previous subsection, see \tht{2.5}-\tht{2.7}, imply that
near $\ze=\frac 12$, if $z\ne z'$ then
$$
\aligned
&\Rout(\ze)=c_1\left(\ze-\tfrac
12\right)^{z}\left(1+O\left(\ze-\tfrac 12\right)\right)+c_2\left(\ze-\tfrac
12\right)^{z'}\left(1+O\left(\ze-\tfrac 12\right)\right), \\&
\Sout(\ze)=c_3\left(\ze-\tfrac
12\right)^{z}\left(1+O\left(\ze-\tfrac 12\right)\right)+c_4\left(\ze-\tfrac
12\right)^{z'}\left(1+O\left(\ze-\tfrac 12\right)\right),\\
&\Rin(\ze)=c_5\left(\ze-\tfrac
12\right)^{-z}\left(1+O\left(\ze-\tfrac 12\right)\right)+c_6\left(\ze-\tfrac
12\right)^{-z'}\left(1+O\left(\ze-\tfrac 12\right)\right), \\&
\Sin(\ze)=c_7\left(\ze-\tfrac
12\right)^{-z}\left(1+O\left(\ze-\tfrac 12\right)\right)+c_8\left(\ze-\tfrac
12\right)^{-z'}\left(1+O\left(\ze-\tfrac 12\right)\right).
\endaligned
$$
Here and below we denote constants by the letters $c_i$, $i=1,2,\dots$. 

If $z=z'$, we have $$
\aligned
&\Rout(\ze)=c_1\left(\ze-\tfrac
12\right)^{z}\left(1+O\left(\ze-\tfrac
12\right)\right)+c_2\ln\left(\ze-\tfrac 12\right)\left(\ze-\tfrac
12\right)^{z}\left(1+O\left(\ze-\tfrac 12\right)\right), \\&
\Sout(\ze)=c_3\left(\ze-\tfrac 12\right)^{z}\left(1+O\left(\ze-\tfrac
12\right)\right)+c_4\ln\left(\ze-\tfrac 12\right)\left(\ze-\tfrac 12\right)^{z}\left(1+O\left(\ze-\tfrac
12\right)\right),\\ 
&\Rin(\ze)=c_5\left(\ze-\tfrac
12\right)^{-z}\left(1+O\left(\ze-\tfrac 12\right)\right)+
c_6\ln\left(\ze-\tfrac 12\right)\left(\ze-\tfrac
12\right)^{-z}\left(1+O\left(\ze-\tfrac 12\right)\right), \\&
\Sin(\ze)=c_7\left(\ze-\tfrac
12\right)^{-z}\left(1+O\left(\ze-\tfrac 12\right)\right)+c_8
\ln\left(\ze-\tfrac 12\right)\left(\ze-\tfrac
12\right)^{-z}\left(1+O\left(\ze-\tfrac 12\right)\right).
\endaligned
$$

Similar formulas hold near $\ze=-\frac 12$ with the parameters $(z,z')$
substituted by $(w,w')$.

Since the Gauss hypergeometric function is holomorphic around the origin, the
definitions of $\Rout$ and $\Sout$ imply that as $\ze\to\infty$,
$$
\alignedat{2}
&\Rout(\ze)=1+O(\ze^{-1}),\quad
&\Sout(\ze)=c_1\,\ze^{-1}(1+O(\ze^{-1})),\\
&\Rout'(\ze)=c_2\,
\ze^{-2}(1+O(\ze^{-1})),\quad 
&\Sout'(\ze)=c_3\,
\ze^{-2}(1+O(\ze^{-1})),\\ 
&\Rout''(\ze)=c_4\,\ze^{-3}(1+O(\ze^{-1})),\quad 
&\Sout''(\ze)=c_5\,
\ze^{-3}(1+O(\ze^{-1})).
\endaligned
\tag 2.8
$$ 
As for $\Rin$ and $\Sin$, the results of the previous subsection, see
\tht{2.7}, imply that, as $\ze\to\infty$,  
$$
\gathered
\Rin(\ze)=c_1(1+O(\ze^{-1}))+c_2\ze^{-1-\s}(1+O(\ze^{-1})),\\
\Sin(\ze)=\cases 
c_2\ze^{-1}(1+O(\ze^{-1}))+c_3\ze^{-\s}(1+O(\ze^{-1})),&\s\ne 1\\
c_4\ze^{-1}(1+O(\ze^{-1}))+c_5\ln(\ze)\ze^{-1}(1+O(\ze^{-1})),&\s=1.
\endcases
\endgathered
\tag 2.9
$$
We will need the exact value of $c_1$ in \tht{2.9} later on. In fact, $c_1=1$,
and 
$$
\Rin(\ze)=1+O(\ze^{-1})+O(\ze^{-1-\s}), \quad \ze\to\infty.
\tag 2.10
$$  
To prove this we do a similar calculation as in the verification of \tht{2.2} above. That
is, we use the relation \tht{2.3} for both hypergeometric functions in the
definition of $\Rin$. Then out of the four summands that arise, the first and
the third summands give contributions of order $\ze^{-1-\s}$ and higher, while 
the second and the fourth ones produce a function
in the variable $(\ze+\frac 12)^{-1}$ holomorphic near the origin  with constant coefficient 1. 

Now we are ready to formulate the problem.
Let
$$
\gathered
J=(a_1,a_2)\sqcup(a_3,a_4)\sqcup\dots\sqcup (a_{2m-1},a_{2m})\subset \R,\\
-\infty\le a_1<a_2<\dots<a_{2m}\le +\infty,
\endgathered
\tag 2.11
$$
be a union of disjoint (possibly infinite) intervals inside the real line such
that the closure of $J$ does not contain the points $\pm\frac 12$. Denote by
$K^J$ the restriction of the continuous ${}_2F_1$ kernel $K(x,y)$ introduced
above to $J$. Our primary goal is to study the Fredholm determinant
$\det(1-K^J)$. 

In the last part of this section we justify the existence of this determinant.

Denote
$$
\gathered
J_{\out}=J\cap \xout, \quad J_{\inr}=J\cap \xin,\\
K^J_{{\out,\out}}=K|_{J_{\out}\times J_{\out}},
 \quad K^J_{{\inr,\inr}}=K|_{J_{\inr}\times J_{\inr}},\\
K^J_{{\out,\inr}}=K|_{J_{\out}\times J_{\inr}},
 \quad K^J_{{\inr,\out}}=K|_{J_{\inr}\times J_{\out}}.
\endgathered
$$

\proclaim{Proposition 2.7} The kernels $K^J_{{\out,\out}}(x,y)$ and
$K^J_{{\inr,\inr}}(x,y)$ define positive trace class operators in
$L^2(J_{\out},dx)$ and $L^2(J_{\inr},dx)$, respectively. 
\endproclaim 
\demo{Proof}
2.1 and 2.3 above imply that the kernels $K^J_{{\out,\out}}(x,y)$ 
and $K^J_{{\inr,\inr}}(x,y)$ are smooth, real-valued, and symmetric. Moreover,
the principal minors of these kernels are always nonnegative, because the
kernel $K$ was obtained as a limit of matrices with nonnegative principal
minors, see \S1. Thus, it remains to prove that the integrals
$$
\int_{J_{\out}}K^J_{{\out,\out}}(x,x)dx, \quad
\int_{J_{\inr}}K^J_{{\inr,\inr}}(x,x)dx $$
converge. For the second integral the claim is obvious since
$\overline{J_{\inr}}\subset(-\frac 12,\frac 12)$, and the integrand is bounded
on $J_{\inr}$. For the first integral we need to control the behavior of the
integrand near infinity (if $J_{\out}$ is not bounded). Since $\psiout(x)
=O(x^{-\s})$ as $x\to\infty$, by 2.1 and \tht{2.8} we see that 
$$
K(x,x)=O(x^{-2-\s}),\quad x\to\infty.
$$
As $\s>-1$, the integral converges.\qed
\enddemo

We will assume that $K^J_{{\out,\inr}}(x,y)=0$ and $K^J_{{\inr,\out}}(x,y)=0$
if $(x,y)$ does not belong to the domain of definition of the corresponding
kernel ($J_{\out}\times J_{\inr}$ for the first kernel and $J_{\inr}\times
J_{\out}$ for the second one).

\proclaim{Proposition 2.8} The kernel $K_0(x,y)=K^J_{{\out,\inr}}(x,y)+
K^J_{{\inr,\out}}(x,y)$ defines a trace-class operator in 
$L^2(J,dx)$. 
\endproclaim
\demo{Proof} 
Consider the operator $-\frac {d^2}{dx^2}$ acting respectively on
$$
\text{(i)} \quad \Cal C_0^\infty(\R);\qquad\qquad 
\text{(ii)} \quad \Cal C_0^{\infty}(J);\qquad\qquad
\text{(iii)}\quad \Cal C_0^\infty(\R\setminus \overline{J}).
$$
In all three cases the operator is essentially self--adjoint, giving rise to the positive self--adjoint operators $H$, $H_J$, and $H_{\R\setminus\overline{J}}$ in $L^2(\R)$, $L^2(J)$, $L^2(\R\setminus\overline{J})$, respectively. It is well known (see e.g.
\cite{RS, Theorem XI.21}) that the operator 
$T=(1+x^2)^{-1}(1+H)^{-1}$ is trace class in $L^2(\R)$. A direct proof can be given as follows. Let $p$ denote the (self--adjoint) closure of 
$-i\frac d{dx}$ acting on $\Cal C_0^\infty$; then $H=p^2$. Commuting
$(1-ix)^{-1}$ and $(1+ip)^{-1}$ in the representation
$$
T=(1+ix)^{-1}(1-ix)^{-1}(1+ip)^{-1}(1-ip)^{-1},
$$
we obtain the formula
$$
\gathered
T=\left((1+ix)^{-1}(1+ip)^{-1}\right)\left((1-ix)^{-1}(1-ip)^{-1}\right)
\\+(1+ix)^{-1}(1+ip)^{-1}(1-ix)^{-1}[x,p]\,(1-ix)^{-1}(1+ip)^{-1}(1-ip)^{-1}.
\endgathered
\tag 2.12
$$
But a simple computation shows that $(1+ix)^{-1}(1+ip)^{-1}$ has kernel
$$
(1+ix)^{-1}\chi_0(y-x)e^{-x-y},
$$
 where $\chi_0$ denotes the
characteristic function of $(0,\infty)$, and as 
$$
\int_{y>x}(1+x^2)^{-1}e^{x-y}dxdy<\infty,
$$
it follows that $(1+ix)^{-1}(1+ip)^{-1}$ is Hilbert--Schmidt. The same is true for $(1-ix)^{-1}(1\pm ip)^{-1}$, and as $[x,p]=i$, the trace class property for $T$ follows immediately from \tht{2.12}.

For $f\in L^2(\R)$, set 
$$
g=\left((1+H)^{-1}-(1+(H_J\oplus H_{\R\setminus
\overline{J}}))^{-1}\right)f.
$$
The function $g$ solves $(-\frac{d^2}{dx^2}+1)g=0$ in the following weak
sense: if $\phi\in \Cal C_0^\infty (\R\setminus \{a_1,\dots, a_{2m}\})$,
then $\int_\R\left((-\frac{d^2}{dx^2} +1)\phi\right)g \,dx=0$. It follows
that in each component of $\R\setminus\{a_1,\dots, a_{2m}\}$, $g$ is
a linear combination of the functions $e^x$ and $e^{-x}$, and hence
the operator $(1+H)^{-1}-(1+(H_J\oplus H_{\R\setminus
\overline{J}}))^{-1}$ is of finite rank. As $T$ is trace class, it follows, in particular, that $(1+x^2)^{-1}(1+H_J)^{-1}$ is trace class in $L^2(J)$. 

Observe that the kernel $K_{\out, \inr}(x,y)$ has the form
$$
\frac{F_1(x)G_1(y)+F_2(x)G_2(y)}{x-y}
$$
for suitable functions $F_i,G_j$. For $x\in\xout$, $y\in\xin$, set
$$
\gathered
K_1(x,y)=K_{\out,\inr}(x,y)-\Biggl(\left(\frac{F_1(x)}x\,G_1(y)+
\frac{F_2(x)}x\,G_2(y)\right)\\+\left(\frac{F_1(x)}{x^2}\,yG_1(y)+
\frac{F_2(x)}{x^2}\,yG_2(y)\right)\Biggr)V(x).
\endgathered
$$
Here $V(x)$ is a smooth function on $\R$ which is zero for $|x|\le L=
\max\{|a_i|:|a_i|<\infty\}$, and $V(x)=1$ for $|x|\ge L+1$.

Finally, for $x\in\xout$, $y\in\xin$ set
$$
K_2(x,y)=K_1(x,y) -\sum_{|a_i|<\infty}\chi_{a_i}(x)K_1(a_i,y),
$$
where the sum is taken over all the finite endpoints of $J$.
Here $\chi_{a_i}(x)$ is a smooth function compactly supported in $\overline{J}$, which equals 1 in a neighborhood of $a_i$, and which vanishes at $a_j$ for $j\ne i$. Clearly $K_2(x,y)=0$ for $x\in\partial J$, which implies that $K_2(\,\cdot\,,y)\in\operatorname{dom} H_J$
for all $y\in\xout$. Using the decay conditions \tht{2.8} (each differentiation with respect to $x$ gives an extra power of decay) it 
follows that $(1+H_J)(1+x^2)K_2(x,y)$ gives rise to a bounded operator
on $L^2(J)$, and hence 
$$
K_2=\left((1+x^2)^{-1}(1+H_J)^{-1}\right)\left(
(1+H_J)(1+x^2)K_2\right)
$$
is trace class. But clearly $K_2$ is a finite rank perturbation of 
$K_{\out,\inr}$. A similar computation is true for $K_{\inr,\out}$, and
we conclude that $K_0$ is trace class on $L^2(J)$. 
\qed
\enddemo
 
Proposition 2.7 and Proposition 2.8 prove that the operator
$$
K^J=\bmatrix
K^{J}_{\out,\out}& K^{J}_{\out,\inr}\\ K^{J}_{\inr,\out}& K^{J}_{\inr,\inr}
\endbmatrix
$$
is trace class. This shows that the determinant $\det(1-K^J)$ is
well-defined.

\head 3. The resolvent kernel and the corresponding Riemann-Hilbert problem. 
\endhead

Starting from this point we assume that the reader is familiar with the
material in the Appendix. 

As was explained in \S1, see Theorem 1.6 et seq., the ${}_2F_1$ kernel
$K$ is a limit of certain discrete kernels which we denoted as
$K^{(N)}$. Moreover, these discrete kernels have rather simple resolvent
kernels $L^{(N)}=K^{(N)}/(1-K^{(N)})$, see Corollary 1.5. The kernels $L^{(N)}$ are {\it integrable}, and, thus, the
kernels $K^{(N)}$ can be found through solving (discrete) Riemann--Hilbert problems, see \cite{Bor2}. 

Our first observation is that the kernel $L^{(N)}$ admits a scaling limit as
$N\to\infty$. Recall that for $x\in \x$ we denote by $x_N$ the 
point of the lattice $\x^{(N)}$ which is closest to $xN$. 

The proof of the following Proposition is straightforward.

\proclaim{Proposition 3.1 \cite{BO5}}
The limit
$$
L(x,y)=\lim_{N\to\infty}N\cdot L^{(N)}(x_N,y_N)\,,\quad 
x,y\in\x,
$$
exists. In the block form corresponding to the splitting 
$\x=\xout\sqcup\xin$, the kernel $L(x,y)$ has the following representation:
$$
L=\left[\matrix 0 & A\\ -{A}^* & 0\endmatrix\right]\,,
$$
where 
$A$ is a kernel on $\xout\times\xin$ of the form
$$
A(x,y)=\frac{\sqrt{\psiout(x)\psiin(y)}}
{x-y}\,,
$$ 
where the functions $\psiout$ and $\psiin$ were introduced at the
beginning of \S2. 
\endproclaim
Now an obvious conjecture would be that $K=L(1+L)^{-1}$, and $K$ can be
obtained through a solution of the corresponding Riemann-Hilbert problem. Both
claims are true, but under certain restrictions on the set of parameters
$(z,z',w,w')$. We begin by showing how to obtain $K$ from a RHP.

Observe that the formulas for the ${}_2F_1$ kernel given in \S2 are identical to \tht{A.2} in the Appendix with
$$
m=\bmatrix m_{11}& m_{12}\\ m_{21}& m_{22}\endbmatrix=\bmatrix \Rout & -\Sin\\ -\Sout &\Rin\endbmatrix, 
\qquad h_I=\sqrt{\psiout},\quad h_{II}=\sqrt{\psiin}.
\tag 3.1
$$
In particular, this means that the ${}_2F_1$ kernel is {\it integrable}. 
Clearly, the matrix-valued function $m$ is holomorphic in $\C\setminus \R$, and as we will see, $\det m(\ze)\equiv 1$ (see proof of Proposition
3.3 below).

\proclaim{Proposition 3.2} The matrix $m$
solves the Riemann--Hilbert problem $(\x, v)$ with
$$
v(x)=\cases \bmatrix1&2\pi i\,\psiout(x)\\0&1\endbmatrix,&
x\in\xout,\\
\bmatrix 1&0\\2\pi i\,\psiin(x)&1\endbmatrix,&x\in\xin.
\endcases
\tag 3.2
$$
If in addition $z+z'+w+w'>0$, then $m(\ze)\sim I$ as $\ze\to\infty$.
\endproclaim
\demo{Proof}
The jump condition $m_+=m_- v$ is equivalent to \tht{2.1}, \tht{2.2}. The
asymptotic relation $m\sim 1$ at infinity follows from \tht{2.8}, \tht{2.9},
\tht{2.10}. \qed
\enddemo

Note that the condition $z+z'+w+w'>0$ is only needed to guarantee the decay of
$m_{12}=-\Sin$ at infinity, see \tht{2.9}. 

Now we investigate the nature of the singularities of $m$ near the points $\pm
\frac 12$ of discontinuity of the jump matrix $v$. We will need this
information further on.

Introduce the matrix 
$$
C(\ze)=\bmatrix \left(\ze-\frac 12\right)^{\frac{z+z'}2}\left(\ze+\frac
12\right)^{\frac{w+w'}2}&0\\ 0& \left(\ze-\frac
12\right)^{-\frac{z+z'}2}\left(\ze+\frac 12\right)^{-\frac{w+w'}2}\endbmatrix.
\tag 3.3
$$

Observe that $C$ is holomorphic in $\C\setminus(-\infty,\frac 12]$.
Furthermore, on $\left(-\infty,\frac 12\right)$
$$
C_-(x)(C_+(x))^{-1}=\cases
\bmatrix e^{-i \pi (z+z')}&0\\ 0& e^{i\pi (z+z')}\endbmatrix,&
x\in\left(-\frac 12,\frac 12\right),\\ 
\bmatrix e^{-i \pi (z+z'+w+w')}&0\\ 0& e^{i\pi (z+z'+w+w')}\endbmatrix,&
x\in\left(-\infty,-\frac 12\right),
\endcases
\tag 3.4
$$
is clearly a piecewise constant matrix. 

\proclaim{Proposition 3.3}
(i) Assume that $z\ne z'$. Then near the point $\ze=\frac 12$
$$
m(\ze)C^{-1}(\ze)=\cases 
H_{1/2}(\ze){\bmatrix
\left(\ze-\frac 12\right)^{\frac{z-z'}2}&0\\0&\left(\ze-\frac
12\right)^{\frac{z'-z}2}\endbmatrix}U_1, &\Im \ze>0,\\
H_{1/2}(\ze){\bmatrix \left(\ze-\frac
12\right)^{\frac{z-z'}2}&0\\0&\left(\ze-\frac
12\right)^{\frac{z'-z}2}\endbmatrix}U_2, &\Im \ze<0
\endcases
$$
for some nondegenerate constant matrices $U_1$ and $U_2$ and
locally holomorphic function $H_{1/2}(\ze)$ such that 
$H_{1/2}(\frac 12)$ is also nondegenerate.

(ii) Assume $z=z'$. Then near the point $\ze=\frac 12$
$$
m(\ze)C^{-1}(\ze)=\cases 
H_{1/2}(\ze){\bmatrix
1&\ln\left(\ze-\frac 12\right)\\0&1\endbmatrix}V_1, &\Im \ze>0,\\
H_{1/2}(\ze){\bmatrix
1&\ln\left(\ze-\frac 12\right)\\0&1\endbmatrix}V_2, &\Im \ze<0
\endcases
$$
for some nondegenerate constant matrices $V_1$ and $V_2$ and
locally holomorphic function $H_{1/2}(\ze)$ such that 
$H_{1/2}(\frac 12)$ is also nondegenerate.
\endproclaim
\demo{Proof}
Let us assume first that $z\ne z'$. Define a new matrix $\wt m(\ze)$ as follows
$$
\wt m(\ze)=\cases m(\ze)C^{-1}(\ze),&\Im \ze>0,\\ m(\ze)C^{-1}(\ze)\bmatrix
1&2\pi i\,C(z,z')\\0&1\endbmatrix,&\Im\ze<0.\endcases
$$
(The constants $C(z,z')$ and $C(w,w')$ were defined at the beginning of \S2.)

By \tht{3.2} we see that the jump matrix $\wt v$ for $\wt m$
locally near the point $\frac 12$ has the form  
$$
\gathered
\wt v=\bmatrix 1&-2\pi i\,C(z,z')\\0&1\endbmatrix C_- v C_+^{-1}
\\=
\cases I, &x>\frac 12,\\ \bmatrix 1&-2\pi
i\,C(z,z')\\0&1\endbmatrix\bmatrix e^{-i\pi(z+z')}&0\\2\pi
i&e^{i\pi(z+z')}\endbmatrix,&x<\frac 12.\endcases
\endgathered
$$
Note that this matrix is piecewise constant. 

An easy computation shows that for a certain nondegenerate matrix $U$,
$$
\bmatrix 1&-2\pi
i\,C(z,z')\\0&1\endbmatrix\bmatrix e^{-i\pi(z+z')}&0\\2\pi
i&e^{i\pi(z+z')}\endbmatrix=U^{-1}\bmatrix e^{i\pi
(z-z')}&0\\0&e^{i\pi(z'-z)}\endbmatrix.
$$
This implies that 
$$
\wt m_0(\ze)=\bmatrix \left(\ze-\frac 12\right)^{\frac
{z-z'}2}&0\\0& \left( \ze-\frac 12\right)^{\frac {z'-z}2}\endbmatrix U 
$$
locally solves the RHP with the jump matrix $\wt v$. 

Our conditions on the parameters $(z,z',w,w')$ imply that $|\Re(z-z')|<1$.
Then the asymptotic formulas of subsection 2.6 imply that $\wt m$ is locally
square integrable near $\ze=\frac 12$, and so are $\wt m_0$ and $\wt
m_0^{-1}$, as follows from the formula above. Since $\wt m$ and $\wt m_0$
locally solve the same RHP, we obtain that  $\wt m\wt m_0^{-1}$ has no jump on
$\R$ near $\ze =\frac 12$, and it is locally integrable as a product of two
locally square integrable functions. Hence, this ratio is a locally
holomorphic function. We denote this holomorphic function by $H_{1/2}(\ze)$,
and set 
$$
U_1=U, \qquad U_2=U\,\bmatrix 1&-2\pi i\,C(z,z')\\0&1\endbmatrix. 
$$
As $v(x)$ in \tht{3.2} has determinant 1, it follows that $\det m_+(x)=
\det m_-(x)$. Also, as above, $\det m(\ze)=\det \wt m(\ze)$ is locally
integrable. Thus, $\det m(\ze)$ is entire. If $z+z'+w+w'>0$, then as noted in Proposition 3.2, $\det m(\ze)\to 1$ as $\ze\to\infty$, and hence, by
Liouville's theorem, $\det m(\ze)\equiv 1$. Analytic continuation
in the parameters $z,z',w,w'$ ensures that the same is true for all
(allowable) values of the parameters. The fact that $H_{1/2}(\frac 12)$
is invertible now follows from the fact that $\det m(\ze)=\det C(\ze)\equiv 1$, and $\det U_1$, $\det U_2$ are nonzero. 
The proof of (i) is complete.

Assume now that $z=z'$. Then there exists a nondegenerate matrix $V$ such that
$$
\bmatrix 1&-2\pi
i\,C(z,z')\\0&1\endbmatrix\bmatrix e^{-i\pi(z+z')}&0\\2\pi
i&e^{i\pi(z+z')}\endbmatrix=V^{-1}\bmatrix 1&1\\0&1\endbmatrix V,
$$
and the local solution of the RHP with the jump matrix $\wt v$ has the form
$$
\wt m_0(\ze)=\bmatrix 1&\ln\left(\ze-\frac 12\right)\\0& 1\endbmatrix V. $$
Repeating word-for-word the argument above we get (ii) with
$$
V_1=V, \quad V_2=V \bmatrix 1&-2\pi
i\,C(z,z')\\0&1\endbmatrix.\qed
$$
\enddemo

Similarly to Proposition 3.3 we have
\proclaim{Proposition 3.4}
(i) Assume that $w\ne w'$. Then near the point $\ze=-\frac 12$
$$
m(\ze)C^{-1}(\ze)=\cases 
H_{-1/2}(\ze){\bmatrix
\left(\ze+\frac 12\right)^{\frac{w-w'}2}&0\\0&\left(\ze+\frac
12\right)^{\frac{w'-w}2}\endbmatrix}U_1, &\Im \ze>0,\\
H_{-1/2}(\ze){\bmatrix \left(\ze+\frac
12\right)^{\frac{w-w'}2}&0\\0&\left(\ze+\frac
12\right)^{\frac{w'-w}2}\endbmatrix}U_2, &\Im \ze<0
\endcases
$$
for some nondegenerate constant matrices $U_1$ and $U_2$ and
locally holomorphic function $H_{-1/2}(\ze)$ such that
$H_{-1/2}(-\frac 12)$ is also nondegenerate.

(ii) Assume $w=w'$. Then near the point $\ze=-\frac 12$
$$
m(\ze)C^{-1}(\ze)=\cases 
H_{-1/2}(\ze){\bmatrix
1&\ln\left(\ze+\frac 12\right)\\0&1\endbmatrix}V_1, &\Im \ze>0,\\
H_{-1/2}(\ze){\bmatrix
1&\ln\left(\ze+\frac 12\right)\\0&1\endbmatrix}V_2, &\Im \ze<0
\endcases
$$
for some nondegenerate constant matrices $V_1$ and $V_2$ and
locally holomorphic function $H_{-1/2}(\ze)$ such that
$H_{-1/2}(-\frac 12)$ is also nondegenerate.
\endproclaim

We now return to the question raised after Proposition 3.1, whether the kernel $L(x,y)$ provides a resolvent operator for the ${}_2F_1$
kernel $K$. The reason why we cannot immediately apply the general theory of the Appendix in
this case is that the functions $f_i$, $g_i$ (or $h_I=\sqrt{\psiout}$,
$h_{II}=\sqrt{\psiin}$) in the notation of the Appendix are not bounded on
the contour as required by \tht{A.1}. We proceed rather by
direct calculation. 

%We are still able to apply the general
%formalism of the Riemann--Hilbert problems for a restricted set of
%parameters, and we do that in Appendix 2. 

First of all, we determine when the operator $L$ is bounded. 

\proclaim{Proposition 3.5}
The kernel $L(x,y)$ defines a bounded operator in $L^2(\x,dx)$ if and only if
$|z+z'|<1$, $|w+w'|<1$.
\endproclaim 
\demo{Proof} It suffices to consider the operator $A:L^2(\xin,dx)\to
L^2(\xout,dx)$ with the kernel $A(x,y)=\sqrt{\psiout(x)\psiin(y)}/(x-y)$.

If $|z+z'|\ge 1$, say, $z+z'\ge 1$, then the restriction of $A(x,y)$ to $(\frac
12, 1)\times (-\frac 12,0)$ is a positive function in 2 variables bounded from
below by  
$$
\multline
\tfrac 23\sqrt{\psiout(x)\psiin(y)}\\ =\tfrac 23\sqrt{C(z,z')}\left|x-\tfrac
12\right|^{-\frac{z+z'}2}\left|x+\tfrac 12\right|^{-\frac{w+w'}2} 
\left|y-\tfrac 12\right|^{\frac{z+z'}2}
\left|y+\tfrac 12\right|^{\frac{w+w'}2}.
\endmultline
$$

This kernel has $(x-\frac 12)^{-\frac{z+z'}2}$ behavior near $x=\frac 12$.
Thus, $A$ is unbounded. Similarly, we see that $A$ is unbounded if $z+z'<-1$
or $|w+w'|\ge 1$. 

Now assume that $|z+z'|<1$, $|w+w'|<1$. Let $\chi$ be the characteristic
function of the set $(-\infty,-\frac 12-\epsilon)\cup(\frac
12+\epsilon,+\infty)$ for some $\epsilon>0$. Then the kernel $\chi(x)A(x,y)$
defines a Hilbert--Schmidt (hence, bounded) operator on $L^2(\x,dx)$. Indeed, 
$$
\multline
\int_{\xout\times\xin} |\chi(x)A(x,y)|^2 dxdy \\
\le
\left(\int_{\frac 12+\epsilon}^{+\infty} \frac{\psiout(x)}{(x-\frac 12)^2}dx+
\int_{-\infty}^{-\frac 12-\epsilon} \frac{\psiout(x)}{(x+\frac 12)^2}dx\right)
\cdot \int_{\xin}\psiin(y)dy<\infty.
\endmultline
$$

Hence, in order to prove that $L$ is bounded, it is enough to show that for any
compactly supported smooth functions $f$ on $\xout$, $\operatorname{supp}
f\subset [-\frac 12-\epsilon, \frac 12)\cup (\frac 12,\frac 12+\epsilon]$, and
$g$ on $\xin$,  
$$ 
\left|\int_{\xin\times\xout} A(x,y)g(x)f(y)\right|\le \const
\Vert f\Vert_2 \Vert g\Vert_2.
\tag 3.5
$$ 
We will assume that $f$ is supported on $(\frac 12, \frac 12+\epsilon]$. The
case when $f$ is supported on $[-\frac 12-\epsilon,-\frac 12)$ is handled
similarly. Assume that $g$ is supported on $[0,\frac 12)$. Let us introduce the polar coordinates $(r,\theta)$ by
$$
x-\tfrac 12=r\cos \theta, \quad \tfrac 12-y=r\sin \theta, \qquad 
0\le \theta \le\tfrac \pi 2,\quad 0\le r\le r(\theta),
$$
for some $r(\theta)\le \const<\infty$.
Then the integral above takes the form
$$
\gathered
\sqrt{C(z,z')}\int_{0}^{\frac \pi 2}\int_{0}^{r(\theta)} 
\frac{(\cos\theta)^{-\frac{z+z'}2} 
(1+r\cos\theta)^{-\frac{w+w'}2} 
(\sin\theta)^{\frac{z+z'}2}
(1-r\sin\theta )^{\frac{w+w'}2}}{\cos\theta+\sin\theta}\\ \times
g\left(r\cos\theta+\tfrac 12\right)f\left(\tfrac 12-r\sin\theta\right)drd\theta.
\endgathered
\tag 3.6
$$
Here $r(\theta)$ is a uniformly bounded continuous function of $\theta$.
Clearly, the factors $|x+\frac 12|^{-\frac{w+w'}2}=(1+r\cos\theta)^{-\frac{w+w'}2}$ and
$|y+\frac 12|^{\frac{w+w'}2}=(1-r\cos\theta)^{\frac{w+w'}2}$
are bounded on the domain of integration. Using the
inequalities
$$
\gathered
\left|\int_{0}^{\infty}g\left(r\cos\theta+\tfrac 12\right)
f\left(\tfrac 12-r\sin\theta\right)dr\right|\le (\cos\theta \sin\theta)^{-\frac
12} \Vert f\Vert_2 \Vert g\Vert_2, \\
\cos\theta+\sin\theta\ge 1,
\endgathered
$$
we see that the integral (3.6) is bounded by
$$
\const \int_0^{\frac \pi 2}
(\cos\theta)^{-\frac{z+z'+1}2}(\sin\theta)^{\frac{z+z'-1}2}d\theta
\cdot\Vert f\Vert_2 \Vert g\Vert_2<\const \Vert f\Vert_2 \Vert g\Vert_2 .
$$

If $f$ is supported on $(\frac 12,\frac 12+\epsilon]$ and $g$ is supported on $(-\frac 12,0]$ then the denominator in $A(x,y)$ is bounded
away from zero, and $A$ is bounded by simple estimates. This completes the
proof of \tht{3.5} in the case that $f$ is supported on
$(\frac 12,\frac 12+\epsilon]$ and $g$ is supported on $(-\frac 12,\frac 12)$.\qed
\enddemo

Since $L^*=-L$, we know that if $L$ is bounded then $(1+L)$ is invertible. 
It seems very plausible that whenever the operator $L$ is bounded, the relation
$K=L(1+L)^{-1}$ should hold. We are able to prove this under the additional
restriction $z+z'+w+w'>0$.

\proclaim{Proposition 3.6} Assume that $z+z'+w+w'>0$, $|z+z'|<1$,
$|w+w'|<1$. Then $K=L(1+L)^{-1}$.
\endproclaim
\demo{Proof}
Since $L$ is bounded and $L=-L^*$, $L$ has a pure imaginary spectrum, and $1+L$ is invertible.
Hence, it is enough to show that $K+KL=L$.
The restrictions on the parameters and the asymptotics of the functions
$\Rout$, $\Sout$, $\Rin$, $\Sin$, from subsection 2.6 imply that the relation
\tht{2.1} and \tht{2.2} can be rewritten in the integral form: 
$$ 
\alignedat{2}
&\int_{\xout}\frac{\psiout(x)\Rout(x)}{x-y}\,dx=-\Sin(y),\quad
&\int_{\xout}\frac{\psiout(x)\Sout(x)}{x-y}\,dx=1-\Rin(y),\\ 
&\int_{\xin}\frac{\psiin(x)\Sin(x)}{x-y}\,dx=1-{\Rout}(y),\quad
&\int_{\xin}\frac{\psiin(x)\Rin(x)}{x-y}\,dx=-{\Sout}(y).
\endalignedat 
\tag 3.7
$$
The 1's on the right-hand side appear because $\Rout(\ze)\sim 1$ and 
$\Rin(\ze)\sim 1$ as $\ze\to\infty$. The restriction $z+z'+w+w'>0$ is needed to
ensure the convergence of the first integral at infinity. Indeed, 
$\psiout(x)\Rout(x)\sim x^{-z-z'-w-w'}$ as $x\to\infty$. 

The identity
$$
K(x,y)+\int_\x L(x,\al)K(\al,y)d\al =L(x,y)
\tag 3.8
$$
for all $x,y\in\x$ follows directly from the relations \tht{3.7} (see
\cite{BO2, Theorem 3.3} for a similar computation). On the other
hand, by \tht{2.8} we see that for any $g\in \Cal C_0^\infty(\x)$,
$$
G(\al)=\int_\x K(\al,y)g(y) dy =Kg(\al)
$$
lies in $L^2(\x,d\al)$. Integrating \tht{3.8} against $g(y)$, we see that
$(1+L)G=Lg$ and hence $Kg=(1+L)^{-1}Lg$ in $L^2(\x)$. It follows
that $K$ extends to a bounded operator $(1+L)^{-1}L=L(1+L)^{-1}$ in
$L^2(\x)$. Conversely, we see that the bounded operator $L(1+L)^{-1}$
has a kernel action given by the ${}_2F_1$ kernel $K(x,y)$.\qed
\enddemo

Proposition 3.6 has the following corollary which will be important for us
later on.

\proclaim{Corollary 3.7} Assume that $z+z'+w+w'>0$, $|z+z'|<1$,
$|w+w'|<1$. Then, in the notation of \S2, the operator $1-K^J$ is invertible.
\endproclaim
\demo{Proof} In the block form corresponding to the splitting $J=J_{\out}\cup
J_{\inr}$, the operator $1-K^J$ has the form
$$
1-K^J=\bmatrix
1-K^{J}_{\out,\out}& K^{J}_{\out,\inr}\\ K^{J}_{\inr,\out}&1-K^{J}_{\inr,\inr} 
\endbmatrix.
$$
But it is easy to see that an operator written in the block form 
$\bmatrix a&b\\c&d\endbmatrix$
is invertible if $a$ is invertible and $(d-ca^{-1}b)$ is invertible.
Therefore, it is enough to prove that 
$$
1-K^{J}_{\out,\out}\quad \text{ and }\quad
(1-K_{\inr,\inr})-K^{J}_{\inr,\out}(1-K^{J}_{\out,\out})^{-1}K^{J}_{\out,\inr}
$$
are invertible.

Proposition 3.6 and the definition of the operator $L$ imply that 
$$
K_{\out,\out}=1-(1+AA^*)^{-1}, \quad K_{\inr,\inr}=1-(1+A^*A)^{-1}.
$$
Hence, $K_{\out,\out}$ and $K_{\inr,\inr}$ are positive operators which are
strictly less then 1. Thus, same is true for $K^J_{\out,\out}$ and 
$K^J_{\inr,\inr}$. In particular, $1-K^{J}_{\out,\out}$ is invertible. Further,
$K^{J}_{\out,\inr}=-(K^{J}_{\inr,\out})^*$. Hence,
$$
(1-K_{\inr,\inr})-K^{J}_{\inr,\out}(1-K^{J}_{\out,\out})^{-1}K^{J}_{\out,\inr}=
(1-K_{\inr,\inr})+\text{bounded positive operator}
$$
is invertible. \qed
\enddemo
 
\example{Remark 3.8} It is plausible that the operator $1-K^J$ is invertible
without any restrictions on the parameters (as opposed to the full operator
$1-K$ which definitely ceases to be invertible if we remove the restrictions
$|z+z'|<1$ and $|w+w'|<1$). However, we do not have a proof of this. In a
similar but simpler situation of the Whittaker kernel we   will prove the
corresponding statement in \S8.2 (see part (3) of Proposition 8.4 below). \endexample

\head 4. System of linear differential equations with rational coefficients
\endhead

Our goal in this section is to show that the kernel of the (trace class and hence Hilbert--Schmidt) operator
$$
R^J=\frac{K^J}{1-K^J}
$$ 
can be expressed through a solution of a system of linear
differential equations with rational coefficients. This result will be crucial 
in our study of the Fredholm determinant $\det(1-K^J)$ in the next section.

In what follows we assume that $\s=z+z'+w+w'>0$.

As noted at the beginning of \S3, $K$ is an integrable kernel:
$$
K(x,y)=\frac{F_1(x)G_1(y)+F_2(x)G_2(y)}{x-y}\,.
$$
Hence, $K^J$ is an integrable kernel. Since $J$ is bounded away from the points
$\pm \frac 12$, it is easy to see that the functions $F_i, G_i$ (which are, in
fact, the functions
$\sqrt{\psiout}\Rout$,
$\sqrt{\psiout}\Sout$, $\sqrt{\psiin}\Rin$, $\sqrt{\psiin}\Sin$  
rearranged in a certain way) belong to $L^p(J,dx)\cap L^\infty(J,dx)$ for any
$p>2\s^{-1}$. This follows from \tht{2.8}, \tht{2.9}. Set 
$$
v_J=I-2\pi i\,FG^t=\bmatrix 1-2\pi i \,F_1G_1& -2\pi i \,F_1G_2\\
-2\pi i \,F_2G_1& 1-2\pi i \,F_2G_2\endbmatrix.
$$
Note that $F^t(x)G(x)=F_1(x)G_1(x)+F_2(x)G_2(x)=0$.

\proclaim{Proposition 4.1} Assume that the operator $1-K^J$ is invertible.
Then there exists a solution $m_J$ of the normalized RHP $(J,v_J)$ such that the kernel
of the operator $R^J=K^J(1-K^J)^{-1}$ has the form $$
\gathered
R^J(x,y)=\frac {\Cal F_1(x)\Cal G_1(y)+\Cal F_2(x)\Cal G_2(y)}{x-y}\,,\\
\Cal F=m_{J+} F=m_{J-}F,\quad  \Cal G=m_{J+} G=m_{J-}G.
\endgathered
$$
The matrix $m_J$ is locally square integrable near the endpoints of $J$.
\endproclaim
\demo{Proof} See Proposition A.2 and the succeeding comment.\qed
\enddemo

Concerning the invertibility of $(1-K^J)$, see Corollary 3.7 and Remark 3.8 above. 

Later on we will need the following property of the decay of $m_J$ at infinity.
\proclaim{Proposition 4.2} As $\ze\to\infty$, $\ze\in\C\setminus \R$,
we have $m_J'(\ze)m_J^{-1}(\ze)=o(|\ze|^{-1}).$
\endproclaim
\demo{Proof} We will give the proof for $J=(s,+\infty)$, $s>\frac 12$. The proof for general $J$ is similar.

Observe that $\det v_J\equiv 1$. Then $\det m_J$ has no jump on $J$. Since
$m_J$ is square integrable near $t$, $\det m_J$ is locally integrable.
Moreover, $\det m_J(\ze)\to 1$ as $\ze\to\infty$, because $m_J(\ze)\to I$. 
Again by Liouville's theorem, $\det m_J\equiv 1$, and $m_J^{-1}$ is bounded near $\ze=\infty$.
Therefore, it suffices to show that $m_J'(\ze)=o(|\ze|^{-1}).$ 

The proof of Proposition A.2 given in \cite{De} implies that for
$\ze\in \C\setminus\R$
$$
m_J(\ze)=I-\int_s^{+\infty}\frac{m_{J+}(t)F(t)G^t(t)}{t-\ze}\,dt =
I-\int_s^{+\infty}\frac{m_{J-}(t)F(t) G^t(t)}{t-\ze}\,dt.
$$
Therefore,
$$
m_J'(\ze)=-\int_s^{+\infty}\frac{m_{J+}(t)F(t)G^t(t)}{(t-\ze)^2}\,dt =
-\int_s^{+\infty}\frac{m_{J-}(t)F(t) G^t(t)}{(t-\ze)^2}\,dt.
$$
If $\Im \ze>\const |\ze|$, then $|t-\ze|>\const|\ze|$. That is, the
distance of the point $\ze$ to the contour of integration is of order
$|\ze|$. Since $m_{J}(t)$ is bounded and $F(t)G^t(t)$ decays at infinity as a positive power of $t$, we see that $m_J'(\ze)=o(|\ze|^{-1}).$

If the point $\ze$ is closer to the real line and, say, $\Im \ze<0$, we can
deform the line of integration up to the line $s+te^{i\theta}$,
$0<\theta<\frac \pi 2$. In other words,
$$
m_J'(\ze)=-\int_0^{+\infty}\frac{m_{J}(s+te^{i\theta})F(s+te^{i\theta})
G^t(s+te^{i\theta}) }{(s+te^{i\theta}-\ze)^2}\,e^{i\theta}dt. 
$$ 
Here it is crucial that the vector--functions $F$ and $G$ (which are 
expressed in terms of the
functions $\sqrt{\psiout}\Rout$ and $\sqrt{\psiout}\Sout$)
have analytic continuations in the sector $0<\operatorname{arg}(\ze-s)<\theta$.
Now the distance of the point $\ze$ to the contour of integration is again of
order $|\ze|$, and the argument above again implies $m_J'(\ze)=o(|\ze|^{-1}).$ If $\Im
\ze>0$, the proof is similar with the line of integration deformed down. \qed
\enddemo

We now describe a general procedure (cf. Steps 1, 2, and 3 in the Introduction) to convert RHP's with "complicated" jump matrices to
RHP's with "simple" jump matrices. The procedure will be used again
in Sections 8 and 9 to analyze a variety of other examples of integrable kernels.

\proclaim{Lemma 4.3}
Suppose $\Sigma=\Sigma_1\cup \Sigma_2$, $\Sigma_1\cap\Sigma _2=\varnothing$, is a decomposition of the oriented contour $\Sigma\subset\C$ into two disjoint parts. Suppose $v$ is a function
on $\Sigma$ with values in $\operatorname{Mat}(k,\C)$. Suppose $m$, $m_1$ solve the RHP's $(\Sigma, v)$, $(\Sigma_1,v)$, respectively. Then if $m^{-1}$ exists, $m_2=m_1m^{-1}$ solves the RHP $(\Sigma_2,v_2)$ where 
$v_2=m_+v^{-1}m_+^{-1}=m_-v^{-1}m_-^{-1}$. Conversely, if $m$, $m_2$
solve the RHP's $(\Sigma, v)$ and $(\Sigma_2,v_2)$, respectively, then 
$m_1=m_2m$ solves the RHP $(\Sigma_1,v)$.
\endproclaim
\demo{Proof} Direct calculation.\qed
\enddemo

Recall that, as noted at the beginning of \S3, the formulas for the
kernel $K$ are identical to \tht{A.2} with $m$, $h_I$, $h_{II}$ given by \tht{3.1}. This, in particular,
means that 
$$
F=m_{\pm}f,\quad G=m_{\pm}^{-t}g
$$
with
$$
f_1(x)=g_2(x)=\cases
\sqrt{\psiout(x)},&x\in\xout,\\0,&x\in\xin,\endcases\ 
f_2(x)=g_1(x)=\cases 0,&x\in\xout,\\\sqrt{\psiin(x)},&x\in\xin.\endcases 
$$
Note that the matrix $v$ in \tht{3.2} has the form $I+2\pi i\, fg^t$. 

\proclaim{Lemma 4.4} If the matrix $v$ in Lemma 4.3 has the form $v=I+2\pi i\,fg^t$ for (arbitrary) $f$ and $g$ with $f^t(x)g(x)=0$, then
$$
v_2=I-2\pi i\,FG^t,
\tag 4.1
$$
where $F=m_+f=m_-f$ and $G=m_+^{-t}g=m_-^{-t}g$.
\endproclaim
\demo{Proof} Have
$$
v_2=m_+(I-2\pi i\,fg^t)m_+^{-1}=I-2\pi i\, (m_+f)(m_+^{-t}g)^t.\qed
$$
\enddemo

Let $\Sigma=\x$, $\Sigma_1=\x\setminus J$, and $\Sigma_2=J$. Now $m_J$ 
solves the RHP $(\Sigma_2,v_2)$ with $v_2=v_J=I-2\pi i\,FG^t$. But as noted above, $F=m_{\pm} f$, $G=m_\pm^{-t}g$, and so it follows by 
Lemmas 4.3 and 4.4 that 
$$
m_1=m_{\x\setminus J}\equiv m_Jm
$$
satisfies the RHP $(\Sigma_1,v)$, where $v=I+2\pi i\, fg^t$ as before. We think of $v_2$ as the ``complicated'' jump matrix and $v$ as
the ``simple'' jump matrix. The formula $m_J=m_{\x\setminus J}m^{-1}$ 
shows that the analysis of the solution of the 
``complicated'' RHP $(J,v_J)$ reduces to the analysis of the solutions
of two ``simple'' RHP's $(\x\setminus J,v)$ and $(\x,v)$. 

These two RHP's are ``simple'' for the following reason.
Recall that in \S3 we have introduced a matrix $C(\ze)$, see \tht{3.3}. 

Set $M=m_{\x\setminus J}\, C^{-1}$. This is a holomorphic function on
$\C\setminus \R$ which has boundary values $M_\pm(x)$ on $\R$. 

\proclaim{Lemma 4.5}
The matrix-valued function $M(\ze)$ satisfies the jump relation $M_+=M_-V$,
where the jump matrix $V$ has the form
$$
V(x)=\cases 
\bmatrix 1& 2\pi i \,C(z,z')\chi(x) \\0&1\endbmatrix,&x>\frac
12,\\ \bmatrix e^{-i\pi(z+z')}& 0\\2\pi i\chi(x)&e^{i\pi(z+z')}\endbmatrix,&\frac
12>x>-\frac 12,\\
\bmatrix e^{-i\pi(z+z'+w+w')}&2\pi i\,
C(w,w')\chi(x)\\ 0&e^{i\pi(z+z'+w+w')}\endbmatrix,&x<-\frac 12,
\endcases
$$
where
$\chi(x)=\chi_{\x\setminus J}(x)$ is the characteristic function of the
set $\x\setminus J$. 
\endproclaim
\demo{Proof}
On $\x\setminus J$ we have
$$
V=M_-^{-1}M_+=C_-  v C_+^{-1}=C_- C_+^{-1}+
2\pi i\, C_-f\, (C_+^{-t}g)^{t},
$$
and on $J$ we have $V=C_-C_+^{-1}$.
The jump relation \tht{3.4} and explicit formulas for $C$, $f$, and $g$
conclude the proof. \qed
\enddemo

The {\it important fact} about the jump matrix $V$ is that it is {\it piecewise constant}. As discussed in the Introduction, this allows us to prove the following central claim.

Recall that $J$ is a union of $m$ intervals with endpoints $\{a_j\}_{j=1}^{2m}$,
see \tht{2.11}.

\proclaim{Theorem 4.6}
The matrix $M$ satisfies the differential equation
$$
M'(\ze)=\left(\frac \ba{\ze-\frac 12}+\frac \bb{\ze +\frac
12}+\sum_{j=1}^{2m}\frac {\bc_j}{\ze-a_j}\right)M(\ze)
$$
with some constant matrices $\ba$, $\bb$, and $\{\bc_j\}_{j=1}^{2m}$. 
If $a_1=-\infty$ then $\bc_1=0$, and if $a_{2m}=+\infty$ then $\bc_{2m}=0$.
Other than that, all matrices $\ba$, $\bb$, $\{\bc_j\}_{j=1}^{2m}$ are
nonzero.

Moreover,
$$
\gathered
\tr\ba=\tr\bb=\tr\bc_j=0,\\
\det\ba=-\left(\frac{z-z'}2\right)^2,\quad 
\det\bb=-\left(\frac{w-w'}2\right)^2,\quad \det\bc_j=0,
\endgathered
$$
for all $j=1,\dots,2m$, and
$$
\ba+\bb+\sum_{j=1}^{2m}\bc_j=-\frac{z+z'+w+w'}2\,\sig, \quad \sig=\bmatrix
1&0\\0&-1\endbmatrix. 
$$ 
\endproclaim
\demo{Proof} Since $M$ satisfies the jump condition with a piecewise constant
jump matrix $V$ (Lemma 4.5), $M'$ satisfies the jump condition with exactly
the same jump matrix. Therefore, $M'M^{-1}$ has no jump across $\x$. 
(Note that 
$$
\det M=\det m_J\det m (\det C)^{-1}\equiv 1,
$$ 
and hence $M^{-1}$ exists.)

Thus, we know that $M'M^{-1}$ is a holomorphic function away from the points $\{\pm \frac 12\}\cup \{a_i\}_{i=1}^{2m}$. We now investigate the
behavior of $M$ near these points.

Near $\ze=\frac 12$, $m_J(\ze)$ is holomorphic, and the behavior of
$m(\ze)C^{-1}(\ze)$ is described by Proposition 3.3. This implies, in the
notation of Proposition 3.3, that for $z\ne z'$
$$
M'(\ze)M^{-1}(\ze)=\frac 1{\ze-\frac
12}\, m_J\left(\tfrac 12\right){H_{1/2}\left(\tfrac 12\right)\bmatrix
\frac{z-z'}2&0\\0&\frac{z'-z}2\endbmatrix H_{1/2}^{-1}\left(\tfrac
12\right)}m_J^{-1}\left(\tfrac 12\right)+O(1),  
$$
and for $z=z'$
$$
M'(\ze)M^{-1}(\ze)=\frac 1{\ze-\frac 12}\, m_J\left(\tfrac 12\right){H_{1/2}\left(\tfrac
12\right)\bmatrix 0&1\\0&0\endbmatrix H_{1/2}^{-1}\left(\tfrac 12\right)m_J^{-1}\left(\tfrac 12\right)}+
O(1). 
$$

Similarly, near $\ze=-\frac 12$ we have the following: for $w\ne w'$ 
$$
\multline
M'(\ze)M^{-1}(\ze)\\=\frac 1{\ze+\frac
12}\, m_J\left(-\tfrac 12\right) {H_{-1/2}\left(-\tfrac 12\right)\bmatrix
\frac{w-w'}2&0\\0&\frac{w'-w}2\endbmatrix H_{-1/2}^{-1}\left(-\tfrac
12\right)}m_J^{-1}\left(-\tfrac 12\right)+O(1),  
\endmultline
$$
and for $w=w'$
$$
M'(\ze)M^{-1}(\ze)=\frac 1{\ze+\frac 12} m_J\left(-\tfrac 12\right){H_{-1/2}\left(-\tfrac
12\right)\bmatrix 0&1\\0&0\endbmatrix H_{-1/2}^{-1}\left(-\tfrac 12\right)}m_J^{-1}\left(-\tfrac 12\right)+
O(1). 
$$
 As for the points $\{a_j\}_{j=1}^{2m}$, we will prove the following claim.
\proclaim{Lemma 4.7} 
In a neighborhood of any finite endpoint $a_j$, $j=1,\dots,2m$,
$$
M'(\ze)M^{-1}(\ze)=\frac{\bc_j}{\ze-a_j}+h_j(\ze),
$$
where $h_j(\ze)$ is holomorphic near $a_j$, and $\bc_j$ is a nonzero nilpotent
matrix. 
\endproclaim 

Let us postpone the proof of this Lemma and proceed with the proof of 
Theorem 4.6. Observe that if we set 
$$
\aligned
&\ba=\cases {m_J\left(\tfrac 12\right)H_{1/2}\left(\frac 12\right)\bmatrix
\frac{z-z'}2&0\\0&\frac{z'-z}2\endbmatrix H_{1/2}^{-1}\left(\frac
12\right)m_J^{-1}\left(\tfrac 12\right)}, &z\ne z',\\ 
{m_J\left(\tfrac 12\right)H_{1/2}\left(\frac
12\right)\bmatrix 0&1\\0&0\endbmatrix H_{1/2}^{-1}\left(\frac 12\right)m_J^{-1}\left(\tfrac 12\right)},&
z=z', \endcases\\ 
&\bb=\cases {m_J\left(-\tfrac 12\right)H_{-1/2}\left(-\frac 12\right)\bmatrix
\frac{w-w'}2&0\\0&\frac{w'-w}2\endbmatrix H_{-1/2}^{-1}\left(-\frac
12\right)m_J^{-1}\left(-\tfrac 12\right)}, &w\ne w',\\ 
{m_J\left(-\tfrac 12\right)H_{-1/2}\left(-\frac
12\right)\bmatrix 0&1\\0&0\endbmatrix H_{-1/2}^{-1}\left(-\frac 12\right)m_J^{-1}\left(-\tfrac 12\right)},&
w=w', \endcases
\endaligned
$$
then the function 
$$
M'(\ze)M^{-1}(\ze)-\left(\frac \ba{\ze-\frac 12}+\frac \bb{\ze +\frac
12}+\sum_{j=1}^{2m}\frac {\bc_j}{\ze-a_j}\right)
\tag 4.2
$$
is entire. At infinity we have, using $m_{\x\setminus J}=m_Jm$,
$$
M'M^{-1}=m_J'm_J^{-1}+ m_Jm'm^{-1}m_J^{-1}+m_Jm(C^{-1})'Cm^{-1}m_J^{-1}.
$$
We know that $m\sim I$, $m_J\sim I$, $m'_Jm_J^{-1}=o(|\ze|^{-1})$ (see
Proposition 4.2), $m'm^{-1}=o(|\ze|^{-1})$ (which follows from \tht{2.8} and
differentiation of \tht{2.9}), and by direct computation 
$$
(C^{-1})'(\ze)C(\ze)=-\frac{z+z'+w+w'}2\,\sig\ze^{-1}.
$$
This implies that 
$$
M'(\ze)M^{-1}(\ze)=-\frac{z+z'+w+w'}2\,\sig\ze^{-1}+o(|\ze^{-1}|).
$$
Then, by Liouville's theorem, the expression \tht{4.2} is identically equal to
zero. Multiplying it by $\ze$ and passing to the limit $\ze\to\infty$ we
see that
$$
\ba+\bb+\sum_{j=1}^{2m}\bc_j+\frac{z+z'+w+w'}2\,\sig=0.
$$
The remaining properties of $\ba$ and $\bb$ follow directly from their 
definitions. 
This concludes the proof of the Theorem modulo Lemma 4.7.

\demo{Proof of the Lemma 4.7}
Let us give a proof for an odd value of $j$. The proof for the even $j$'s is obtained by
changing the sign of $\ze$. We will omit the subscript ``$j$'' in $a_j$ and
$\bc_j$. 

Near the point $a$ the jump matrix for $M$ has the form (Lemma 4.5)
$$
(M_-(x))^{-1}M_+(x)=\cases C^o+2\pi i \, f^o(g^o)^t,&
x<a,\\                            C^o,&x>a, \endcases      
$$
where $f^o=C_-f$, $g^o=C_+^{-t}g$ are locally constant vectors, and
$C^o=C_-C_+^{-1}$ is a locally constant matrix. Note that 
$((C^o)^{-1}f^o)^tg^o=f^tg=0.$

Set 
$$
\wt  M(\ze)=\cases M(\ze),&\Im \ze>0,\\ M(\ze)C^o,& \Im \ze<0.\endcases
$$  
Then 
$$
(\wt M_-(x))^{-1}\wt M_+(x)=\cases I+2\pi i \, (C^o)^{-1}f^o(g^o)^t,&
x<a.\\                            I,&x>a, \endcases      
$$
and we note that
$$
\gathered
\wt M^o=\exp\left(\frac 1{2\pi i}\,\ln(I+2\pi i \,
(C^o)^{-1}f^o(g^o)^t)\ln(\ze-a)\right)\\
=\exp((C^o)^{-1}f^o(g^o)^t\,\ln(\ze-a))=I+
(C^o)^{-1}f^o(g^o)^t\,\ln(\ze-a)
\endgathered
$$
is also a solution of this local RHP. (Here we use the fact that
$\left((C^o)^{-1}f^o(g^o)^t\right)^2=0$.) Hence,
$$
M^o(\ze)=\cases \wt M^o(\ze),&\Im \ze>0,\\ \wt M^o(\ze)(C^o)^{-1},& \Im
\ze<0,\endcases 
$$  
is a local solution of the RHP for $M$ near $\ze=a$.

Since $M=m_JmC^{-1}$, $m$ and $C^{-1}$ are bounded near $a$, and $m_J$ is
square integrable near $a$ (Proposition 4.1), we conclude that $M$ is square
integrable near $a$. Clearly, $M^o$ is also locally square integrable, and  
$\det M^o\equiv 1$. Hence,
$H_a(\ze)\equiv M(\ze)(M^o(\ze))^{-1}$ is locally integrable and does not have
any jump across $\R$ near $a$. Therefore, $H_a(\ze)$ is holomorphic near $a$.
Since $\det M\equiv 1$, $H_a$ is nonsingular. We obtain 
$$
M(\ze)=H_a(\ze)M^o(\ze).
\tag 4.3
$$ 
Computing $M'M^{-1}$ explicitly we arrive at the
desired claim with 
$$
h=H'_a H_a^{-1}\quad\text{and}\quad
\bc=H_a(a)\, (C^o)^{-1}f^og^o\, H_a^{-1}(a).
$$
Since $(C^o)^{-1}f^og^o$ is nilpotent and nonzero, the proof of Lemma 4.7 and
Theorem 4.6 is complete.\qed
\enddemo
\enddemo

\example{Remark 4.8} Arguing in exactly the same way as we did in the proof of
Theorem 4.6 above, it is not hard to prove the equation
$$
(m(\ze)C^{-1}(\ze))'=\left(\frac {\ba_0}{\ze-\frac
12}+\frac{\bb_0}{\ze+\frac 12}\right)m(\ze)C^{-1}(\ze)
$$
with some constant matrices $\ba_0$, $\bb_0$. These matrices can be explicitly
computed (as opposed to the matrices $\ba$, $\bb$, $\{\bc_j\}$ in Theorem
4.6!). The resulting system of differential equations is equivalent to
\tht{2.5}, \tht{2.6}, \tht{2.7}.
\endexample

\head 5. General setting
\endhead

As noted earlier, the arguments which we used to derive Theorem 4.6
and which we will use to derive further results, can be applied to other
kernels as well (see \S8 for examples). In this short section we place
the results of the previous section in a general framework. 

Let $K(x,y)$ be a smooth integrable kernel 
$$
K(x,y)=\frac{\sum_{j=1}^N F_j(x)G_j(y)}{x-y}\,, \quad
\sum_{j=1}^NF_j(x)G_j(x)=0,
$$
defined on a subset $\x$ of the real line. Let us assume that $\x$ is finite
union of (possibly infinite) disjoint intervals. 

We list the conditions on the kernel $K$ that we need.

\medskip
\noindent (1) Assume we are given functions $\{f_j,g_j\}_{j=1}^N$
on $\x$ for which there exists a solution $m$ of the RHP $(\x,v)$ with the jump
matrix  
$$
v=I+2\pi i\,
fg^t=[\delta_{kl}+2\pi i\,f_kg_l]_{k,l=1}^N
$$ 
such that the relations $F=m_{+}f=m_{-}f$, $G=m_+^{-t}g=m_-^{-t}g$ are satisfied. 
\medskip
Then necessarily
$$
\sum_{j=1}^N f_j(x)g_j(x)=f^t(x)g(x)=(mf)^t(x)(m^{-t}g)(x)=F^t(x)G(x)=0.
$$

For such functions $\{f_j,g_j\}_{j=1}^N$  the kernel
$L(x,y)$ defined by 
$$
L(x,y)=\frac{\sum_{j=1}^N f_j(x)g_j(y)}{x-y}\,,
$$
formally satisfies the relation $K=L(1+L)^{-1}$, see Proposition A.2. As we
have seen in \S3, it can happen that the solution $m$ of the RHP is defined but
the integral operator $L$ in $L^2(\x,dx)$ given by the kernel $L(x,y)$ is
unbounded. In such cases, greater care must be taken in assigning
a meaning to $L(1+L)^{-1}$.   

Let $J$ be a subset of $\x$ formed by a union of finitely many possibly
infinite disjoint intervals:
$$
\gathered
J=(a_1,a_2)\cup \dots\cup (a_{2m-1},a_{2m})\subset \x,\\
-\infty\le a_1<a_2<\dots<a_{2m}\le +\infty.
\endgathered
$$

The endpoints $\{a_j\}$ of $J$ are allowed to coincide with the endpoints of
$\x$. 
\medskip
\noindent (2) Assume that the kernel $K^J=K|_{J}$ defines a trace class
integral operator in $L^2(J,dx)$.
\medskip
\noindent (3) Assume also that the operator $1-K^J$ is invertible in
$L^2(J,dx)$. 
\medskip
Further,
\medskip
\noindent (4) Assume that the restrictions of the functions $F_j, G_j$ to $J$
lie in $L^p(J,dx)\cap L^\infty(J,dx)$ for some $p$, $1<p<\infty$.
\medskip
Then by Proposition A.2 there exists a solution $m_J$ of the normalized RHP $(J,v_J)$ 
with $v_J=I-2\pi iFG^t$, and the kernel of the operator $R^J=K^J(1-K^J)^{-1}$
has the form
$$
\gathered
R^J(x,y)=\frac{\sum_{j=1}^N \Cal F_j(x)\Cal G_j(y)}{x-y}\,,\\
\Cal F=(m_J)_\pm F=(m_J)_\pm m_\pm f, \quad \Cal
G=(m_J^{-t})_\pm G=(m_J^{-t})_\pm m^{-t}_\pm g. \endgathered
$$

Set $m_{\x\setminus J}=m_Jm$. As in \S4, we see that $m_{\x\setminus J}$ satisfies the
RHP $(\x\setminus J, v)$. The crucial condition is that
this RHP can be reduced to a RHP with a piecewise constant jump matrix. We
formulate this more precisely as follows.
\medskip
\noindent (5) Assume that there exists a matrix valued holomorphic function
$C:\C\setminus\s\to \Ma(N,\C)$ such that

(a) $C$ is invertible;
\medskip
(b) $f^o=C_-f$ is a piecewise constant vector on $\x$;
\medskip
(c) $g^o=C_+^{-t}g$ is a piecewise constant vector on $\x$;
\medskip
(d) $C^o=C_-C_+^{-1}$ is an invertible piecewise constant matrix on $\x$;
\medskip
(e) $(C^{-1})'(\ze)C(\ze)=D\,\ze^{-1}+o(|\ze|^{-1})$ as $\ze\to\infty$,
where $D$ is a constant matrix.
\medskip

Now form the matrix $M=m_{\x\setminus J} C^{-1}=m_J m C^{-1}$. Condition (5) implies
that the jump matrix for $M$, which is equal to 
$$
V=M_-^{-1}M_+=C^o+2\pi i f^o(g^o)^t \chi_{\x\setminus J}
$$
(cf. Lemma 4.5) is piecewise constant.

Now in order to ensure the existence of a differential equation for $M$ with respect to $\ze$ we
need to know something about the local behavior of $M$ near the points of
discontinuity of $V$ and near infinity. 

To state the condition on the local behavior of $M$ we have to be sure that the matrix $M^{-1}$ exists.
Note that the determinants of $v$ and $v_J$ are identically equal to one,
because both $v$ and $v_J$ are equal to the identity plus a nilpotent matrix.
This means that the scalar functions $\det m$ and $\det m_J$ have no jump
across $\x$. As $m$ and $m_J$ tend to $I$ at infinity, $\det m$ and $\det m_J$
tend to 1 at infinity. Modulo certain regularity conditions on $\det m$ and
$\det m_J$ near the endpoints of $\x$ and $J$ (which are always satisfied in
the applications), Liouville's theorem implies that $\det m=\det m_J\equiv 1$,
and the matrices $m$, $m_J$, and $M$ are invertible.
\medskip
\noindent (6) Assume that $M'(\ze)M^{-1}(\ze)=O(|\ze-a|^{-1})$ at any
endpoint $a$ of $\x$. 
\medskip

\noindent (7) Assume that $m'_{\x\setminus J}(\ze) m_{\x\setminus J}^{-1}(\ze)=o(|\ze|^{-1})$ as
$\ze\to\infty$ (recall that $m_{\x\setminus J}=m_Jm$).
\medskip

Before going any further, we indicate where we proved that the conditions (1)--(7) hold for the ${}_2F_1$ kernel. 

The condition (1) is verified in Proposition 3.2; (2) follows from Propositions 2.7 and 2.8; (3) is Corollary 3.7 (here we needed  additional restrictions on the parameters $z,z',w,w'$); (4) is a corollary of \tht{2.8}, \tht{2.9}; (5) consists of obvious properties of the matrix \tht{3.3}; (6) follows from Proposition 3.3 and 3.4; (7) is a corollary of \tht{2.8}, \tht{2.9}, and Proposition 4.2.

Denote by $\{b_j\}_{j=1}^n\subset \C$ all finite endpoints of $\x$ and $J$. 

\proclaim{Theorem 5.1} Under the conditions \tht{1}-\tht{7} above, there exist
constant matrices $\{\bb_j\}_{j=1}^n$ such that the matrix $M$ satisfies the
following linear differential equation:
$$
M'(\ze)=\sum_{j=1}^n\frac{\bb_j}{\ze-b_j}\,\cdot M(\ze).
\tag 5.1
$$
If $b_j$ is an endpoint of $J$ but not an endpoint of $\x$ then the
corresponding matrix $\bb_j$ is nilpotent and nonzero. Moreover, $\sum_{j=1}^n
\bb_j=D$, where the constant matrix $D$ is given in \tht{5e}.
\endproclaim

\demo{Proof} We will follow the proof of Theorem 4.6. Since $M$ has a constant
jump matrix, the matrix $M'M^{-1}$ has no jump across $\x$. If $b_j$ is an
endpoint of $\x$ then by (6), $b_j$ is either a regular point or a
first order pole of $M'M^{-1}$. If $b_j$ is an endpoint of $J$ and not an
endpoint of $\x$ then the proof of Lemma 4.7 (which can be repeated
word-for-word in the general setting) shows that near $b_j$ 
$$
M'(\ze)M^{-1}(\ze)=\frac{\bb_j}{\ze-b_j}+\text{a locally holomorphic
function} 
$$
with a nilpotent constant matrix $\bb_j$. Thus,
$$
M'(\ze)M^{-1}(\ze)-\sum_{j=1}^n \frac{\bb_j}{\ze-b_j}
\tag 5.2
$$
is an entire function for the (constant) matrices
$\{\bb_j\}_{j=1}^n$.

Near $\ze=\infty$, 
$$
M'M^{-1}=m'_{\x\setminus J}m_{\x\setminus J}^{-1}+m_{\x\setminus J} (C^{-1})'C m_{\x\setminus J}^{-1}=D\,\ze^{-1}+o(|\ze|^{-1}),
$$
as follows from (5e) and (7). 
Hence, by Liouville's theorem, the function \tht{5.2} is identically zero, and
computing the terms of order $\ze^{-1}$ at infinity we see that
$\sum_{j=1}^n\bb_j=D$. \qed
\enddemo

\example{Remark 5.2} Arguing as above and replacing the condition (7) by
the estimate $m'm^{-1}=o(|\ze|^{-1})$ as $\ze\to\infty$, one can easily prove
that $$ (m(\ze)C^{-1}(\ze))'=\sum_{j=1}^l\frac{\bb_J^0}{\ze-b_j^0}\,\cdot
m(\ze)C^{-1}(\ze),
$$
where $\{b_j^0\}_{j=1}^l$ are the endpoints of $\x$, and $\{\bb_j^0\}$ are some
constant matrices, cf. Remark 4.8.   
\endexample

If we allow the differential equation to have an irregular singularity at
infinity, then the condition (5e) on the matrix $C$ can be relaxed. Let us
introduce the condition
\medskip
\noindent (5e') $(C^{-1})'(\ze)C(\ze)=D+o(1)$ as $\ze\to\infty$,
where $D$ is a constant matrix.
\medskip

We can then relax the condition (7) to 
\medskip
\noindent (7') Assume that $m'_{\x\setminus J}(\ze) m_{\x\setminus J}^{-1}(\ze)=o(1)$ as
$\ze\to\infty$.
\medskip

The following claim is proved in exactly the same way as Theorem 5.1.
\proclaim{Theorem 5.3}  Under the conditions \tht{1}-\tht{4}, \tht{5a-d},
\tht{5e'}, \tht{6}, \tht{7'} above,
there exist constant matrices $\{\bb_j\}_{j=1}^n$ such that the matrix $M$
satisfies the following linear differential equation: 
$$
M'(\ze)=\left(\sum_{j=1}^n\frac{\bb_j}{\ze-b_j}+D\right) M(\ze).
\tag 5.3
$$
If $b_j$ is an endpoint of $J$ but not an endpoint of $\x$ then the
corresponding matrix $\bb_j$ is nilpotent and nonzero. 
\endproclaim

\example{Remark 5.4} Once again, if $m'm^{-1}\to 0$ as $\ze\to\infty$, then 
$$ 
(m(\ze)C^{-1}(\ze))'=\left(\sum_{j=1}^l\frac{\bb_J^0}{\ze-b_j^0}+D\right)
m(\ze)C^{-1}(\ze),
$$
where $\{b_j^0\}_{j=1}^l$ are the endpoint of $\x$, and $\{\bb_j^0\}$ are some
constant matrices, cf. Remarks 4.8, 5.2.  
\endexample

\head 6. Isomonodromy deformations. Jimbo-Miwa-Ueno $\tau$-function
\endhead
 
Let $M(\ze)$ be a matrix-valued function on the complex $\ze$--plane satisfying a
linear differential equation of the form $M'(\ze)=B(\ze)M(\ze)$, where $B(\ze)$
is a rational matrix. 

Fix a fundamental solution $M$ of this equation. In
general, $M(\ze)$ is a multivalued function. If $\{b_1,\dots,b_n\}$ are the
poles of $B$, then $\{b_1,\dots, b_n, \infty$\} are the branch points for
$M$. When we continue $M$ along a closed path $\gamma$ avoiding the branch
points, the column vectors of $M$ are changed into some linear combinations of the columns of the original matrix:
$M(\ze)\mapsto M(\ze) X_\ga$. Here $X_\ga$ is a constant invertible matrix
depending on the homotopy class $[\ga]$ of the path $\gamma$. Thus, $X_\ga$'s
provide a ``monodromy representation'' of the fundamental group of $\C\setminus
\{b_1,\dots,b_n\}$:
$$
X:\pi_1(\C\setminus \{b_1,\dots,b_n\})\to GL(N,\C),\qquad [\gamma]\mapsto
X_\ga.
$$

Now view the singular points $\{b_1,\dots, b_n\}$ as variables. It may
happen that moving these points a little and changing the rational matrix
$B(\ze)$ in an appropriate way, we do not change the monodromy representation. In such a
case we say that we have an {\it isomonodromy deformation} of the initial
differential equation. 

For general information on isomonodromy deformations we refer the reader to \cite{IN},
\cite{JMU}.

Without loss of generality, we can assume that, in the notation of \S5, 
the first $k\le n$ points $\{b_1,\dots,b_k\}$ of the set
$\{b_j\}_{j=1}^{n}$ are exactly those endpoints of $J$ which are not the
endpoints of $\x$. Clearly, $\{b_j\}_{j=1}^k\subset \{a_j\}_{j=1}^{2m}$. 

The following statement is immediate.

\proclaim{Proposition 6.1} Under the assumptions of Theorem 5.1 (or Theorem 5.3),
there exists $\epsilon>0$ with the property that moving the points $b_1,\dots,b_k$ within
their $\epsilon$-neighborhoods inside $\R$ provides an isomonodromy
deformation of the equation \tht{5.1} (or of the equation \tht{5.3},
respectively).   
\endproclaim

Note that the matrices $\{\bb_j\}_{j=1}^n$ are now functions of
$b_1,\dots, b_k$.

\demo{Proof} Choose $\epsilon>0$ so that the points
$b_1,\dots, b_k$ cannot collide between themselves or with the other
endpoints $b_{k+1}, \dots, b_n$. Since the matrix 
$M=m_JmC^{-1}$ has nonzero determinant, this matrix can be viewed as a fundamental
solution of \tht{5.1}. The monodromy of this solution, as we go along any
closed curve which avoids the singular points, is equal to the product of the
values of the jump matrix $V$  or their inverses at the points where the curve
meets $\x$. Since $V$ does not depend on $b_1,\dots,b_k$, the proof is
complete. 
\qed \enddemo

In 1912, Schlesinger realized that if the matrix $B(\ze)$ has
simple poles then a deformation of $b_j$'s preserves monodromy if
and only if the residues $\{\bb_j\}$ of $B$ at the singular points, as functions
of $b_j$'s, satisfy a certain system of nonlinear partial differential
equations. These equations are called the {\it Schlesinger equations}. The
analogs of the Schlesinger equations in the case when $B$ has higher order
poles were derived in \cite{JMU}. 

In what follows we will use the Schlesinger equations arising from
the isomonodromy deformation described in Proposition 6.1. Since our situation
is simpler than the general case in \cite{JMU}, it is more instructive to rederive the equations that we need, rather than to refer to the general theory.

\proclaim{Proposition 6.2 (Schlesinger equations)} (i) The matrices
$\{\bb_j\}_{j=1}^n$ from \tht{5.1}, as functions in $b_1,\dots, b_k$, satisfy
the equations 
$$
\frac{\partial \bb_l}{\partial b_j}=\frac
{[\bb_j,\bb_l]}{b_j-b_l}\,,\qquad   \frac{\partial \bb_j}{\partial
b_j}=\sum_{\Sb 1\le l\le n\\ l\ne j\endSb}\frac {[\bb_j,\bb_l]}{b_l-b_j}\,,
\tag 6.1
$$
where $j=1,\dots,k$, $l=1,\dots,n$.

(ii) The matrices
$\{\bb_j\}_{j=1}^n$ from \tht{5.3}, as functions in $b_1,\dots, b_k$, satisfy
the equations 
$$
\frac{\partial \bb_l}{\partial b_j}=\frac
{[\bb_j,\bb_l]}{b_j-b_l}\,,\qquad   \frac{\partial \bb_j}{\partial
b_j}=\sum_{\Sb 1\le l\le n\\ l\ne j\endSb}\frac
{[\bb_j,\bb_l]}{b_l-b_j}-[\bb_j,D]\,, 
\tag 6.2
$$
where $j=1,\dots,k$, $l=1,\dots,n$.
\endproclaim
\demo{Sketch of the proof} Since $M$ satisfies a RHP with a constant jump matrix
$V$, the derivative $M_{b_j}=\frac {\partial M}{\partial b_j}$ satisfies the
same jump condition, $j=1,\dots,k$. Hence, the matrix $M_{b_j} \, M^{-1}$
has no jump across $\x$. Thus, it is holomorphic in $\C\setminus \{b_j\}$. 
As was shown in the proof of Lemma 4.7, locally near $\ze=b_j$  we have  
$$
\multline
M(\ze)=H(\ze)\exp((C^o)^{-1}f^o(g^o)^t\,\ln(\ze-b_j))\\
=
H(\ze)\left(I+(C^o)^{-1}f^o(g^o)^t\,\ln(\ze-b_j)\right),
\endmultline
$$
where $H$ is holomorphic. With some additional effort, one can show that $H$ is differentiable with respect to $b_j$, and differentiating with respect to $b_j$ we see
that
$$
M_{b_j}(\ze)M^{-1}(\ze)=-\frac{H(b_j)\, (C^o)^{-1}f^og^o\,
H^{-1}(b_j)}{\ze-b_j}+O(1)=-\frac {\bb_j}{z-b_j}+O(1). 
$$
Since
$M\sim I$ at $\ze=\infty$, one can show that $M_{b_j}M^{-1}\to 0$ as $\ze\to\infty$. By Liouville's theorem, 
$M_{b_j}M^{-1}+{\bb_j}/({z-b_j})\equiv 0 $, and
$$
M_{b_j}=-\frac {\bb_j}{\ze-b_j}\,M.
\tag 6.3
$$
The linear equations \tht{5.1} and \tht{6.3} form a {\it Lax
pair} for \tht{6.1}. 

Differentiating \tht{5.1} with respect to $b_j$ and \tht{6.3} with respect to
$\ze$, subtracting the results, and multiplying the difference by $M^{-1}$ on
the right, we obtain
$$
\sum_{l=1}^n \frac{\partial \bb_l}{\partial b_j}\,\frac 1{\ze-b_l}=
\frac 1{\ze-b_j}\sum_{\Sb 1\le l\le n\\ l\ne j\endSb}
\frac{[\bb_l,\bb_j]}{\ze-b_l}. 
$$
The equality of residues at the points $\{b_l\}_{l=1}^n$ on both sides of this
identity gives \tht{6.1}. The equations \tht{6.2} are proved in exactly the
same way. \qed 
\enddemo

\proclaim{Corollary 6.3}
In the notation of Theorem 4.6,
$$
\gather
\frac{\partial \ba}{\partial a_j}=\sum_{j=1}^{2m}\frac{[\bc_j,\ba]}{a_j-\frac
12}\,,\qquad  
\frac{\partial \bb}{\partial a_j}=\sum_{j=1}^{2m}
\frac{[\bc_j,\bb]}{a_j+\frac 12}\,,\quad
\tag 6.4\\
\frac{\partial \bc_l}{\partial a_j}=\frac{[\bc_j,\bc_l]}{a_j-a_l},
\qquad
\frac{\partial \bc_j}{\partial a_j}=-\frac{[\bc_j,\ba]}{a_j-\frac 12}-
\frac{[\bc_j,\bb]}{a_j+\frac 12}-\sum_{\Sb 1\le l\le 2m\\ l\ne j\endSb}
\frac{[\bc_j,\bc_l]}{a_j-a_l}
\,.
\tag 6.5
\endgather
$$ 
Here $j,l=1,2,\dots,2m$, and if $a_1=-\infty$ or $a_{2m}=+\infty$ then the
corresponding terms and equations are removed.
\endproclaim
\demo{Proof} Direct application of Proposition 6.2. \qed
\enddemo

It is known that for any solution of Schlesinger equations there exists 
an associated
remarkable 1-form $\om$ which is closed, see \cite{SMJ}, \cite{JMU}. For the
equations \tht{6.1}, the form of $\om$ is as follows:
$$
\om=\sum_{j=1}^k\sum_{\Sb 1\le l\le n\\ l\ne
j\endSb}\frac{\tr(\bb_j\bb_l)}{b_j-b_l}\,db_j,
\tag 6.6
$$
while for the equations \tht{6.2} the form is different:
$$
\om=\sum_{j=1}^k
\left(\sum_{\Sb 1\le l\le n\\ l\ne
j\endSb}\frac{\tr(\bb_j\bb_l)}{b_j-b_l}+\tr(\bb_jD)\right)\,db_j.
\tag 6.7
$$
\proclaim{Definition 6.4 \cite{JMU}} A function $\tau=\tau(b_1,\dots,b_k)$ is
called a $\tau$-function for the system of Schlesinger equations \tht{6.1} (or
\tht{6.2}) if 
$$
d\ln\tau=\om
$$ 
with $\om$ given by \tht{6.6} {\rm(}or \tht{6.7},
respectively{\rm \,)}.
\endproclaim

The definition can be extended to the most general case of an arbitrary
rational matrix $B(\ze)$, see \cite{JMU}. 

Since $d\om=0$, the $\tau$-function is defined at least locally. Clearly, the
$\tau$-function is unique up to a multiplicative constant.

The following claim is a corollary of much more general statements proved in \cite{Miw} and \cite{Mal}. 
\proclaim{Painlev\'e property}
Any solutions $\{\bb_j\}_{j=1}^n$ of the Schlesinger equations \tht{6.1} or
\tht{6.2} are analytic function in $(b_1,\dots,b_k)$ which have at most poles
in addition to the fixed singularities $b_j=b_l$ for some $j\ne l$. 

The corresponding $\tau$-function
is holomorphic everywhere on the universal covering manifold of  $$
\C^k\setminus \{(b_1,\dots,b_k)\,|\,b_j=b_l\text{  for  some  }j\ne l,\,
j=1,\dots,k,\ l=1,\dots,n\}. $$
\endproclaim 

Let us now return to the general setting of \S5. The next statement is our main result in this section.

\proclaim{Theorem 6.5} Under the assumptions of Theorem 5.1 {\rm (}or Theorem 5.3{\rm \,)}, the Fredholm determinant $\det(1-K^J)$ is the $\tau$-function  for the system of Schlesinger equations \tht{6.1} {\rm(}or \tht{6.2}, respectively{\rm \,)}.
\endproclaim
\demo{Proof} We will give a proof under the assumptions of Theorem 5.1, the case of Theorem 5.3 is handled similarly. 

First of all, by condition (2) of \S5 the operator $K^J$ is trace class. Hence, $\det(1-K^J)$ is well-defined. Note that $(1-K^J)$ is invertible by condition (3). 
By a well--known formula from functional analysis, we have that
$$
\frac{\partial \ln\det(1-K^J)}{\partial b_j}= \pm R^J(b_j,b_j),\quad j=1,\dots,k,
$$
where $R^J=K^J(1-K^J)^{-1}$, the sign ``$+$'' is chosen if $b_j$ is
a left endpoint of $J$, and the sign ``$-$'' is chosen is a right endpoint of $J$. Thus, in order to verify that $d\ln\det(1-K^J)=\om$
we must prove that
$$
R^J(b_j,b_j)=\sum_{\Sb 1\le l\le n\\ l\ne
j\endSb}\frac{\tr(\bb_j\bb_l)}{b_j-b_l}\,,\quad j=1,\dots,k.
\tag 6.8
$$

We give a proof when $b_j$ is a left endpoint of an interval from $J$. The proof for the right endpoints is obtained by changing the sign of $\ze$. 

We have
$$
\multline
R^J(b_j,b_j)=
\lim_{x,y\to b_j}\frac { \G^t(y)\F(x) }{x-y}
=\lim_{x,y\to b_j}\frac{ ((m_{\x\setminus J}^{-t})_+g(y))^t  (m_{\x\setminus J})_- (x)f(x) }{x-y}
\\
=\lim_{x,y\to b_j}\frac { ((M_+C_+)^{-t}g(y))^tM_-C_-f(x) }{x-y}
=\lim_{x,y\to b_j}\frac { (g^o)^tM^{-1}_+(y)M_-(x)f^o }{x-y}
\\
=(g^o)^t(M^{-1}_+M'_-)(b_j)f^o.
\endmultline
$$

The local representation \tht{4.3} of the matrix $M(\ze)$ near the point $\ze=b_j$ implies that
$$
M(\ze)=\cases H_{b_j}(\ze)\exp\left((C^o)^{-1}f^o(g^o)^t\ln(\ze-b_j)\right),&\Im \ze>0,\\
H_{b_j}(\ze)\exp\left((C^o)^{-1}f^o(g^o)^t\ln(\ze-b_j)\right)(C^o)^{-1}, &\Im \ze<0.
\endcases
$$
Hence, for $x\in \x$ near $b_j$, 
$$
\gathered
M^{-1}_+=\exp\left(-(C^o)^{-1}f^o(g^o)^t\ln(x-b_j)\right)H^{-1}_{b_j},\\
M'_-=H'_{b_j}\exp\left((C^o)^{-1}f^o(g^o)^t\ln(x-b_j)\right)(C^o)^{-1}\\+H_{b_j}\,\frac{(C^o)^{-1}f^o(g^o)^t}{x-b_j}
\exp\left((C^o)^{-1}f^o(g^o)^t\ln(x-b_j)\right)(C^o)^{-1},\\
M^{-1}_+M'_-=\frac{(C^o)^{-1}f^o(g^o)^t(C^o)^{-1}}{x-b_j}\\+\exp\left(-(C^o)^{-1}f^o(g^o)^t\ln(x-b_j)\right)H^{-1}_{b_j}H_{b_j}'
\exp\left((C^o)^{-1}f^o(g^o)^t\ln(x-b_j)\right)(C^o)^{-1}
\endgathered
$$
where $H_{b_j}$ is a function holomorphic near $b_j$, and $\det H_{b_j}(b_j)\ne 0$. 

Since $(g^o)^t(C^o)^{-1}f^o=((C^o)^{-1}f^o)^tg^o=0$, we have
$$
\gathered
(g^o)^t\frac{(C^o)^{-1}f^o(g^o)^t(C^o)^{-1}}{x-b_j}=0,\\
(g^o)^t\exp\left(-(C^o)^{-1}f^o(g^o)^t\ln(x-b_j)\right)=(g^o)^t,\\
\exp\left((C^o)^{-1}f^o(g^o)^t\ln(x-b_j)\right)(C^o)^{-1}f^o=(C^o)^{-1}f^o.
\endgathered
$$

Therefore, 
$$
R(b_j,b_j)=(g^o)^t(M^{-1}_+M'_-)(b_j)f^o=(g^o)^t(H_{b_j}^{-1}H'_{b_j})(b_j)(C^o)^{-1}f^o.
\tag 6.9
$$

On the other hand, let us compute the right-hand side of \tht{6.8} through $C^o$, $f^o$, $g^o$, and $H_{b_j}$. As above, locally near $b_j$ we have
$$
M'(\ze)M^{-1}(\ze)=H_{b_j}'(\ze)H_{b_j}^{-1}(\ze)+\frac{H_{b_j}(\ze) (C^o)^{-1}f^o(g^o)^tH_{b_j}^{-1}(\ze)}{\ze-b_j}\,.
$$
Comparing with \tht{5.1}, we conclude that
$$
\bb_j=H_{b_j}(b_j)(C^o)^{-1}f^o(g^o)^tH_{b_j}^{-1}(b_j)
$$
and
$$
\gathered
\sum_{\Sb 1\le l\le n\\l\ne j\endSb}\frac{\bb_l}{b_j-b_l}
=
H'_{b_j}(b_j)H_{b_j}^{-1}(b_j)+H'_{b_j}(b_j)(C^o)^{-1}f^o(g^o)^t
H_{b_j}^{-1}(b_j)\\ -
H_{b_j}(b_j) (C^o)^{-1}f^o(g^o)^tH_{b_j}^{-1}(b_j)H_{b_j}'(b_j)
H^{-1}_{b_j}(b_j).
\endgathered
$$

Multiplying these two relations, taking the trace of both sides, and
using the fact that $(g^o)^t(C^o)^{-1}f^o=0$, we obtain
$$
\tr(H'_{b_j}(b_j)(C^o)^{-1}f^o(g^o)^tH_{b_j}^{-1}(b_j))=\sum_{\Sb 1\le l\le n\\l\ne j\endSb}\frac{\tr\bb_l\bb_j}{b_j-b_l}\,.
$$
But the left-hand side of the last equality equals the right-hand side of
\tht{6.9}. This concludes the proof of \tht{6.8}. \qed 

\enddemo

\proclaim{Corollary 6.6} Let $K$ be the continuous ${}_2F_1$ kernel of \S2 and assume that 
$$
z+z'+w+w'>0,\quad |z+z'|<1,\quad |w+w'|<1.
$$
 Then, in the notation of Theorem 4.6, $\det(1-K^J)$ is the $\tau$-function of the Schle\-sin\-ger equations \tht{6.4}, \tht{6.5}, where the matrices $\ba$, $\bb$, $\{\bc_j\}_{j=1}^{2m}$ satisfy the conditions stated in Theorem 4.6.
\endproclaim
\demo{Proof} Direct application of Theorem 6.5. \qed
\enddemo

Note that the restrictions on the parameters $z,z',w,w'$ come from Corollary 3.7 (see also Remark 3.8).

\head 7. Painlev\'e VI
\endhead

In this section we consider the case of the ${}_2F_1$ kernel acting on $J=(s,+\infty)$ for
$s>\frac 12$. We will show that the Fredholm
determinant $\det(1-K_s)=\det (1-K|_{(s,+\infty)})$  can be expressed through a solution of the
Painlev\'e VI equation. The appearance of the PVI equation is to be
expected from the general results of \cite{JMU}; the precise form of the
equation is not clear in general, and requires considerable calculations,
as we now show.

Our goal is to prove the following claim.

\proclaim{Theorem 7.1} Let $K_s$ be the restriction of the continuous
${}_2F_1$ kernel to the interval $(s,+\infty)$, $s>\frac 12$. Assume that
$\s=z+z'+w+w'>0$, $|z+z'|<1$, $|w+w'|<1$. Then the function
$$
\si(s)=\left(s-\tfrac 12\right)\left(s+\tfrac 12\right)\,\frac
{d\ln\det(1-K_s)}{ds}-
\nu_1^2 s+\frac{\nu_3\nu_4}2 
$$
satisfies the differential equation
$$
\gathered
-\si'\left(\left(s-\tfrac 12\right)\left(s+\tfrac
12\right)\si''\right)^2=\left(2\left(s\si'-\si\right)\si'
-\nu_1\nu_2\nu_3\nu_4\right)^2
\\-
(\si'+\nu_1^2)(\si'+\nu_2^2)(\si'+\nu_3^2)(\si'+\nu_4^2),
\endgathered
\tag 7.1
$$ 
where
$$
\nu_1=\nu_2=\frac{z+z'+w+w'}2\,,\quad \nu_3=\frac{z-z'+w-w'}2\,,\quad
\nu_4=\frac{z-z'-w+w'}2\,.
$$
\endproclaim
\example{Remarks 7.2} 1. The equation \tht{7.1} is the so-called
Jimbo-Miwa $\si$-version of the Painlev\'e VI equation, see \cite{JM, Appendix C}.  
It is easily reduced to the standard form of the Painlev\'e VI, see \cite{JM},
\cite{Mah}.
\medskip
\noindent 2. As $s\to+\infty$,  
$$
\frac
{d\ln\det(1-K_s)}{ds}\sim
K(s,s)=\psiout(s)(\Rout'(s)\Sout(s)-\Sout'(s)\Rout(s)). 
\tag 7.2
$$
The error term in this asymptotic relation is of order
$\int_{s}^{+\infty}K(s,y)K(y,s)dy$. 
Using the leading asymptotic terms
$$
\psiout(s)\sim \frac{\sin\pi z\sin\pi z'}{\pi^2}\, s^{-\s},\quad \Rout(s)\sim 1, \quad \Sout(s)\sim \const
s^{-1},
$$
we see that $K(s,s)=O(s^{-\s-2})$, and  
$\int_{s}^{+\infty}K(s,y)K(y,s)dy=O(s^{-2\s-3})$. Hence,
$$
\si(s)=-\nu_1^2s +\frac{\nu_3\nu_4}2+\frac{\sin\pi z\sin\pi z'}{\pi^2}\,s^{-2\nu_1}+o(s^{-2\nu_1}).
$$
This expansion determines $\si(s)$ uniquely as a solution of \tht{7.1} by a result of O.~Costin and R.~D.~Costin \cite{Cos}.

\medskip
\noindent 3. The restrictions $\s>0$, $|z+z'|<1$, and $|w+w'|<1$ are taken from
Corollary 3.7. Most likely, they can be removed from Corollary 3.7, and
hence from Theorem 7.1, see Remark 3.8. Another possible way of removing these
restrictions from Theorem 7.1 is to prove that the Fredholm determinant
$\det(1-K_s)$ and its derivatives with respect to $s$, which are well-defined
for all admissible sets of parameters (see the end of \S2) are real-analytic
function of the parameters. Then the result would follow by analytic continuation.  

\medskip
\noindent 4. The equation \tht{7.1} depends only on 3 independent parameters:
the shifts
$$
z\mapsto z+\alpha, \quad z'\mapsto z'+\al, \quad w\mapsto w-\al, \quad
w\mapsto w'-\al
$$
do not change the values of $\nu_1,\dots,\nu_4$. However, the solution of \tht{7.1} which is of interest here, depends nontrivially on all four
parameters, as can be seen from the above asymptotic expansion. 

\medskip
\noindent 5. The proof of Theorem 7.1 follows the derivation of the Painlev\'e
VI equation from Schlesinger equations given in \cite{JM, Appendix C}, see also
\cite{Mah} for a more detailed description.

\endexample

\demo{Proof of Theorem 7.1}
By Theorem 4.6, the matrix $M$ satisfies a differential equation 
$$
\frac {d}{d\ze}M(\ze)=\left(\frac \ba{\ze-\frac 12}+\frac \bb{\ze +\frac
12}+\frac \bc{\ze-s}\right)M(\ze)
$$
with some constant matrices $\ba$, $\bb$, and $\bc$, such that
$$
\gathered
\tr\ba=\tr\bb=\tr\bc=0,\\
\det\ba=-\left(\frac{z-z'}2\right)^2,\quad 
\det\bb=-\left(\frac{w-w'}2\right)^2,\quad \det\bc=0,
\endgathered
$$
and
$$
\ba+\bb+\bc=-\frac{\s}2\,\sig, \quad \sig=\bmatrix
1&0\\0&-1\endbmatrix. 
$$

By Corollary 6.3, the
matrices $\ba$, $\bb$, $\bc$ satisfy the Schlesinger equations 
$$
\gather
\frac{\partial \ba}{\partial s}=\frac{[\bc,\ba]}{s-\frac 12}\,,\quad 
\frac{\partial \bb}{\partial s}=\frac{[\bc,\bb]}{s+\frac 12}\,,\quad
\tag 7.3\\
\frac{\partial \bc}{\partial s}=-\frac{[\bc,\ba]}{s-\frac 12}-
\frac{[\bc,\bb]}{s+\frac 12}\,.
\tag 7.4
\endgather
$$ 

Introduce the notation
$$
\tha=\frac{z-z'}2\,,\quad \thb=\frac{w-w'}2\,.
$$
Set
$$
\hsi(s)=\left(s-\tfrac
12\right)\left(s+\tfrac
12\right)\frac{d\ln\det(1-K_s)}{ds}=\si(s)+\frac{\s^2}4s-\frac{\tha^2-\thb^2}2\,.
$$ 
\proclaim{Lemma 7.3} $\hsi(s)=\tr\left[\left(\left(s+\frac
12\right)\ba+\left(s-\frac 12\right)\bb\right)\bc\right].$
\endproclaim
\demo{Proof} Follows from \tht{6.6}, Definition 6.4, and Corollary 6.6.\qed
\enddemo

Write the matrices $\ba$ and $\bc$ in the form
$$
\ba=\bmatrix \za&\xa\\ \ya&-\za\endbmatrix,\quad \bc=\bmatrix \zc&\xc\\
\yc&-\zc\endbmatrix
\tag 7.5
$$
with
$$
\xa\ya=-\det\ba-\za^2=\tha^2-\za^2,\quad \xc\yc=-\det \bc-\zc^2=-\zc^2.
\tag 7.6
$$
\proclaim{Lemma 7.4}
$ \hsi'=-\s\zc.$
\endproclaim
\demo{Proof} Lemma 7.3 implies
$$
\multline
\hsi'(s)=\tr((\ba+\bb)\bc)+\left(s+\tfrac 12\right)\tr(\ba'\bc)\\+\left(s-\tfrac
12\right)\tr(\bb'\bc)+\tr\left[\left(\left(s+\tfrac 12\right)\ba+\left(s-\tfrac
12\right)\bb\right)\bc'\right].
\endmultline
$$
The Schlesinger equations \tht{7.3}, \tht{7.4} imply that last three terms
vanish due to the identity $\tr([X,Y]X)=0$. 
Further, since $\ba+\bb=-\frac\s
2\sig-\bc$ and $\bc^2=0$, we have
$$
\hsi'(s)=\tr\left(\left(-\frac\s2\sig-\bc\right)\bc\right)=-\frac\s
2\tr(\sig \bc)=-\s \zc.\qed
$$
\enddemo

\proclaim{Lemma 7.5} 
$$
\left(s-\tfrac 12\right)\left(s+\tfrac
12\right)\hsi''(s)=-\dfrac \s 2\tr(\sig[\ba,\bc])=\s(\xc\ya-\xa\yc).
$$
\endproclaim
\demo{Proof} Differentiating the equality $\hsi'(s)=-\frac\s 2\tr(\sig \bc)$
and using the equation \tht{7.4} we get
$$
%\multline
\left(s-\tfrac 12\right)\left(s+\tfrac
12\right)\hsi''(s)=-\frac\s 2\tr\left(\sig\left[\left(s+\tfrac
12\right)\ba+\left(s-\tfrac 12\right)\bb,\bc\right]\right). 
%\endmultline
$$
Substituting $\bb=-\frac \s 2\,\sig-\ba-\bc$ and simplifying we arrive at
the first equality. The second equality follows from the explicit form of
matrices $\ba$ and $\bc$, see \tht{7.5}.\qed 
\enddemo
\proclaim{Lemma 7.6} $(s-\frac
12)\hsi'(s)-\hsi(s)=-\tr(\ba\bc)=-(\xa\yc+\xc\ya+2\za\zc)$. 
\endproclaim
\demo{Proof} We have
$$
\gather
\left(s-\tfrac 12\right)\hsi'(s)-\hsi(s)=-\frac\s 2\left(s-\tfrac 12\right)
\tr(\sig \bc)-\tr\left[\left(\left(s+\tfrac
12\right)\ba+\left(s-\tfrac 12\right)\bb\right)\bc\right]\\
=-\frac\s 2\left(s-\tfrac 12\right)
\tr(\sig \bc)-\tr\left[\left(\left(s+\tfrac
12\right)\ba+\left(s-\tfrac 12\right)\left(-\frac\s
2\,\sig-\ba-\bc\right)\right)\bc\right]\\=-\tr(\ba\bc), 
\endgather
$$
where we used Lemmas 7.3, 7.4 and the relations $\bb=-\dfrac \s
2\,\sig-\ba-\bc$ and $\bc^2=0$. The second equality follows from \tht{7.5}.
\qed
\enddemo
\proclaim{Lemma 7.7} $(s+\frac 12)\hsi'(s)-\hsi(s)=\s
\za+\tha^2-\thb^2+\s^2/4.$ \endproclaim
\demo{Proof} We have
$$
\gather
-\tr(\ba\bc)=\tr\left(\ba\left(\frac \s 2 \,\sig +\bb+\ba\right)\right)\\=
\frac 12\,\left(\tr(\ba+\bb)^2+\tr\ba^2-\tr\bb^2\right)+\frac \s
2\tr(\ba\sig)\\ =\frac 12\left(\tr\left(\frac \s
2\,\sig+\bc\right)^2+\tr\ba^2-\tr\bb^2\right)+\s \za\\
=\s\za+\s\zc+\tha^2-\thb^2+\s^2/4,
\endgather
$$
where we used the equalities 
$$
\tr\ba^2=2\tha^2,\quad \tr\bb^2=2\thb^2,\quad \bc^2=0,\quad \tr\sig^2=2.
$$
Lemmas 7.4 and 7.6 conclude the proof. \qed
\enddemo

Now we use the following trick to derive the differential equation for $\si$.
We learned this trick from \cite{JM} in which the authors refer further to \cite{Oka}.

From Lemmas 7.5 and 7.6 we know that
$$
\align
\xc\ya-\xa\yc&=\frac 1\s\left(s-\tfrac 12\right)\left(s+\tfrac
12\right)\hsi''(s),\\
-(\xa\yc+\xc\ya)&=\left(s-\tfrac
12\right)\hsi'(s)-\hsi(s)+2\za\zc.
\endalign
$$
Squaring these equalities and then subtracting the first one from the
second one, we obtain
$$
\gather
4\xa\xc\ya\yc=\left(\left(s-\tfrac
12\right)\hsi'(s)-\hsi(s)+2\za\zc\right)^2\\-\frac 1{\s^2}\left(\left(s-\tfrac
12\right)\left(s+\tfrac 12\right)\hsi''(s)\right)^2.
\endgather
$$
But \tht{7.6} implies that $\xa\xc\ya\yc=(\za^2-\tha^2)\zc^2.$ This gives
$$
\gathered
4(\za^2-\tha^2)\zc^2=\left(\left(s-\tfrac
12\right)\hsi'(s)-\hsi(s)+2\za\zc\right)^2\\-\frac 1{\s^2}\left(\left(s-\tfrac
12\right)\left(s+\tfrac 12\right)\hsi''(s)\right)^2.
\endgathered
\tag 7.7
$$
Next, Lemmas 7.4 and 7.7 provide expressions for $\zc$ and $\za$ via
$\hsi(s)$. Namely,
$$
\zc=-\frac {\hsi'(s)}\s, \quad \za=\frac 1\s\left(\left(s+\tfrac
12\right)\hsi'(s)-\hsi(s)-\tha^2+\thb^2-\frac {\s^2} 4\right). 
$$
Substituting these relations into \tht{7.7} we can obtain a differential
equation for $\hsi$. But we can also rewrite everything in terms of $\si(s)$. 
We have
$$
\gathered
\hsi''(s)=\si''(s),\\
\left(s-\tfrac
12\right)\hsi'(s)-\hsi(s)=\left(s-\tfrac
12\right)\si'(s)-\si(s)+\frac{\tha^2-\thb^2}2-\frac{\s^2}8,\\
\zc=-\frac 1\s\left(\si'(s)+\frac{\s^2}4\right),\\
\za=\frac 1\s\left(\left(s+\tfrac
12\right)\si'(s)-\si(s)-\frac{\tha^2-\thb^2}2-\frac {\s^2} 8\right).
\endgathered
$$
Substituting this into \tht{7.7} we have
$$
\gathered
-\frac 1{\s^2}\left(\left(s-\tfrac
12\right)\left(s+\tfrac 12\right)\si''(s)\right)^2\\=\frac 4{\s^2}\left(\frac
1{\s^2}\left(\left(s+\tfrac 12\right)\si'(s)-\si(s)-\frac{\tha^2-\thb^2}2-\frac
{\s^2} 8\right)^2-\tha^2\right)\left(\si'(s)+\frac{\s^2}4\right)^2\\
-\Biggl(\left(s-\tfrac
12\right)\si'(s)-\si(s)+\frac{\tha^2-\thb^2}2-\frac{\s^2}8\\
-\frac 2{\s^2}\left(\left(s+\tfrac
12\right)\si'(s)-\si(s)-\frac{\tha^2-\thb^2}2-\frac {\s^2} 8\right)
\left(\si'(s)+\frac{\s^2}4\right)
\Biggr)^2.
\endgathered 
$$
Purely algebraic manipulations show that the equation above after the
multiplication by $\s^2\si'(s)$ turns into the equation \tht{7.1}. Note that in this notation 
$$
\nu_1=\nu_2=\frac{\s}2,\quad \nu_3=\tha+\thb,\quad \nu_4=\tha-\thb.\qed
$$
\enddemo

\head 8. Other kernels
\endhead

\subhead 8.1. The Jacobi kernel \endsubhead
We introduce some notation related to the Jacobi polynomials. Our notation follows \cite{Er, 10.8}. 

Let $\{P_n=P_n^{(\al,\be)}(x)\}_{n=0}^\infty$ be the system of orthogonal polynomials on $(-1,1)$,
$\deg P_n=n$, with respect to the weight function $w(x)=(1-x)^\al(1+x)^\be$,
where $\al$ and $\be$ are real constants, $\al,\be>-1$. The normalization is
determined from the relation  $$
P_n(1)=\binom{n+\al}n=\frac{(\al+1)_n}{n!},
$$
where $(a)_k=\Ga(a+k)/\Ga(a)$ is the Pochhammer symbol. The $P_n$'s are  the
{\it Jacobi polynomials} with parameters $\al$ and $\be$. Let us denote by
$h_n$ the square of the norm of $P_n$ in $L^2((-1,1),w(x)dx)$ and by $k_n>0$ the
highest coefficient of $P_n$:
$$
h_n=\int_{-1}^1 P_n^2(x)w(x)dx,\qquad P_n(x)=k_nx^n+\{\text{ lower order terms
}\}. 
$$
The explicit form of these constant is known, see \cite{Er, 10.8},
$$
h_n=\frac{2^{\al+\be+1}\Ga(n+\al+1)\Ga(n+\be+1)}{(2n+\al+\be+1)n!
\Ga(n+\al+\be+1)}\,,\quad k_n=2^{-n}\binom{2n+\al+\be}{n}\,.
$$
The Jacobi polynomials are expressed through the Gauss hypergeometric
function ${}_2F_1$
$$
P_n(x)=\binom{n+\al}{n}\,{}_2F_1\left[\matrix -n,\, n+\al+\be+1\\
\al+1\endmatrix\,\Biggl|\,\frac{1-x}2\right].
$$
The {\it Jacobi functions of the second kind} $Q_n(x)=Q_n^{(\al,\be)}(x)$ are
defined by the formula
$$
\multline
Q_n(x)=\frac {2^{n+\al+\be}\Ga(n+\al+1)\Ga(n+\be+1)}{\Ga(2n+\al+\be+2)}\,
\\ \times (x-1)^{-n-\al-1}(x+1)^{-\be}\,{}_2F_1\left[\matrix n+1,\, n+\al+1\\
2n+\al+\be+2\endmatrix\,\Biggl|\,\frac 2{1-x}\right].
\endmultline
$$
$Q_n$ satisfies the same second order differential equation as $P_n$. The
Jacobi functions of the second kind are related to the Jacobi polynomials by a
number of well--known formulas, see \cite{Er, 6.8}, \cite{Sz, \S4.6} for details.

\proclaim{Proposition 8.1} For any $n=1,2,\dots$, take two arbitrary integers
$k$ and $l$ such that $k+l=n$, and set
$$
z=k,\quad z'=k+\al,\quad w=l, \quad w'=l+\be.
$$
Then 
$$
\gathered
\psiout(x)\equiv 0,\quad
\psiin(x)=\frac{(1-2x)^{2k+\al}(1+2x)^{2l+\be}}{2^{2n+\al+\be}}=
\frac{(1-2x)^{2k}(1+2x)^{2l}w(2x)}{2^{2n+\al+\be}}\,,\\
\Rin(x)=\frac{(-1)^{k+1}2^n
n!\Ga(n+\al+\be+1)}{\Ga(2n+\al+\be+1)}\,(1-2x)^{-k} (1+2x)^{-l}P_n(2x),\\
\Sin(x)=\frac{(-1)^{k+1}2^n\Ga(2n+\al+\be)}{\Ga(n+\al)\Ga(n+\be)}\,(1-2x)^{-k}
(1+2x)^{-l}P_{n-1}(2x). \endgathered
$$
\endproclaim
\demo{Proof} Follows from the direct comparison of formulas. The relation
$$
\frac{\sin(\pi z')\Ga(z-z')}{\pi}=\frac{(-1)^{k}\sin(\pi
(z'-z))\Ga(z-z')}{\pi}=\frac{(-1)^{k}}{\Ga(1+z'-z)}=\frac{(-1)^{k}}{\Ga(1+\al)}
$$
should be used along the way. \qed
\enddemo
\example{Remark 8.2} The functions $\Rout$ and $\Sout$ can be similarly
expressed through $Q_{n-1}$ and $Q_n$, respectively. Since, we do not use
the corresponding formulas below, we leave their derivation to the
interested reader. 
\endexample

The {\it $n$th Christoffel-Darboux kernel} for the Jacobi polynomials is
given by the formula
$$
K_N^{CD}(x,y)=\sum_{j=0}^{n-1}\frac{P_j(x)P_j(y)}{h_j}
=\frac{k_{n-1}}{k_nh_{n-1}}\,
\frac{P_n(x)P_{n-1}(y)-P_{n-1}(x)P_n(y)}{x-y}\,.
$$
We define the {\it $n$th Jacobi kernel} on the interval $(-\frac 12,\frac 12)$
by the formula
$$
K_n^{\text{Jac}}(x,y)=2 K_n^{CD}(2x,2y)\sqrt{w(2x)w(2y)},\qquad
x,y\in\left(-\tfrac 12,\tfrac 12\right). 
$$
The corresponding integral operator $K_n^{\text{Jac}}$ is the orthogonal
projection in $L^2((-\frac 12,\frac 12),dx)$ onto the $n$-dimensional 
subspace spanned by 
$$
\left(\tfrac 12-x\right)^{\frac \al2}\left(\tfrac 12+x\right)^{\frac \be2},\ x\left(\tfrac
12-x\right)^{\frac \al2}\left(\tfrac 12+x\right)^{\frac \be2},\ \dots,\ x^{n-1}\left(\tfrac
12-x\right)^{\frac \al2}\left(\tfrac 12+x\right)^{\frac \be2}.  
$$
\proclaim{Proposition 8.3}
Under the assumptions of Proposition 8.1,
$$
\gathered
K_{\out,\out}=K_{\out,\inr}=K_{\inr,\out}=0,\quad
K_{\inr,\inr}=K_n^{\text{Jac}},
\endgathered
$$
where $K$ is the continuous ${}_2F_1$ kernel.
\endproclaim
\demo{Proof} The vanishing follows from the vanishing of $\psiout$, which, in
turn, follows from the vanishing of $\sin\pi z$ and $\sin\pi w$. The equality
$K_{\inr,\inr}=K_n^{\text{Jac}}$ follows from the definition of both kernels
and Proposition 8.1. \qed \enddemo 

Thus, the Jacobi kernel can be viewed as a special case of the ${}_2F_1$
kernel. Our next step is to extend the results of \S6 and \S7 to this kernel. 

Let $J=(a_1,a_2)\cup\dots\cup(a_{2m-1},a_{2m})$ be a finite union of disjoint
intervals inside $(-\frac 12, \frac 12)$. It may happen that $a_1=-\frac 12$
or $a_{2m}=\frac 12$. However, we require $J$ to be a proper subset of
$(-\frac 12, \frac 12)$.

\proclaim{Proposition 8.4} Assume that $0<\al<\frac 12$, $0<\be<\frac 12$.
Then the Jacobi kernel satisfies the conditions \tht{1}--\tht{7} of \S5.
\endproclaim
\demo{Proof}
\tht{1} follows from the fact that $K_n^{\text{Jac}}$ coincides with
the ${}_2F_1$ kernel for a specific set of parameters (Proposition 8.3), and
for that kernel the condition was verified in Proposition 3.2.  

\tht{2} is obvious, since $K_n^{\text{Jac}}$ is a finite rank operator.

\tht{3} follows from the fact that $K_n^{\text{Jac}}$ is a projection on a
finite-dimensional space, and the range of this projection
intersects $L^2(J,dx)$ trivially (here we used the condition that $J$ is a
proper subset of $(-\frac 12,\frac 12)$). 

\tht{4} follows from the explicit form of the kernel (here we use the
condition $\al,\be>0$, which guarantees the boundedness near the points
$\pm\frac 12$).

\tht{5} is exactly the same as for the ${}_2F_1$ kernel.

\tht{6} is the only nontrivial condition. If $a_1\ne -\frac 12$ and $a_{2m}\ne
\frac 12$ then the claim follows from Propositions 3.3 and 3.4, as for
the ${}_2F_1$ kernel. Now assume that $a_1=-\frac 12$. Since $0<\be<\frac 12$,
Proposition 3.4 implies that $m(\ze)C^{-1}(\ze)$ is locally in $L^4$ on
any smooth curve passing through $\ze=-\frac 12$. By Proposition A.2, $m_J$ is
locally $L^2$, hence, $M=m_Jm {C^{-1}}$ is locally in $L^{4/3}$. 

The jump
matrix $V$ for $M$ locally near $-\frac 12$ coincides with the jump matrix
$C^o=C_-C_+^{-1}=C_+^{-1}C_-$ for $C^{-1}$, see \S5. This means that
$H(\ze)C^{-1}(\ze)$ is a local solution of the RHP for $M$ for any locally
holomorphic $H(\ze)$. Set $M^o=HC^{-1}$ with
$$
H(\ze)=\bmatrix (\ze+\frac 12)^w&0\\0&(\ze+\frac 12)^{-w}\endbmatrix.
$$
Note that $H$ has no branch at $-\frac 12$, because $w=l\in\Z$. Then
$$
M^o(\ze)=\bmatrix (\ze-\frac 12)^{-\frac{z+z'}2}(\ze+\frac
12)^{\frac{w-w'}2}&0\\0&(\ze-\frac 12)^{\frac{z+z'}2}(\ze+\frac
12)^{\frac{w'-w}2}\endbmatrix.
$$
Hence, $(M^o(\ze))^{-1}$ (as well as $M^o(\ze)$) is locally in $L^4$, because $w'-w=\be\in (0,\frac 12)$. Thus, $M(M^o)^{-1}$ is a locally
$L^1$ function with no jump across $\R$. This means that near $\ze=\frac 12$
$$
M(\ze)=H_{-\frac 12}(\ze)\bmatrix (\ze+\frac
12)^{-\frac{\be}2}&0\\0&(\ze+\frac
12)^{\frac{\be}2}\endbmatrix
$$
for some locally holomorphic function $H_{-\frac 12}(\ze)$ such that
$H_{-\frac 12}(-\frac 12)$ is nonsingular. Hence
$$
M'M^{-1}(\ze)=\frac \bb{\ze+\frac 12}+\text{ a locally holomorphic function },
$$
where $\bb$ has eigenvalues $\beta/2$ and $-\beta/2$.

The argument in the case $a_{2m}=\frac 12$ is similar, and the eigenvalues of
the residue $\ba$ of $M'M^{-1}$ at $\ze=\frac 12$ are equal to $\al/2$ and
$-\al/2$.

Finally, the condition \tht{7} for the Jacobi kernel follows from that for the
${}_2F_1$ kernel. \qed
\enddemo

Now, by Theorem 5.1, for $\al,\be\in (0,\frac 12)$, the matrix $M$
corresponding to the Jacobi kernel, satisfies the differential equation (cf.
Theorem 4.6)
$$
M'(\ze)=\left(\frac \ba{\ze-\frac 12}+\frac \bb{\ze +\frac
12}+\sum_{j=1}^{2m}\frac {\bc_j}{\ze-a_j}\right)M(\ze)
$$
for some constant matrices $\ba$, $\bb$, and $\{\bc_j\}_{j=1}^{2m}$. 
If $a_1=-\frac 12$ then $\bc_1=0$, and if $a_{2m}=\frac 12$ then $\bc_{2m}=0$.

Moreover,
$$
\gathered
\tr\ba=\tr\bb=\tr\bc_j=0,\\
\det\ba=-{\al^2}/4,\quad 
\det\bb=-{\be^2}/4,\quad \det\bc_j=0,
\endgathered
$$
for all $j=1,\dots,2m$, and
$$
\ba+\bb+\sum_{j=1}^{2m}\bc_j=-n-\frac{\al+\be}2\,\sig, \quad \sig=\bmatrix
1&0\\0&-1\endbmatrix. 
$$ 

Further, by Proposition 6.2 the matrices $\ba,\, \bb,\, \{\bc_j\}_{j=1}^{2m}$,
satisfy the Schlesinger equations \tht{6.4} and \tht{6.5}. Finally, Theorem
6.5 implies
\proclaim{Theorem 8.5} Assume that $0<\al,\,\be<\frac 12$. Then the Fredholm
determinant $\det(1-K_n^{\text{Jac}}|_J)$, where $K_n^{\text{Jac}}$ is the
Jacobi kernel, is the $\tau$-function for the system of Schlesinger equations
\tht{6.4}, \tht{6.5} with matrices $\ba,\, \bb,\, \{\bc_j\}_{j=1}^{2m}$
satisfying the conditions stated above. \endproclaim

Similarly to the ${}_2F_1$ kernel, the cases when $J=(-\frac 12,s)$ or $J=(s,\frac 12)$
lead to the Painlev\'e VI equation. Note that there are no restrictions on $\al$ and $\be$.

\proclaim{Theorem 8.6 \cite{HS}} Let $K_s$ be the restriction of the
Jacobi kernel $K_n^{\text{Jac}}$ kernel to either the interval $(-\frac 12,s)$, $s<1$,
or to the interval $(s,\frac 12)$, $s>-1$. Then the function 
$$
\si(s)=\left(s-\tfrac 12\right)\left(s+\tfrac 12\right)\frac
{d\ln\det(1-K_s)}{ds}-
\left(n+\frac{\al+\be}2\right)^2\,s+\frac{\al^2-\be^2}8
$$
satisfies the differential equation
$$
\gathered
-\si'\left(\left(s-\tfrac 12\right)\left(s+\tfrac
12\right)\si''\right)^2=\left(2\left(s\si'-\si\right)\si'
-\nu_1\nu_2\nu_3\nu_4\right)^2
\\-
(\si'+\nu_1^2)(\si'+\nu_2^2)(\si'+\nu_3^2)(\si'+\nu_4^2),
\endgathered
$$ 
where
$$
\nu_1=\nu_2=n+\frac{\al+\be}2\,,\quad \nu_3=\frac{\al+\be}2\,,\quad
\nu_4=\frac{\al-\be}2\,.
$$
\endproclaim 
\demo{Proof} Simply repeat the proof of Theorem 7.1. Note that
in this way we only prove the theorem for $\al,\be\in(0,\frac 12)$. But
for the finite-dimensional Jacobi kernel it is obvious that the determinant
$\det(I-K_s)$ and all its derivatives depend on the parameters $\al$ and $\be$
analytically. That is the reason why we can remove the additional restrictions
on $\al$ and $\be$. \qed
\enddemo

\subhead 8.2. The Whittaker kernel \endsubhead
The Whittaker kernel, which we are about to introduce, plays the same role in harmonic analysis on the infinite symmetric group as the ${}_2F_1$
kernel plays in the harmonic analysis on the infinite-dimensional unitary
group, see \S1. The problem for the infinite symmetric group was 
investigated by G.~Olshanski and one of the authors in a series of papers, see
\cite{P.I-P.V}, \cite{BO1--3}, \cite{Bor1}. 
For a brief summary we refer the reader to \cite{BO1}, \cite{BO3, \S3}, \cite{Bor1, Introduction}.

Split the space $\x=\R\setminus\{0\}$ into two parts 
$$
\x=\x_+\sqcup \x_-,\qquad \x_+=\R_+,\quad \x_-=\R_-.
$$
Let $z,z'$ be two complex nonintegral numbers such that either $z'=\bar{z}$, or
$z$ and $z'$ are both real and $k<z,z'<k+1$ for some $k\in \Z$.

The functions
$$
\psi_+:\x_+\to\R_+,\quad \psi_-:\x_-\to\R_+
$$
are defined by the formulas
$$
\psi_+(x)=C(z,z') x^{-z-z'} e^{-x},\quad \psi_-(x)=(-x)^{z+z'} e^x,
$$
where $C(z,z')=\sin\pi z\sin\pi z'/\pi^2$, as before.
The Whittaker kernel is a kernel on $\x$, which in block form
$$
K=\bmatrix K_{+,+}&K_{+,-}\\ K_{-,+}& K_{-,-}\endbmatrix
$$
corresponding to the splitting $\x=\x_+\sqcup \x_-$, is given by:
 $$
\gathered
K_{+,+}(x,y)=\sqrt{\psi_+(x)\psi_+(y)}
\,\frac {R_+(x)S_+(y)-S_+(x)R_+(y)}{x-y}\,,\\
K_{+,-}(x,y)=\sqrt{\psi_+(x)\psi_-(y)}
\,\frac {R_+(x)R_-(y)-S_+(x)S_-(y)}{x-y}\,,\\
K_{-,+}(x,y)=\sqrt{\psi_-(x)\psi_+(y)}
\,\frac {R_-(x)R_+(y)-S_-(x)S_+(y)}{x-y}\,,\\
K_{-,-}(x,y)=\sqrt{\psi_-(x)\psi_-(y)}
\,\frac {R_-(x)S_-(y)-S_-(x)R_-(y)}{x-y}\,,
\endgathered
$$
where
$$
\aligned
R_+(x)&=x^{\frac{z+z'-1}2}e^{\frac x2}\,W_{\frac{-z-z'+1}2,\frac
{z-z'}2}(x)\,,\\ 
S_+(x)&=\Gamma(z+1)\Gamma(z'+1)\,x^{\frac{z+z'-1}2}e^{\frac
x2}\,W_{\frac{-z-z'-1}2,\frac {z-z'}2}(x)\,,\\
R_-(x)&=(-x)^{\frac{-z-z'-1}2}e^{-\frac x2}\,W_{\frac{z+z'+1}2,\frac
{z-z'}2}(-x)\,,
\\ 
S_-(x)&=-\frac 1{\Gamma(z)\Gamma(z')}(-x)^{\frac{-z-z'-1}2}e^{-\frac
x2}\,W_{\frac{z+z'-1}2,\frac {z-z'}2}(-x)\,.
\endaligned
$$
Here $W_{\kappa,\mu}(x)$ is the Whittaker function, see \cite{Er, 6.9}.

In the definition of the Whittaker kernel above we have switched the signs of the
parameters $z$ and $z'$, compared to the standard notation.
The reason for the switch is the following.

\proclaim{Proposition 8.7} The Whittaker kernel $K^W$ can be realized as a
scaling limit of the ${}_2F_1$ kernel $K^{F}$, 
$$
K^W(x,y)=\lim_{\varepsilon\to +0}\varepsilon \cdot K^F\left(\tfrac
12+\varepsilon x,\tfrac 12+\varepsilon y\right), \quad x,\,y\in \R\setminus
\{0\}, 
$$
where for the ${}_2F_1$ kernel we set $w=\varepsilon^{-1}$, $w'=0$, and
the parameters $(z,z')$ for both kernels are the same.
\endproclaim
\demo{Proof} Using the well-known limit relations ($a,b,c\in \C$)
$$
\gathered
\lim_{\varepsilon\to +0}
(1+a\varepsilon)^{1/\varepsilon}=e^a,\quad
\lim_{\varepsilon\to +0}
\varepsilon^{a-b}\frac{\Ga(a+\varepsilon^{-1})}{\Ga(b+\varepsilon^{-1})}=1,
\\ 
\lim_{\varepsilon\to +0} {}_2F_1\left[ \matrix a,\, b\\
1/\varepsilon +c\endmatrix \, \Biggl |\, -\frac {1}{\varepsilon
x}\right]=x^{\frac{a+b-1}2}e^{\frac x2}W_{\frac {-a-b+1}2\,\frac
{a-b}2}(x),\quad x\notin \R_-, 
\endgathered
$$ 
we see that
(remember  
$w=\varepsilon^{-1}$, $w'=0$)
$$
\gathered 
\lim_{\varepsilon\to +0}\varepsilon^{z+z'}\psiout\left(\tfrac 12+\varepsilon
x\right)=\psi_+(x),\quad x>0,\\
\lim_{\varepsilon\to +0}\varepsilon^{-z-z'}\psiin\left(\tfrac 12+\varepsilon
x\right)=\psi_-(x),\quad x<0,\\
\lim_{\varepsilon\to +0}\Rout\left(\tfrac 12+\varepsilon
x\right)=R_+(x),\quad \lim_{\varepsilon\to
+0}\varepsilon^{-z-z'}\Sout\left(\tfrac 12+\varepsilon x\right)=S_+(x),\quad
x>0. 
\endgathered
$$

Further, if we identify $R_+$ and $S_+$ with their analytic continuations, then
on $\R_-$ we have
$$
\frac 1{\psi_-}\,\frac{S_+^--S_-^+}{2\pi i}=R_-\,,
\quad \frac1{\psi_-}\,\frac {R_+^--R_+^+}{2\pi i}=S_-\,,
$$
where for we denote by
$F^+$ and $F^-$ the boundary values of a function $F$:
$$
F^+(x)=F(x+i0), \qquad F^-(x)=F(x-i0),
$$
see \cite{Er, 6.5(7), 6.8(15), 6.9(4)}.
Comparing these relations with \tht{2.1}, we conclude that, for $x<0$,
$$
\lim_{\varepsilon\to +0}\Rin\left(\tfrac 12+\varepsilon
x\right)=R_-(x),\quad \lim_{\varepsilon\to
+0}\varepsilon^{z+z'}\Sin\left(\tfrac 12+\varepsilon x\right)=S_-(x).
$$
The result now follows from the explicit form of the kernels. \qed
\enddemo

Let $J=(a_1,a_2)\cup\dots\cup(a_{2m-1},a_{2m})$ be a union of disjoint,
possibly infinite, intervals such that the closure of $J$ does not contain the origin. 

\proclaim{Proposition 8.8} The Whittaker kernel satisfies the conditions
\tht{1}-\tht{4}, \tht{5a-d}, \tht{5e'}, \tht{6}, \tht{7} of \S5 with the
matrices $$
C(\ze)=\bmatrix
\ze^{\frac{z+z'}2}e^{\ze/2}&0\\0&\ze^{-\frac{z+z'}2}e^{-\ze/2}\endbmatrix,\quad 
D=-\frac 12\bmatrix 1&0\\0&-1\endbmatrix=-\frac \sig 2\,. 
$$
\endproclaim
\demo{Proof} The proof of (1) is very similar to the case of the ${}_2F_1$ kernel. The kernel $L(x,y)$
has the form
$$
L=\left[\matrix 0 & A\\ -A^* & 0\endmatrix\right]\,,
$$
where 
$A$ is a kernel on $\R_+\times\R_-$ of the form
$$
A(x,y)=\frac{\sqrt{\psi_+(x)\psi_-(y)}}{x-y}=\frac{C(z,z')x^{-\frac{z+z'}2}e^{-x}
(-y)^{\frac{z+z'}2}e^{y}}
{x-y}\,.
$$
The jump condition is verified using the formulas 
\cite{Er, 6.5(7), 6.8(15), 6.9(4)}.
\medskip
\noindent (2) can be either verified in the same way as for the ${}_2F_1$
kernel, or it can be deduced from the fact that the Whittaker kernel is the
correlation kernel of a determinantal point process which has finitely many
particles in $J$ almost surely (see \cite{So, Theorem 4} for the general
theorem about determinantal point processes, and \cite{BO1}, \cite{Bor1}
for the needed property of the Whittaker kernel).

\medskip
\noindent (3) If $|z+z'|<1$ then the kernel $L$ introduced above defines a skew,
bounded operator in $L^2(\x,dx)$, and $K=L(1+L)^{-1}$, see \cite{P.V},
\cite{BO1}. Then, similarly to Corollary 3.7, we can prove that $1-K^J$ is
invertible. 

However, for the restricted operator $K^J$, we can prove the
invertibility of $1-K^J$ for all admissible values of $(z,z')$. The following
argument is due to G.~Olshanski. 

Write $K^J$ in the block form  
$$
K^J=\bmatrix K^J_{+,+}&K^J_{+,-}\\ K^J_{-,+}& K^J_{-,-}\endbmatrix
$$
corresponding to the splitting $J=(J\cap \R_+)\sqcup (J\cap \R_-)$. Since $K$
is a correlation kernel, $K^J_{+,+}$ and $K^J_{-,-}$ are positive definite.
Moreover, $K^J_{-,+}(x,y)=-K^J_{+,-}(y,x)$ by definition of the Whittaker
kernel. Thus, it is enough to prove the invertibility of $1-K^J_{+,+}$ and
$1-K^J_{-,-}$ (see proof of Corollary 3.7). 

We consider $K^J_{+,+}$; the proof for $K^J_{-,-}$ is similar. 
By \cite{So, Theorem 3}, $K_{+,+}\le 1$ and $K^J_{+,+}=K_{+,+}|_J\le 1$. The
only way $K_{+,+}$ can have norm 1 (remember that $K^J_{+,+}$ is of trace
class, hence, it is compact) is that $K_{+,+}$ has an eigenfunction with
eigenvalue 1 which is supported on $J\cap \R_+$. By \cite{P.V, Proposition
3.1}, see also \cite{BO1}, $K_{+,+}=K_{+,+}(x,y)$ commutes with a Sturm-Liouville
operator  
$$
D_x=-\frac d{dx} \, x^2\frac d{dx} +\frac{(z+z'+x)^2}4
$$
in the sense that 
$$
K_{+,+}(x,y)D_y=D_xK_{+,+}(x,y)
$$ 
for all $x,y>0$. Suppose $f\in L^2(\R_+)$ is an eigenfunction of $K_{+,+}$ with eigenvalue 1 and supported in $J\cap \R_+$, i.~e.,
$$
\int_{\R_+}K_{+,+}(x,y)f(y)dy=\int_{J\cap \R_+}K_{+,+}(x,y)f(y)dy=f(x),\quad x>0.
$$
Then using the decay and smoothness properties of $K_{+,+}(x,y)$, which follow easily from the known properties of the Whittaker function, one sees that $D_xf$ also belongs to $L^2(\R_+)$ and
$$
\int_{\R_+} K_{+,+}(x,y)D_yf(y)dy=D_xf(x).
$$
Thus
$$
V=\operatorname{Span}\{D_x^kf\,:\, k\ge 0\}\subset \operatorname{Ker}(1-K_{+,+}^J)\subset L^2(\R_+).
$$
But as $K_{+,+}^J$ is compact, $\dim V<\infty$, and hence $V$ is a finite dimensional invariant subspace for $D_x$. It follows that there exists a nonzero $v\in V$ such that $D_xv=\lambda v$ for some scalar $\lambda$. But as $v\in V$, it must vanish in a neighborhood of $x=0$, which is not possible for nontrivial solutions $v(x)$ of the differential equation
$D_xv=\lambda v$. Thus, we obtain a contradiction, and hence $\Vert K_{+,+}^J\Vert <1$ and $1-K_{+,+}^J$ is invertible. The proof of (3) is now complete.

\medskip
\noindent (4) and (5a-d), (5e') are easily verified. The proofs of \tht{6}
and \tht{7} are similar to the case of the ${}_2F_1$ kernel, and
we do not reproduce them here. \qed
\enddemo

By Theorem 5.3, the matrix $M$ for the Whittaker kernel satisfies the
differential equation 
$$
M'(\ze)=\left(\frac {\ba}{\ze} +\sum_{j=1}^{2m}\frac
{\bc_j}{\ze-a_j}-\frac{\sig}2\right) M(\ze).
$$
The matrices $\{\bc_j\}_{j=1}^{2m}$ are nilpotent (if $a_1=-\infty$ or
$a_{2m}=+\infty$, then $\bc_1=0$ or $\bc_{2m}=0$, respectively), and 
an analog of Proposition 3.3 shows that 
$$
\tr \ba=0, \quad \det \ba=-\left(\frac{z-z'}2\right)^2.
$$  

\proclaim{Theorem 8.9} The Fredholm determinant $\det(1-K|_J)$, where $K$ is
the Whittaker kernel, is the $\tau$-function for the system of Schlesinger
equations
$$
\frac{\partial \ba}{\partial a_j}=\sum_{j=1}^{2m}\frac{[\bc_j,\ba]}{a_j-\frac
12}\,,\qquad  
\frac{\partial \bc_l}{\partial a_j}=\frac{[\ba,\bc_j]}{a_j}-
\sum_{\Sb 1\le l\le 2m\\ l\ne j\endSb}
\frac{[\bc_j,\bc_l]}{a_j-a_l}+\frac {[\bc_j,\sig]}{2}
\,.
\tag 8.1
$$ 
The matrices $\{\bc_j\}_{j=1}^{2m}$ are 
nilpotent (if $a_1=-\infty$ or $a_{2m}=+\infty$, then $\bc_1=0$ or
$\bc_{2m}=0$, respectively), and   
$$
\tr \ba=0, \quad \det \ba=-\left(\frac{z-z'}2\right)^2.
$$  
\endproclaim
\demo{Proof} These results follow from Theorem 6.5. \qed
\enddemo

The next step is to consider $J=(s,+\infty)$, $s>0$. It turns
out that in this case the Schlesinger equations reduce to the $\si$-form of
the Painlev\'e V equation. This reduction can be performed in the spirit of
\S7, following the corresponding part of \cite{JM, Appendix C}. Although we do
not perform the computation here, let us state the result.

\proclaim{Theorem 8.10 \cite{Tr}} Assume that $s>0$. Then the function 
$$
\sigma(s)=s\,\frac{d\ln\det(1-K|_{(s,+\infty)})}{ds}
$$
satisfies the differential equation
$$
(s\si'')^2=(2(\si')^2-s\si'+\si+
(\nu_1+\nu_2+\nu_3+\nu_4)\si')^2-
4(\si'+\nu_1)(\si'+\nu_2)(\si'+\nu_3)(\si'+\nu_4),
\tag 8.2
$$ 
where
$$
\nu_1=\nu_2=0,\quad \nu_3=-z,\quad
\nu_4=-z'\,.
$$
\endproclaim

This result can of course also be obtained from Theorem 7.1 via the limit transition
discussed in Proposition 8.8.

Very much in the same way as the ${}_2F_1$ kernel becomes the Jacobi kernel at
integral values of $z$ and $w$, the Whittaker kernel becomes the Laguerre
kernel if one of the parameters $z$, $z'$ is an integer, see \cite{P.III,
Remark 2.4}. Without giving any details, we formulate the results which can be
obtained using this specialization.

Let $J$ be a proper subset of $\R_+$, whose left endpoint is allowed to
coincide with $0$, and whose right endpoint is allowed to coincide with
$+\infty$. 

\proclaim{Theorem 8.11} Assume that $0<\al<\frac 12$. Then the Fredholm
determinant $\det(1-K_n^{\text{Lag}}|_J)$, where $K_n^{\text{Lag}}$ is the
$n$th Laguerre kernel with parameter $\al$, is the $\tau$-function for the
system of Schlesinger equations \tht{8.1}. The matrices 
$\{\bc_j\}_{j=1}^{2m}$ are nilpotent {\rm(}if $a_1=0$ or $a_{2m}=+\infty$ then
$\bc_1=0$ or $\bc_{2m}=0$, respectively{\rm\,)}, and the eigenvalues of $\ba$ are
equal to $\pm \al/2$. 
\endproclaim

\proclaim{Theorem 8.12 \cite{TW4}} Assume that $s>0$, $\al>-1$. 
Let $K_s$ be the $n$th Laguerre kernel with parameter $\al$ restricted to
either $(0,s)$ or $(s,+\infty)$. Then the function  
$$ 
\sigma(s)=s\,\frac{d\ln\det(1-K_s)}{ds}
$$
satisfies the differential equation \tht{8.2} with
$\nu_1=\nu_2=0$, $\nu_3=n,$ $\nu_4=n+\al$.
\endproclaim

\subhead 8.3. The confluent hypergeometric kernel
\endsubhead
This subsection is based on the following observation.
\proclaim{Proposition 8.13} Set
$$
z=z_0+i\up,\quad z'=\overline{z}_0-i\up,\quad w=w_0+i\up,\quad
w'=\overline{w}_0-i\up.
\tag 8.3 
$$
Then the ${}_2F_1$ kernel $K^F$ has the following scaling limit:
$$
\Cal K(x,y)=\lim_{\up\to+\infty}\up\cdot K^F\left(\frac\up x, \frac \up
y\right),\quad x,y\ne 0, 
$$
where the limit kernel depends on 1 complex parameter
$r={z}_0+\overline{w}_0$, $\Re r>-\frac 12$, and has the form
$$
\gathered
\Cal K(x,y)=\frac 1{2\pi}\,\Gamma\left[\matrix  r+1,\,\bar 
r+1\\
2\Re r+1,\, 2\Re r +2\endmatrix\right]\frac{Q(x)P(y)-P(x)Q(y)}{x-y}\,,
\\
P(x)=\left|2x\right|^{\Re r}
{e^{-ix+\pi\Im r\cdot\sgn(x)/2}}\,{}_1F_1\left[\matrix  r\\ 2\Re r\endmatrix
\,  \Biggl|\,{2i}x\right],\\
Q(x)=2x\,\left|2x\right|^{\Re r}
{e^{-ix+\pi\Im r\cdot\sgn(x)/2}}\,{}_1F_1\left[\matrix  r+1\\ 2\Re
r+2\endmatrix \,  \Biggl|\,{2i}x\right].
\endgathered
$$
\endproclaim
\noindent Here ${}_1F_1\left[\matrix a\\c\endmatrix\Bigl|\, x\,\right]$ is the
confluent hypergeometric function also denoted as
$\Phi(a,c;x)$, see \cite{Er, 6.1}.

The determinantal point process with the correlation kernel $\Cal K(x,y)$
describes the decomposition of a remarkable family of measures on
infinite Hermitian matrices on the ergodic (with respect to the $U(\infty)$
action) measures, see \cite{BO4}. We will call $\Cal K(x,y)$ the {\it
confluent hypergeometric kernel}.

%It can also be expressed through the
%M-Whittaker functions (see \cite{Er, 6.9}) using the formulas
%$$
%P(x)=e^{-\frac{i\pi\bar{s}\,\sgn(x)}2}M_{\frac{\bar{s}-s}2,\frac{s+\bar{s}-1}2} 
%\left({2i}x\right),\quad 
%Q(x)=e^{-\frac{i\pi(\bar{s}+1)\sgn(x)}2}M_{\frac{\bar{s}-s}2,\frac{s+\bar{s}+1}2}
%\left({2i}x\right).
%$$

For real values of $r$ this kernel was obtained in \cite{WF} as a scaling
limit of Christoffel--Darboux kernels for a certain system of orthogonal
polynomials (called the {\it pseudo-Jacobi polynomials}). For complex values of
$r$ such limit transition can be carried out as well, see \cite{BO4, \S2}.

\demo{Proof of Proposition 8.13} This is a direct computation. The relevant limit relation
for the hypergeometric functions in this case,
has the form
$$
\lim_{\up\to+\infty} {}_2F_1\left[\matrix a,\,b+2i\up\\ c\endmatrix
\,\Biggl|\,  \left(\frac 12-\frac \up x\right)^{-1}\right] =
{}_1F_1\left[\matrix a\\c\endmatrix\,\Bigl| 2ix\right]. \qed
$$
\enddemo

The determinantal point process defined by $\Cal K$ has locally finite point
configurations almost surely, see \cite{BO4}. Hence (\cite{So, Theorem
4}), the restriction  $\Cal K_t=\Cal K|_{(0,t)}$ to any finite interval $(0,t)$
defines an operator of trace class, and $\det(1-\Cal K_t)$ is well-defined. It
is natural to conjecture that this Fredholm determinant satisfies a differential
equation obtained by taking the corresponding scaling limit of the Painlev\'e
VI equation of Theorem 7.1. 

In the Proposition below we check that the limit of the differential
equation exists, and we observe that it is a $\si$-form of the Painlev\'e V
equation. In \cite{WF} it was proved that for the real values of $r$ the
determinant  $\det(1-\Cal K_t)$ does indeed satisfy this equation. The justification
of this statement for all values of $r$ in our setup requires a proof that the
corresponding restriction of the ${}_2F_1$ kernel converges to $\Cal K_t$ in
trace norm, and we leave this technical issue aside here. 

\proclaim{Proposition 8.14} Under the change of parameters \tht{8.3}, and the
change of the independent variable $s=\up/t$, equation \tht{7.1}
converges to the following $\si$-version of the Painlev\'e V:  
$$
-(t\wt\si'')^2=(2(t\wt\si'-\wt\si)+(\wt\si')^2+i(\bar{r}-r)\wt\si')^2-
(\wt\si')^2(\wt\si'-2ir)(\wt\si'+2i\bar{r}),
\tag 8.4
$$
where $r=z_0+\overline{w}_0$ and
$$
\si(s)=-\frac
\up t\left(\wt\si(t)-\frac{i(r-\bar{r})}{2}\,t+\frac{(r+\bar{r})^2}4\right).
$$
\endproclaim
\demo{Proof} We derive \tht{8.4} from \tht{7.1}, assuming that certain limits exist, as noted above. Keeping in mind the relation $s=\up/t$, we have
$$
\gathered
\si(s)=\left(\frac\up t-\frac 12\right)\left(\frac\up t+\frac 12\right)
\left(-\frac {t^2}{\up}\frac
d{dt}\right)\ln\det\left(1-K^{F}_{(\up/t,+\infty)}\right)\\-
\frac{(r+\bar{r})^2}4\frac \up t+\frac
12\left(2i\up+\frac{z_0+w_0-\overline{z}_0-\overline{w}_0}2\right)
\frac{r-\bar{r}}2\\
=\up\left(-\frac {d\ln\det(1-\Cal K|_t)}{dt}-
\frac{(r+\bar{r})^2}{4t}+\frac{i(r-\bar{r})}2\right)+O(1).
\endgathered 
$$
Hence, anticipating that $\wt \si(t)=t\dfrac {d\ln\det(1-\Cal K|_t)}{dt}+o(1)$
as $\up\to+\infty$, we define $\wt\si(t)$ by the relation
$$
\si(s)=-\frac
\up
t\left(\wt\si(t)-\frac{i(r-\bar{r})}{2}\,t+\frac{(r+\bar{r})^2}4\right),
$$ 
which leads to
$$
\gathered
\frac {d\si(s)}{ds}=\frac{t^2}{\up}\frac d{dt}\left(\frac
\up
t\left(\wt\si(t)-\frac{i(r-\bar{r})}{2}\,t
+\frac{(r+\bar{r})^2}4\right)\right)
=t\frac{d\wt\si(t)}{dt}-\si(t)-\frac{(r+\bar{r})^2}4,\\
s\frac
{d\si(s)}{ds}-\si(s)=\up\left(\frac{d\wt\si(t)}{dt}-\frac{i(r-\bar{r})}2\right),
\endgathered
$$
$$
\gathered
\frac {d^2\si(s)}{ds^2}=-\frac{t^2}{\up}\frac d{dt}
\left(t\frac{d\wt\si(t)}{dt}-\wt\si(t)-\frac{(r+\bar{r})^2}4\right)
=-
\frac{t^3}{\up}\frac {d^2\wt\si(t)}{dt^2},\\
\left(s-\frac 12\right)\left(s+\frac 12\right)\frac {d^2\si(s)}{ds^2}=
-\up t \frac {d^2\wt\si(t)}{dt^2} +O(1).
\endgathered  
$$
Substituting these relations into the equation \tht{7.1} and passing to the limit
$\up\to+\infty$ we obtain 
$$
\gathered
-\left(t\frac{d\wt\si(t)}{dt}-\wt\si(t)-\frac{(r+\bar{r})^2}4\right)
\left(t \frac {d^2\wt\si(t)}{dt^2}\right)^2\\=
\Biggl(2\left(\frac{d\wt\si(t)}{dt}-\frac{i(r-\bar{r})}2\right)
\left(t\frac{d\wt\si(t)}{dt}-\wt\si(t)-\frac{(r+\bar{r})^2}4\right)
-2i\,\frac{(r+\bar{r})^2}4\frac{(r-\bar{r})}2\Biggr)^2\\
+4\left(t\frac{d\wt\si(t)}{dt}-\wt\si(t)\right)^2
\left(t\frac{d\wt\si(t)}{dt}-\wt\si(t)-\frac{(r+\bar{r})^2}4+
\frac{(r-\bar{r})^2}4\right).
\endgathered
$$
It is readily seen that the right-hand side of this equation is
divisible by
$$\left(t\frac{d\wt\si(t)}{dt}-\wt\si(t)-\frac{(r+\bar{r})^2}4\right),$$ and
after this cancellation the equation exactly coincides with \tht{8.4}.
\qed 
\enddemo 

\example{Remark 8.15}

For $r=0$ the kernel $\Cal K$ becomes the sine kernel
${\sin(x-y)}/{\pi(x-y)}\,,$
see \cite{BO4, \S2}. 
Accordingly, equation \tht{8.4} takes the form 
$$
-(t\wt\si'')^2=4(t\wt\si'-\wt\si)(t\wt\si'-\wt\si+(\wt\si')^2).
\tag{8.5}
$$
This agrees with the celebrated result of
\cite{JMMS}, which states that equation \tht{8.5} is satisfied by the
function
$$
\wt\si(t)=t\,\frac
d{dt}\,\ln\det\left(1-\frac{\sin(x-y)}{\pi(x-y)}\Biggr|_{(0,t)}\right).
$$
\endexample
\head 9. Differential equations: a general approach
\endhead

Lemmas 4.3, 4.4 point to a general method for proving that a wide class of determinants satisfy Painlev\'e equations (see the Introduction and \cite{TW1--4}). We illustrate the method in the case of the Airy kernel
$$
A(x,y)=\frac{Ai(x)Ai'(y)-Ai(x)Ai'(y)}{x-y}
$$
where $Ai(x)$ is the well-known Airy function. This kernel arises in random matrix theory (\cite{F}, \cite{TW2}) and plays a central role in the interaction of combinatorics and random matrix theory
(see e.g. \cite{BDJ1}, \cite{BDJ2}, \cite{Ok}, \cite{BOO}, \cite{J}).

For $s\in \R$, let $A_s$ denote the operator obtained by restricting the kernel $A(x,y)$ to $L^2(s,+\infty)$. The basic result of Tracy and Widom \cite{TW2} is that 
$$
-\frac{d^2}{ds^2}\ln\det(1-A_s)=u^2(s),
$$
 where $u(s)$ solves the Painlev\'e II equation
$$
u''=2u^3+su
$$
with initial conditions
$$
u(s)\sim -Ai(s) \ \text{  as  } s\to+\infty.
$$ 

We now outline a proof of this fact using Lemmas 4.3, 4.4. It will be clear to the reader that the method extends, in particular, to the general class of kernels considered in \cite{TW4}.

In the notation of Lemmas 4.3, 4.4, let $\Sigma=\R$, $\Sigma_2=J=(s,+\infty)$, and $\Sigma_1=\Sigma\setminus \Sigma_2=(-\infty,s]$. Let $B(\ze)$ be a $2\times 2$ fundamental solution of the differential equation
$$
\frac{dB(\ze)}{d\ze}=\bmatrix 0&\ze\\1&0\endbmatrix B(\ze),\qquad \det B(\ze)\equiv 1,
$$
with $B_{11}(\ze)=Ai'(\ze)$, $B_{21}(\ze)=Ai(\ze)$. Set
$$
m(\zeta)=\cases B(\zeta),&\Im\ze>0,\\B(\zeta)
\bmatrix 1&-2\pi i\\0&1\endbmatrix,&\Im \ze<0.
\endcases
$$
Then $m$ satisfy the jump relation $m_+(x)=m_-(x)v(x)$, $x\in \R$, where
$$
v(x)\equiv \bmatrix 1&2\pi i\\0&1\endbmatrix.
$$
Set 
$$
f(x)\equiv\bmatrix 1\\0\endbmatrix,\quad g(x)\equiv\bmatrix 0\\1\endbmatrix,\qquad
x\in\R,
$$
and note that $v=I+2\pi i fg^t$. Also 
$$
F(x)=m(x)f(x)=\bmatrix Ai'(x)\\ Ai(x)\endbmatrix,\quad
G(x)=m^{-t}(x)g(x)=\bmatrix -Ai(x)\\Ai'(x)\endbmatrix,
$$
and we set 
$$
v_2(x)=I-2\pi i F(x) G^t(x),\qquad x\in\Sigma_2=(s,+\infty).
$$
Let $m_s$ solve the normalized RHP $(\Sigma_2,v_2)$, $m_s(\ze)\to I$ as $\ze\to\infty$. By Lemma 4.4,
$$
M(\ze)\equiv m_s(\ze)m(\ze)
$$
solves the ``simple'' jump relation $M_+=M_-v$ on $\Sigma_1$. Standard arguments as in Theorem 5.1 and Proposition 6.2 now imply that $M$ satisfies the Lax pair
$$
\gather
\frac{dM(\ze)}{d\ze}=\left(
\bmatrix a&\ze+b\\1&-1\endbmatrix +\frac 1{\ze-s}\bmatrix p&q\\r&-p\endbmatrix \right)M(\ze),\tag 9.1\\
\frac{dM(\ze)}{ds}=-\frac{1}{\ze-s}\bmatrix p&q\\r&-p\endbmatrix M(\ze),
\tag 9.2
\endgather
$$
where $\bmatrix p&q\\r&-p\endbmatrix$ is nilpotent. Here $a,\,b,\,p,\,q,\,r$ are suitable constants which depend only on $s$.
By \tht{6.9},
$$
\frac d{ds}\ln\det(1-A_s)=(H^{-1}(s)H'(s))_{21},
$$
where the prime refers to the derivative with respect to $\ze$, and
$$
M(\ze)=H(\ze)\bmatrix 1 & 2\pi i\ln(\ze-s)\\ 0&1\endbmatrix.
\tag 9.3
$$
As noted, in \S4 (see \tht{4.3}), $\det H(\ze)\equiv 1$ and $H(\ze)$ is analytic near $\ze=s$ (in fact, $H(\ze)$ is entire). Using \tht{9.1} and \tht{9.3}, we find
$$
M'M^{-1}=H'H^{-1}+\frac 1{\ze-s}\,H\bmatrix 0&1\\0&0\endbmatrix H^{-1}=
\bmatrix a&\ze+b\\1&-1\endbmatrix +\frac 1{\ze-s}\bmatrix p&q\\r&-p\endbmatrix
$$
which leads to the relation
$$
(H^{-1}(s)H'(s))_{21}=2ap+q+(s+b)r.
$$ 
The compatibility of the Lax-pair equations \tht{9.1}, \tht{9.2} yields
(\cite{KH, \S2}) the relations $2ap+q+(s+b)r=a$, $\frac {da}{ds}=r$, where
$r$ solves the Painlev\'e 34 equation
$$
\frac {d^2r}{ds^2}=\frac{1}{2r}\left(\frac {dr}{ds}\right)^2-4r^2+2sr.
$$
Writing $r=-u^2$, a simple calculation shows that $u$ solves the Painlev\'e II equation. This verifies the above claim for
$$
-\frac {d^2}{ds^2}\ln\det(1-A_s)=-\frac{da}{ds}=-r=u^2.
$$

Note that one can also show that $\det(1-A_s)$ is the $\tau$-function for the isomonodromy deformation described by \tht{9.1}, \tht{9.2}.

\head Appendix. Integrable operators and Riemann-Hilbert problems
\endhead

This appendix contains a brief summary of results on integrable operators and
corresponding Riemann-Hilbert problems which can be found in \cite{IIKS},
\cite{KBI}, \cite{De}.

Let $\Sigma$ be an oriented contour in $\C$. We call an operator $L$ acting 
in $L^2(\Sigma,|d\zeta|)$ {\it integrable} if its kernel has the form
$$
L(\zeta,\zeta')=\frac {\sum_{j=1}^N f_j(\zeta)g_j(\zeta')}{\zeta-\zeta'},\quad 
\zeta,\ \zeta'\in\Sigma, 
$$
for some functions $f_j,\ g_j$, $j=1,\dots,N$. We shall always assume that 
$$
\sum_{j=1}^N f_j(\zeta)g_j(\zeta)=0,\quad \zeta\in\Sigma,
$$ 
so that the
kernel $L(\zeta,\zeta')$ is nonsingular (this assumption is not necessary for
the general theory).

We do not impose here any restrictions on the functions $f_i,g_i$ and on the contour 
$\Sigma$. For our
purposes it suffices to assume that $\Sigma$ is a finite union of
disjoint (possibly infinite) intervals on the real line, $f_i, g_i$ are
smooth functions inside $\s$, and  
$$
f_i,g_i\in L^p(\Sigma,|d\ze|)\cap
L^{\infty}(\Sigma,|d\ze|) \quad\text{  for some  }\  p,\  1<p<+\infty.
\tag A.1
$$
These restrictions guarantee, in particular, that $L$ is a bounded operator in $L^2(\Sigma)$.

Particular examples of integrable operators appeared in the mathematical phy\-sics literature a long time ago. However,
integrable operators were first singled out as a distinguished class in  \cite{IIKS}.

It turns out that for an integrable operator $L$ such that $(1+L)^{-1}$ exists,
the operator  $K=L(1+L)^{-1}$ is also integrable.

\proclaim{Proposition A.1 [IIKS]} Let $L$ be an integrable operator as
described above and $K=L(1+L)^{-1}$. Then the kernel $K(\zeta,\zeta')$ has
the form 
$$
K(\zeta,\zeta')=\frac {\sum_{j=1}^N F_j(\zeta)G_j(\zeta')}{\zeta-\zeta'},
\quad \zeta,\zeta'\in\Sigma, 
$$
where
$$
F_j=(1+L)^{-1}f_j,\qquad G_j=(1+L^t)^{-1}g_j,\quad j=1,\dots,N.
$$
If $\sum_{j=1}^N f_j(\zeta)g_j(\zeta)=0$ on $\Sigma$, then 
$\sum_{j=1}^N F_j(\zeta)G_j(\zeta)=0$ on $\Sigma$ as well.
\endproclaim

A remarkable fact is that $F_j$ and $G_j$ can be expressed through a solution of
an associated Riemann--Hilbert problem (RHP, for short). 

Let $v$ be a map from $\Sigma$ to $\Ma(k,\C)$, where $k$ is a fixed integer.

We say that a matrix function $m:\C\setminus\Sigma\to \Ma(k,\C)$ is a
solution of the RHP $(\Sigma,v)$ if the following conditions are satisfied 
$$
\align
&\bullet\quad  m(\zeta)\text{ is analytic in }\Bbb C\setminus \Sigma,\\
&\bullet\quad m_+(\zeta)=m_-(\zeta)v(\zeta),\  \zeta\in\Sigma,\ \text{where }
m_\pm(\zeta)=\lim_{\Sb \zeta'\to \zeta\\ \zeta'\in
(\pm)\text{-side}\endSb}m(\zeta'),
\endalign
$$
If in addition
$$
\bullet\quad m(\zeta)\to I \text{ as }
\zeta\to \infty,\qquad\qquad\qquad\qquad\qquad\qquad\qquad\qquad\qquad
$$
we say that $m$ solve the {\it normalized} RHP $(\Sigma, v)$.

The matrix $v(\zeta)$ is called the {\it jump matrix}.

\proclaim{Proposition A.2 [IIKS]} Let $L$ be an integrable operator as 
described above such that the operator $1+L$ is invertible. 
Then there exists a unique solution $m(\zeta)$ of the normalized RHP
$(\Sigma, v)$ with 
$$ 
v(\zeta)=I+2\pi i\,f(\zeta)g(\zeta)^t\in \Ma(N,\C),
$$
where
$$
f=\left(f_1,\dots,f_N\right)^t,\qquad
g=\left(g_1,\dots,g_N\right)^t,
$$
and the kernel of the operator $K=L(1+L)^{-1}$ has the form
$$
K(\zeta,\zeta')=\frac { G^t(\zeta')F(\zeta)}{\zeta-\zeta'},\quad
\zeta,\zeta'\in\Sigma, 
$$
where 
$$
F=\left(F_1,\dots,F_N\right)^t,\qquad
G=\left(G_1,\dots,G_N\right)^t
$$
are given by
$$
F(\zeta)=m_+(\zeta)\, f(\zeta)=m_-(\zeta)\, f(\zeta),
\qquad 
G(\zeta)=m_+^{-t}(\zeta)\, g(\zeta)=m_-^{-t}(\zeta)\, g(\zeta).
$$
\endproclaim

In other words, the inverse $(1+L)^{-1}$ of $1$ plus an integrable operator can be expressed in terms of the solution of an associated problem in complex variables.

The function $m(\ze)$ may have singularities at the points of discontinuity of
the jump matrix $v$ (e.g., at the endpoints of $\s$). Unless specified
otherwise, we assume that $m(\ze)$ belongs to the $L^2$-space locally on any 
smooth curve passing through the singular point. Under our restrictions on
$f_i, g_i,$ and $\s$, see above, the solution $m(\ze)$ in Proposition A.2
satisfies this condition. 

A discrete version of the Propositions A.1 and A.2 is given in \cite{Bor2}. 

Let now $\Sigma=\Sigma_I\cup\Sigma_{II}$ be a union of two contours. Assume that
the operator $L$ in the block form corresponding to this splitting is as
follows 
$$ 
L(x,y)=\left[\matrix 0&\frac {h_{I}(x)h_{II}(y)}{x-y}\\
\frac {h_{I}(y)h_{II}(x)}{x-y}&0\endmatrix\right]
$$
for some functions $h_I(\,\cdot\,)$ and $h_{II}(\,\cdot\,)$ defined on 
$\Sigma_I$ and $\Sigma_{II}$, respectively.

Then the operator $L$ is integrable with $N=2$. Indeed,
$$
L(x,y)=\frac{f_1(x)g_1(y)+f_2(x)g_2(y)}{x-y},\qquad x,y\in \Sigma,
$$
where
$$
f_1(x)=g_2(x)=\cases h_I(x),&x\in\Sigma_I,\\0,&x\in\Sigma_{II},\endcases\quad
f_2(x)=g_1(x)=\cases 0,&x\in\Sigma_I,\\h_{II}(x),&x\in\Sigma_{II}.\endcases
$$
The jump matrix $v(x)$ of the corresponding RHP has the form
$$
v(x)=\cases \bmatrix1&2\pi i\,h_I^2(x)\\0&1\endbmatrix,&
x\in\Sigma_I,\\
\bmatrix 1&0\\2\pi i\, h_{II}^2(x)&1\endbmatrix,&x\in\Sigma_{II}.
\endcases
$$
It can be easily seen that the RHP in such a situation is equivalent to the 
following set of conditions:

\noindent$\bullet$  matrix elements $m_{11}$ and $m_{21}$ are 
holomorphic in $\C\setminus\Sigma_{II}$; 

\noindent$\bullet$  matrix elements $m_{12}$ and $m_{22}$ are 
holomorphic in $\C\setminus\Sigma_{I}$; 

\noindent$\bullet$  on $\Sigma_{II}$ the following relations hold
$$
\gathered
{m_{11}}_+(x)-{m_{11}}_-(x)=2\pi i\,h_I^2(x)m_{12}(x),\\
{m_{21}}_+(x)-{m_{21}}_-(x)=2\pi i\,h_I^2(x)m_{22}(x);
\endgathered
$$
\noindent$\bullet$  on $\Sigma_{I}$ the following relations hold
$$
\gathered
{m_{12}}_+(x)-{m_{12}}_-(x)=2\pi i\,h_{II}^2(x)m_{11}(x),\\
{m_{22}}_+(x)-{m_{22}}_-(x)=2\pi i\,h_{II}^2(x)m_{21}(x);
\endgathered
$$

\noindent$\bullet$ $m(x)\sim I$ as $x\to\infty$.

According to Proposition A.2, the kernel $K(x,y)$ in block form 
corresponding to the splitting $\Sigma=\Sigma_I\cup\Sigma_{II}$ is given by
$$
\gathered
K(x,y)\\ =\bmatrix
\frac{h_I(x)h_I(y)(-m_{11}(x)m_{21}(y)+m_{21}(x)m_{11}(y))}
{x-y}&
\frac{h_I(x)h_{II}(y)(m_{11}(x)m_{22}(y)-m_{21}(x)m_{12}(y))}
{x-y}\\
\frac{h_{II}(x)h_I(y)(m_{22}(x)m_{11}(y)-m_{12}(x)m_{21}(y))}
{x-y}&
\frac{h_{II}(x)h_{II}(y)(-m_{22}(x)m_{12}(y)+m_{12}(x)m_{22}(y))}
{x-y}
\endbmatrix.
\endgathered
\tag A.2
$$

\enddocument